\input harvmac

\def\Title#1#2{\rightline{#1}\ifx\answ\bigans\nopagenumbers\pageno0\vskip1in
\else\pageno1\vskip.8in\fi \centerline{\titlefont #2}\vskip .5in}

scaled\magstep3 
 
scaled\magstep3 
 
scaled\magstep3 

\font\bbbi=msbm10
\def\bbb#1{\hbox{\bbbi #1}}
%
%
\font\ticp=cmcsc10
\def\undertext#1{$\underline{\smash{\hbox{#1}}}$}
\def\sq{{\vbox {\hrule height 0.6pt\hbox{\vrule width 0.6pt\hskip 3pt
   \vbox{\vskip 6pt}\hskip 3pt \vrule width 0.6pt}\hrule height 0.6pt}}}
\def\kten{\kappa_{10}}
\def\subsubsec#1{\noindent{\undertext { #1}}}

\def\tg{ { \tilde{g} } }
\def\tgmn{ { \tilde{g}_{mn} } }
\def\d{ {\delta} }

\def\tnab{ {\tilde{\nabla}} }

\def\tstar{ {\tilde{*}} }
\def\hf{{1\over 2}}
\def\varI{\delta_I}
\def\calL{{\cal L}}

\def\cald{{\cal D}}
\def\hf{{1\over 2}}

\def\charge{ {4 \pi  N\alpha^{\prime 2}} }
\def\chargeb{ {48 \pi  N\alpha^{\prime 2}} }
\def\calo{{\cal O}}
\def\tJ{{\tilde J}}

\overfullrule=0pt
%
%
\lref\DvTy{
  G.~R.~Dvali and S.~H.~H.~Tye,
  ``Brane inflation,''
  Phys.\ Lett.\ B {\bf 450}, 72 (1999)
  [arXiv:hep-ph/9812483].
}
\lref\CQS{
  J.~P.~Conlon, F.~Quevedo and K.~Suruliz,
  ``Large-volume flux compactifications: Moduli spectrum and D3/D7 soft
  supersymmetry breaking,''
  arXiv:hep-th/0505076.
}
\lref\Saulina{
  N.~Saulina,
  ``Topological constraints on stabilized flux vacua,''
  arXiv:hep-th/0503125.
}
\lref\Kalloshetal{
  R.~Kallosh, A.~K.~Kashani-Poor and A.~Tomasiello,
  ``Counting fermionic zero modes on M5 with fluxes,''
  arXiv:hep-th/0503138.
}
\lref\Kachruetal{
  L.~Gorlich, S.~Kachru, P.~K.~Tripathy and S.~P.~Trivedi,
  ``Gaugino condensation and nonperturbative superpotentials in flux
  compactifications,''
  arXiv:hep-th/0407130.
}
\lref\Wittinst{
  E.~Witten,
  ``Non-Perturbative Superpotentials In String Theory,''
  Nucl.\ Phys.\ B {\bf 474}, 343 (1996)
  [arXiv:hep-th/9604030].
}
\lref\KPV{
  S.~Kachru, J.~Pearson and H.~L.~Verlinde,
  ``Brane/flux annihilation and the string dual of a non-supersymmetric  field
  theory,''
  JHEP {\bf 0206}, 021 (2002)
  [arXiv:hep-th/0112197].
}
\lref\FLW{
  A.~R.~Frey, M.~Lippert and B.~Williams,
  ``The fall of stringy de Sitter,''
  Phys.\ Rev.\ D {\bf 68}, 046008 (2003)
  [arXiv:hep-th/0305018].
}
\lref\BHL{
  K.~Becker, M.~Becker, M.~Haack and J.~Louis,
  ``Supersymmetry breaking and alpha'-corrections to flux induced
  potentials,''
  JHEP {\bf 0206}, 060 (2002)
  [arXiv:hep-th/0204254].
}
\lref\GiMe{
  S.~B.~Giddings and R.~C.~Myers,
  ``Spontaneous decompactification,''
  Phys.\ Rev.\ D {\bf 70}, 046005 (2004)
  [arXiv:hep-th/0404220].
}
\lref\fate{
  S.~B.~Giddings,
  ``The fate of four dimensions,''
  Phys.\ Rev.\ D {\bf 68}, 026006 (2003)
  [arXiv:hep-th/0303031].
}
\lref\LMW{
  F.~Leblond, R.~C.~Myers and D.~J.~Winters,
  ``Consistency conditions for brane worlds in arbitrary dimensions,''
  JHEP {\bf 0107}, 031 (2001)
  [arXiv:hep-th/0106140].
}
\lref\KlSt{
  I.~R.~Klebanov and M.~J.~Strassler,
  ``Supergravity and a confining gauge theory: Duality cascades and
  chiSB-resolution of naked singularities,''
  JHEP {\bf 0008}, 052 (2000)
  [arXiv:hep-th/0007191].
}
\lref\CKLT{
  G.~Curio, A.~Klemm, D.~Lust and S.~Theisen,
  ``On the vacuum structure of type II string compactifications on  Calabi-Yau
  spaces with H-fluxes,''
  Nucl.\ Phys.\ B {\bf 609}, 3 (2001)
  [arXiv:hep-th/0012213].
}
\lref\Quev{
  F.~Quevedo,
  ``Lectures on string / brane cosmology,''
  Class.\ Quant.\ Grav.\  {\bf 19}, 5721 (2002)
  [arXiv:hep-th/0210292].
}
\lref\TaVa{
  T.~R.~Taylor and C.~Vafa,
  ``RR flux on Calabi-Yau and partial supersymmetry breaking,''
  Phys.\ Lett.\ B {\bf 474}, 130 (2000)
  [arXiv:hep-th/9912152].
}
\lref\GVW{
  S.~Gukov, C.~Vafa and E.~Witten,
  ``CFT's from Calabi-Yau four-folds,''
  Nucl.\ Phys.\ B {\bf 584}, 69 (2000)
  [Erratum-ibid.\ B {\bf 608}, 477 (2001)]
  [arXiv:hep-th/9906070].
}
\lref\CPV{
  C.~S.~Chan, P.~L.~Paul and H.~L.~Verlinde,
  ``A note on warped string compactification,''
  Nucl.\ Phys.\ B {\bf 581}, 156 (2000)
  [arXiv:hep-th/0003236].
}
\lref\HVer{
  H.~L.~Verlinde,
  ``Holography and compactification,''
  Nucl.\ Phys.\ B {\bf 580}, 264 (2000)
  [arXiv:hep-th/9906182].
}
\lref\GHK{
  S.~S.~Gubser, C.~P.~Herzog and I.~R.~Klebanov,
  ``Symmetry breaking and axionic strings in the warped deformed conifold,''
  JHEP {\bf 0409}, 036 (2004)
  [arXiv:hep-th/0405282].
}
\lref\Kalloshtalk{R. Kallosh's talk at the workshop {\it Superstring cosmology}, KITP, 2003,\hfil\break http://online.itp.ucsb.edu/online/strings03/kallosh/ .
}
\lref\Kalloshrho{
  J.~P.~Hsu, R.~Kallosh and S.~Prokushkin,
  ``On brane inflation with volume stabilization,''
  JCAP {\bf 0312}, 009 (2003)
  [arXiv:hep-th/0311077].
}
\lref\Dealwistwo{
  S.~P.~de Alwis,
  ``Brane worlds in 5D and warped compactifications in IIB,''
  Phys.\ Lett.\ B {\bf 603}, 230 (2004)
  [arXiv:hep-th/0407126].
}
\lref\Dealwisone{
  S.~P.~de Alwis,
  ``On potentials from fluxes,''
  Phys.\ Rev.\ D {\bf 68}, 126001 (2003)
  [arXiv:hep-th/0307084].
}
\lref\MaNu{
  J.~M.~Maldacena and C.~Nunez,
  ``Supergravity description of field theories on curved manifolds and a no  go
  theorem,''
  Int.\ J.\ Mod.\ Phys.\ A {\bf 16}, 822 (2001)
  [arXiv:hep-th/0007018].
}
\lref\DSH{
  B.~de Wit, D.~J.~Smit and N.~D.~Hari Dass,
  Nucl.\ Phys.\ B {\bf 283}, 165 (1987).
}
\lref\DeGi{
  O.~DeWolfe and S.~B.~Giddings,
  ``Scales and hierarchies in warped compactifications and brane worlds,''
  Phys.\ Rev.\ D {\bf 67}, 066008 (2003)
  [arXiv:hep-th/0208123].
}
\lref\DiLa{
  S.~Dimopoulos and G.~Landsberg,
  ``Black holes at the LHC,''
  Phys.\ Rev.\ Lett.\  {\bf 87}, 161602 (2001)
  [arXiv:hep-ph/0106295].
}
\lref\GiTh{
  S.~B.~Giddings and S.~Thomas,
  ``High energy colliders as black hole factories: The end of short  distance
  physics,''
  Phys.\ Rev.\ D {\bf 65}, 056010 (2002)
  [arXiv:hep-ph/0106219].
}
\lref\CMP{
  E.~J.~Copeland, R.~C.~Myers and J.~Polchinski,
  ``Cosmic F- and D-strings,''
  JHEP {\bf 0406}, 013 (2004)
  [arXiv:hep-th/0312067].
}
\lref\SussL{
  L.~Susskind,
  ``The anthropic landscape of string theory,''
  arXiv:hep-th/0302219.
}
\lref\BoPo{
  R.~Bousso and J.~Polchinski,
  ``Quantization of four-form fluxes and dynamical neutralization of the
  cosmological constant,''
  JHEP {\bf 0006}, 006 (2000)
  [arXiv:hep-th/0004134].
}
\lref\GKP{
  S.~B.~Giddings, S.~Kachru and J.~Polchinski,
  ``Hierarchies from fluxes in string compactifications,''
  Phys.\ Rev.\ D {\bf 66}, 106006 (2002)
  [arXiv:hep-th/0105097].
}
\lref\KKLT{
  S.~Kachru, R.~Kallosh, A.~Linde and S.~P.~Trivedi,
  ``De Sitter vacua in string theory,''
  Phys.\ Rev.\ D {\bf 68}, 046005 (2003)
  [arXiv:hep-th/0301240].
}
\lref\KKLMMT{
  S.~Kachru, R.~Kallosh, A.~Linde, J.~Maldacena, L.~McAllister and S.~P.~Trivedi,
  ``Towards inflation in string theory,''
  JCAP {\bf 0310}, 013 (2003)
  [arXiv:hep-th/0308055].
}
\lref\GrayVW{
J.~Gray and A.~Lukas,
``Gauge five brane moduli in four-dimensional heterotic models,''
arXiv:hep-th/0309096.
}
\lref\GrimmUQ{
T.~W.~Grimm and J.~Louis,
``The effective action of N = 1 Calabi-Yau orientifolds,''
arXiv:hep-th/0403067.
}
\lref\GiddingsVR{
S.~B.~Giddings and R.~C.~Myers,
``Spontaneous decompactification,''
arXiv:hep-th/0404220.
}
\lref\BrandtUW{
F.~Brandt,
``New N = 2 supersymmetric gauge theories: The double tensor multiplet  and its
interactions,''
Nucl.\ Phys.\ B {\bf 587}, 543 (2000)
[arXiv:hep-th/0005086].
}
\lref\TheisJJ{
U.~Theis and S.~Vandoren,
``N = 2 supersymmetric scalar-tensor couplings,''
JHEP {\bf 0304}, 042 (2003)
[arXiv:hep-th/0303048].
}
\lref\GrimmUQ{
T.~W.~Grimm and J.~Louis,
``The effective action of N = 1 Calabi-Yau orientifolds,''
arXiv:hep-th/0403067.
}
\lref\DallAgataYR{
G.~Dall'Agata, R.~D'Auria, L.~Sommovigo and S.~Vaula,
``D = 4, N = 2 gauged supergravity in the presence of tensor multiplets,''
Nucl.\ Phys.\ B {\bf 682}, 243 (2004)
[arXiv:hep-th/0312210].
}
\lref\HerzogXK{
  C.~P.~Herzog, I.~R.~Klebanov and P.~Ouyang,
  ``Remarks on the warped deformed conifold,''
  arXiv:hep-th/0108101.
}
\lref\CveticDB{
  M.~Cvetic, G.~W.~Gibbons, H.~Lu and C.~N.~Pope,
  ``Ricci-flat metrics, harmonic forms and brane resolutions,''
  Commun.\ Math.\ Phys.\  {\bf 232}, 457 (2003)
  [arXiv:hep-th/0012011].
}
\lref\MDone{
  M.~R.~Douglas,
  ``The statistics of string / M theory vacua,''
  JHEP {\bf 0305}, 046 (2003)
  [arXiv:hep-th/0303194].
}
\lref\MDtwo{
  S.~Ashok and M.~R.~Douglas,
  ``Counting flux vacua,''
  JHEP {\bf 0401}, 060 (2004)
  [arXiv:hep-th/0307049].
}
\lref\MDthree{
  F.~Denef and M.~R.~Douglas,
  ``Distributions of flux vacua,''
  JHEP {\bf 0405}, 072 (2004)
  [arXiv:hep-th/0404116].
}
\lref\MDfour{
  M.~R.~Douglas,
  ``Statistical analysis of the supersymmetry breaking scale,''
  arXiv:hep-th/0405279.
}
\lref\MDfive{
  F.~Denef and M.~R.~Douglas,
  ``Distributions of nonsupersymmetric flux vacua,''
  JHEP {\bf 0503}, 061 (2005)
  [arXiv:hep-th/0411183].
}
\lref\BHK{
  M.~Berg, M.~Haack and B.~Kors,
  ``Loop corrections to volume moduli and inflation in string theory,''
  Phys.\ Rev.\ D {\bf 71}, 026005 (2005)
  [arXiv:hep-th/0404087].
}
\lref\BHKIP{
  M.~Berg, M.~Haack and B.~Kors, to appear.
}
\lref\BHKtwo{
  M.~Berg, M.~Haack and B.~Kors,
  ``On the moduli dependence of nonperturbative superpotentials in brane
  inflation,''
  arXiv:hep-th/0409282.
}
\lref\BuRo{
  A.~Buchel and R.~Roiban,
  ``Inflation in warped geometries,''
  Phys.\ Lett.\ B {\bf 590}, 284 (2004)
  [arXiv:hep-th/0311154].
}
\lref\McAl{
  L.~McAllister,
  ``An inflaton mass problem in string inflation from threshold corrections to
  volume stabilization,''
  arXiv:hep-th/0502001.
}
\lref\Kalloshshift{
  J.~P.~Hsu and R.~Kallosh,
  ``Volume stabilization and the origin of the inflaton shift symmetry in
  string theory,''
  JHEP {\bf 0404}, 042 (2004)
  [arXiv:hep-th/0402047].
}
\lref\Gano{
  O.~J.~Ganor,
  ``A note on zeroes of superpotentials in F-theory,''
  Nucl.\ Phys.\ B {\bf 499}, 55 (1997)
  [arXiv:hep-th/9612077].
}
\lref\Shandera{
  S.~E.~Shandera,
  ``Slow roll in brane inflation,''
  JCAP {\bf 0504}, 011 (2005)
  [arXiv:hep-th/0412077].
}
\lref\LMRS{
  S.~M.~Lee, S.~Minwalla, M.~Rangamani and N.~Seiberg,
  ``Three-point functions of chiral operators in D = 4, N = 4 SYM at  large
  N,''
  Adv.\ Theor.\ Math.\ Phys.\  {\bf 2}, 697 (1998)
  [arXiv:hep-th/9806074].
}
\lref\DLS{
  G.~Dall'Agata, K.~Lechner and D.~P.~Sorokin,
  ``Covariant actions for the bosonic sector of D = 10 IIB supergravity,''
  Class.\ Quant.\ Grav.\  {\bf 14}, L195 (1997)
  [arXiv:hep-th/9707044].
}
\lref\chendasgupta{
  P.~Chen, K.~Dasgupta, K.~Narayan, M.~Shmakova and M.~Zagermann,
  ``Brane inflation, solitons and cosmological solutions: I,''
  arXiv:hep-th/0501185.
}
\lref\Buchel{
  A.~Buchel,
  ``On effective action of string theory flux compactifications,''
  Phys.\ Rev.\ D {\bf 69}, 106004 (2004)
  [arXiv:hep-th/0312076].
}
\lref\AganagicNN{
  M.~Aganagic, C.~Popescu and J.~H.~Schwarz,
  ``Gauge-invariant and gauge-fixed D-brane actions,''
  Nucl.\ Phys.\ B {\bf 495}, 99 (1997)
  [arXiv:hep-th/9612080].
}
\lref\CederwallRI{
  M.~Cederwall, A.~von Gussich, B.~E.~W.~Nilsson, P.~Sundell and A.~Westerberg,
  ``The Dirichlet super-p-branes in ten-dimensional type IIA and IIB
  supergravity,''
  Nucl.\ Phys.\ B {\bf 490}, 179 (1997)
  [arXiv:hep-th/9611159].
}
\lref\BergshoeffTU{
  E.~Bergshoeff and P.~K.~Townsend,
  ``Super D-branes,''
  Nucl.\ Phys.\ B {\bf 490}, 145 (1997)
  [arXiv:hep-th/9611173].
}
\lref\BanksES{
  T.~Banks, M.~Dine and E.~Gorbatov,
  ``Is there a string theory landscape?,''
  JHEP {\bf 0408}, 058 (2004)
  [arXiv:hep-th/0309170].
}
\lref\BanksXH{
  T.~Banks,
  ``Landskepticism or why effective potentials don't count string models,''
  arXiv:hep-th/0412129.
}
\lref\early{A.~Strominger,
  ``Superstrings With Torsion,''
  Nucl.\ Phys.\ B {\bf 274}, 253 (1986)\semi
  J.~Polchinski and A.~Strominger,
  ``New Vacua for Type II String Theory,''
  Phys.\ Lett.\ B {\bf 388}, 736 (1996)
  [arXiv:hep-th/9510227]\semi
  K.~Becker and M.~Becker,
  ``M-Theory on Eight-Manifolds,''
  Nucl.\ Phys.\ B {\bf 477}, 155 (1996)
  [arXiv:hep-th/9605053]\semi
  J.~Michelson,
  ``Compactifications of type IIB strings to four dimensions with  non-trivial
  classical potential,''
  Nucl.\ Phys.\ B {\bf 495}, 127 (1997)
  [arXiv:hep-th/9610151]\semi
  B.~R.~Greene, K.~Schalm and G.~Shiu,
  ``Warped compactifications in M and F theory,''
  Nucl.\ Phys.\ B {\bf 584}, 480 (2000)
  [arXiv:hep-th/0004103]\semi
  M.~Grana and J.~Polchinski,
  ``Supersymmetric three-form flux perturbations on AdS(5),''
  Phys.\ Rev.\ D {\bf 63}, 026001 (2001)
  [arXiv:hep-th/0009211]\semi
  G.~Curio and A.~Krause,
  ``Four-flux and warped heterotic M-theory compactifications,''
  Nucl.\ Phys.\ B {\bf 602}, 172 (2001)
  [arXiv:hep-th/0012152]\semi
  K.~Becker and M.~Becker,
  ``Supersymmetry breaking, M-theory and fluxes,''
  JHEP {\bf 0107}, 038 (2001)
  [arXiv:hep-th/0107044]\semi
  M.~Haack and J.~Louis,
  ``M-theory compactified on Calabi-Yau fourfolds with background flux,''
  Phys.\ Lett.\ B {\bf 507}, 296 (2001)
  [arXiv:hep-th/0103068]\semi
  J.~Louis and A.~Micu,
  ``Type II theories compactified on Calabi-Yau threefolds in the presence  of
  background fluxes,''
  Nucl.\ Phys.\ B {\bf 635}, 395 (2002)
  [arXiv:hep-th/0202168]\semi
  P.~Mayr,
  ``On supersymmetry breaking in string theory and its realization in brane
  worlds,''
  Nucl.\ Phys.\ B {\bf 593}, 99 (2001)
  [arXiv:hep-th/0003198]\semi
   K.~Dasgupta, G.~Rajesh and S.~Sethi,
  ``M theory, orientifolds and G-flux,''
  JHEP {\bf 9908}, 023 (1999)
  [arXiv:hep-th/9908088]\semi
  S.~Kachru, M.~B.~Schulz and S.~Trivedi,
  ``Moduli stabilization from fluxes in a simple IIB orientifold,''
  JHEP {\bf 0310}, 007 (2003)
  [arXiv:hep-th/0201028]\semi
  A.~R.~Frey and J.~Polchinski,
  ``N = 3 warped compactifications,''
  Phys.\ Rev.\ D {\bf 65}, 126009 (2002)
  [arXiv:hep-th/0201029]\semi
  P.~K.~Tripathy and S.~P.~Trivedi,
  ``Compactification with flux on K3 and tori,''
  JHEP {\bf 0303}, 028 (2003)
  [arXiv:hep-th/0301139]\semi
  A.~Giryavets, S.~Kachru, P.~K.~Tripathy and S.~P.~Trivedi,
  ``Flux compactifications on Calabi-Yau threefolds,''
  JHEP {\bf 0404}, 003 (2004)
  [arXiv:hep-th/0312104].
}
\lref\CandelasPI{
  P.~Candelas and X.~de la Ossa,
  ``Moduli Space Of Calabi-Yau Manifolds,''
  Nucl.\ Phys.\ B {\bf 355}, 455 (1991).
}
\lref\morerefs{R.~D'Auria, S.~Ferrara and S.~Vaula,
  ``N = 4 gauged supergravity and a IIB orientifold with fluxes,''
  New J.\ Phys.\  {\bf 4}, 71 (2002)
  [arXiv:hep-th/0206241]\semi
  S.~Ferrara and M.~Porrati,
  ``N = 1 no-scale supergravity from IIB orientifolds,''
  Phys.\ Lett.\ B {\bf 545}, 411 (2002)
  [arXiv:hep-th/0207135]\semi
  R.~D'Auria, S.~Ferrara, M.~A.~Lledo and S.~Vaula,
  ``No-scale N = 4 supergravity coupled to Yang-Mills: The scalar potential and
  super Higgs effect,''
  Phys.\ Lett.\ B {\bf 557}, 278 (2003)
  [arXiv:hep-th/0211027]\semi
  R.~D'Auria, S.~Ferrara, F.~Gargiulo, M.~Trigiante and S.~Vaula,
  ``N = 4 supergravity Lagrangian for type IIB on $T^6/Z(2)$ in presence of
  fluxes and D3-branes,''
  JHEP {\bf 0306}, 045 (2003)
  [arXiv:hep-th/0303049]\semi
  L.~Andrianopoli, S.~Ferrara and M.~Trigiante,
  ``Fluxes, supersymmetry breaking and gauged supergravity,''
  arXiv:hep-th/0307139\semi
    B.~de Wit, H.~Samtleben and M.~Trigiante,
  ``Maximal supergravity from IIB flux compactifications,''
  Phys.\ Lett.\ B {\bf 583}, 338 (2004)
  [arXiv:hep-th/0311224]\semi
  V.~Balasubramanian and P.~Berglund,
  ``Stringy corrections to Kaehler potentials, SUSY breaking, and the
  cosmological constant problem,''
  JHEP {\bf 0411}, 085 (2004)
  [arXiv:hep-th/0408054]\semi
  M.~Becker, G.~Curio and A.~Krause,
  ``De Sitter vacua from heterotic M-theory,''
  Nucl.\ Phys.\ B {\bf 693}, 223 (2004)
  [arXiv:hep-th/0403027]\semi
  A.~Font,
  ``Z(N) orientifolds with flux,''
  JHEP {\bf 0411}, 077 (2004)
  [arXiv:hep-th/0410206]. 
}
\lref\FreyTF{
  A.~R.~Frey,
  ``Warped strings: Self-dual flux and contemporary compactifications,''
  arXiv:hep-th/0308156.
  }
\lref\SilversteinID{
  E.~Silverstein,
  ``TASI / PiTP / ISS lectures on moduli and microphysics,''
  arXiv:hep-th/0405068.
}
\lref\Seth{S. Sethi, private communication.}

\Title{\vbox{\baselineskip12pt
\hbox{hep-th/0507158}
}}
{\vbox{\centerline{Dynamics of warped compactifications and } \vbox{\centerline{the shape of the warped landscape }}
}}
\centerline{{\ticp Steven B. Giddings}\footnote{$^\dagger$}
{Email address:
giddings@physics.ucsb.edu} and {\ticp Anshuman Maharana}\footnote{$^\ddagger$}
{Email address:
anshuman@physics.ucsb.edu} }
\bigskip\centerline{ {\sl Department of Physics}}
\centerline{\sl University of California}
\centerline{\sl Santa Barbara, CA 93106-9530}
\bigskip\bigskip
\centerline{\bf Abstract}
The dynamics of warped/flux compactifications is studied, including warping effects,  providing a firmer footing for investigation of the ``landscape."  We present a general formula for the four-dimensional potential of  warped compactifications in terms of ten-dimensional quantities.  This allows a systematic investigation of moduli-fixing effects and potentials for mobile branes.  We provide a necessary criterion, ``slope-dominance," for evading ``no-go" results for de Sitter vacua.  We outline the ten-dimensional derivation of the non-perturbative effects that should accomplish this in KKLT examples, and
outline a systematic discussion of their corrections.  We show that potentials for mobile branes receive generic contributions inhibiting slow-roll inflation.  We give a linearized analysis of general scalar perturbations of  warped  IIB compactifications, revealing new features for both time independent and dependent moduli, and new aspects of the kinetic part of the four-dimensional effective action.  The universal Kahler modulus is found {\it not} to be a simple scaling of the internal metric, and a prescription is given for defining holomorphic Kahler moduli, including warping effects.  In the presence of mobile branes, this prescription elucidates couplings between bulk and brane fields.  Our results are thus relevant to investigations of the existence of de Sitter vacua in string theory, and of their phenomenology, cosmology, and statistics.

\Date{}

 \newsec{ Introduction}

                     Compactifications of string theory in the
presence of fluxes have provided us with phenomenologically
attractive vacua of string theory and at the same time have
significantly advanced our thinking about the space of string theoretic vacua.  In the string revolution of 1984, many phenomenologically important features were found to emerge from compactification on a Calabi-Yau manifold -- low energy supersymmetry, generations, mechanisms for GUT breaking, etc.  However, one of the thorny problems remaining was that of the moduli of these compactifications, giving phenomenologically unacceptable light scalar fields.

Recent developments in warped flux compactifications have shown
how these moduli can be fixed\refs{\GKP,\KKLT}, have indicated the
possibility of finding vacua with positive cosmological
constant\refs{\KKLT}, and have pointed towards a new view of the
space of string vacua in which it may be that the only way to
determine details of much of low energy physics is through the
principle of environmental selection\refs{\BoPo,\SussL}. Moreover,
these compactifications have suggested new possible mechanisms for
inflation\refs{\KKLMMT}, and have suggested that the fundamental
scale of physics could be unexpectedly low (for general discussion
of scales in these compactifications, see \DeGi), raising the
possibility that superstrings could be detected as cosmic
strings\refs{\CMP}, or black holes and strings could even be
studied at accelerators\refs{\GiTh,\DiLa}.

Given the possibly profound implications arising from the study of flux vacua and warped compactifications (a partial list references includes \refs{\early\GVW\TaVa\CKLT\DeGi\morerefs-\BHK}; useful reviews include\refs{\SilversteinID,\FreyTF}), it is particularly important to understand the space of such compactifications and their dynamics.  While there have been many advances, the inherently greater complexity of these compactifications has left the subject in a significantly more primitive state than that of the traditional Kaluza-Klein compactifications of string/M theory.

Many questions have received at best partial answers in the literature.  A proper description of space-time dependent warped compactifications has not been given, and indeed puzzles have remained regarding even the moduli of static solutions.  One would expect dynamic solutions to be governed by a four-dimensional effective action, but there are various subtleties in deriving such an action from a more fundamental ten-dimensional formulation.  Moreover, in order to verify the existence of the de Sitter vacua proposed in \KKLT, one needs to carefully understand the sources and impact of possible corrections to these solutions in a systematic analysis.  The possibility of achieving slow-roll inflation has been raised, but found difficult to achieve\refs{\KKLMMT}; a more systematic understanding would be desireable.  And treatment of environmental selection has indicated the need to understand the systematics of such vacua, but again a careful understanding of their construction is needed first.  In short, many aspects of both cosmology and phenomenology require dynamical knowledge of the four-dimensional physics of these compactifications, and only pieces of this are known so far.

This paper will take steps towards providing some of these missing pieces.  One of the results of this paper is a systematic  linearized treatment of dynamic  perturbations of the moduli in such warped compactifications.  This treatment, which is needed for a proper understanding of the reduction from the ten-dimensional theory to a four-dimensional effective theory, reveals some features that some may consider unexpected both in the description of the moduli and in the form of the corresponding time-dependent solutions of the ten-dimensional equations of motion.  These include proper treatment of the universal Kahler deformation -- which is not a simple scaling of the internal metric -- and extra ``compensator" deformations needed to obtain a consistent ten-dimensional solution.  While this work is a prelude to a systematic derivation of a four-dimensional effective action in the presence of warping, and provides some important clues as to its structure, we are not yet able to give such a derivation, primarily due to issues with kinetic terms.

However, we are able to give a very general formula for the four-dimensional potential in a general warped compactification.  This formula, which provides the potential in terms of ten-dimensional quantities, and explicitly includes warping effects, arises from investigation of the ten-dimensional equations of motion and provides a link between ten- and four- dimensional analysis.  One can check that it yields the familiar results for flux and brane generated potentials on a compact space, together with corrections due to warping.  It also gives a systematic rederivation of the potential for complex structure perturbations given in \refs{\GKP,\DeGi}.  Moreover, this formula allows a systematic derivation of the potential once other effects, such as anti branes or non-perturbative effects are included, as in recent work on constructing de Sitter vacua\refs{\KKLT}.  It also extends to treat the case of mobile branes, and thus provides a controlled derivation of interbrane potentials relevant for investigating the possibility of inflation in these and other models.

We are also able to improve understanding of several issues in
warped compactification dynamics.  Our discussion clarifies the
origin of  ``no-go" theorems for de Sitter
compactifications\refs{\DSH,\MaNu}, and formulates a  necessary
condition, which we refer to as ``slope dominance," for their
evasion such as in \KKLT.  In the process, we also explain the
relative role of other competing formulas for the potential with
less acceptable features; for example, \refs{\Dealwisone} showed
that a na\"\i ve derivation gives a negative definite potential.
We outline a ten-dimensional description of the origin of the
non-perturbative effects (euclidean D3 branes, gaugino
condensation on D7 branes) that should give the necessary ``slope
dominance" in KKLT models, and we outline a systematic discussion
of the corrections to the potential for these models which is
relevant to checking their validity.   Moreover, our discussion of
potentials for mobile branes elaborates the origin of troubling
terms -- noticed in \refs{\KKLMMT} -- that generically spoil
slow-roll inflation.   Our treatment of non-perturbative effects
combined with our earlier discussion of proper treatment of moduli
of solutions also suggests a useful definition of the holomorphic
Kahler moduli with mobile branes present, that yields a clear and
general resolution to puzzles over couplings between these fields
and gauge fields on D7 branes (the ``rho
problem"\refs{\Kalloshrho,\Kalloshtalk,\BHK}).

In outline, the next section reviews the IIB warped solutions of \GKP.  Section three then describes the origin of the light spectrum from the underlying Calabi-Yau, and describes the form of static deformations of the GKP solutions both before and after three-flux has lifted complex structure moduli.  A particular focus is the parametrization of the universal Kahler modulus, which is {\it not} a simple rescaling of the compact geometry.  Section four gives an overview of dynamic moduli perturbations, and in particular the role of ``compensator" fields in lifting to ten-dimensional solutions.  It also describes some of the progress and issues with deriving the full four-dimensional effective action, and outlines the basics of the Kaluza-Klein expansion.

Section five derives a general formula for the four-dimensional potential of a general warped compactification, with full account of warping contributions: the central formula is (5.8).  Simple examples of potentials due to branes and fluxes are given, along with a discussion of radial dilaton dynamics that elucidates the relative role of other expressions for the potential.   This formula is also applied to the GKP solutions\GKP, with generalization to more generic sources such as anti-D3 branes.  Finally, de Sitter ``no-go" results are investigated, and our necessary condition for their violation, ``slope dominance," is formulated.

Section six describes potential terms used to lift Kahler moduli directions -- both non-perturbative potentials and their ten-dimensional origin, and those due to anti D3 branes.  Section seven then provides a general discussion of the systematics of moduli-lifting effects, such as in \KKLT, and the various sources of corrections and their importance.  The interplay with the slope dominance condition yields the fine tuning conditions on the superpotential and the redshifted tension of the anti D3 branes.  We also discuss effects of warping, which could modify vacuum statistics results or even the presence of a controlled approximation in which to derive de Sitter vacua.

Section eight treats dynamics of mobile D3 branes, using the general potential formula.  Potentials for such motion are derived in terms of geometrical quantities, and the origin of a  generic problem for slow-roll in models with fixed moduli is described.  This section also clarifies the definition of holomorphic Kahler moduli when mobile branes are present, thus giving a general resolution to the rho problem.

Appendix A gives a ten-dimensional treatment of linearized perturbations of warped compactifications, with particular emphasis on those of \GKP.  We provide expressions for the perturbed metric and curvatures for a general warped compactification, and for perturbed stresses, and outline the solution of the equations of motion with both zero and non-zero three-form flux.  This clarifies the role of ``compensators'' -- extra field components in time dependent situations -- and the need for Kaluza-Klein excitations, and provides some discussion of the problem of deriving the four-dimensional effective action.   Appendix A also presents an argument that the remaining flat directions after fluxes are turned on are indeed purely Kahler, with no admixture of complex structure/axidilaton deformations, resolving questions raised previously\refs{\Seth}.  Appendix B closes with brief comments on relations to other work on dynamical warped compactifications.

We end this introduction with a few comments on use of the effective action, which has been in particular criticized in \refs{\BanksES,\BanksXH}.  Use of the effective action, and the resulting equations of motion, is central to the approach of the present work.   We do not presently know the fundamental formulation of string/M theory.  Since actions are a particularly economical way of summarizing quantum dynamics, we might expect that  string/M theory could be described by functional integrals of the form
\eqn\Mint{\int {\cal D}M e^{iS[M]}(\cdots)\ ,}
where $M$ represents a dynamical M-theory configuration, although certainly some more fundamental formulation of quantum dynamics may be discovered.  In certain limits, the configurations $M$ can be well approximated by familiar objects -- supergravity fields, strings, D-branes, etc., although ultimately we anticipate that the description of these configurations respects a reduction of the number of degrees of freedom arising from holographic or related considerations.  However, we believe it is sensible to use the effective action in regions where the dynamics is well approximated by these familiar configurations, as long as we don't follow it into a region where it predicts its own demise, for example through evolution to black holes or other strong gravity/string effects.\foot{For many purposes there appears to even be an effective field-theory description of black holes; the effects of holography are only believed to be revealed when computing particular classes of observables.}  It is in this spirit that we use the effective action, and take it seriously as long as it stays out of trouble, although it is not inconceivable that its use is vitiated by some more subtle fundamental flaw.

\newsec{Review }

We begin by reviewing the construction of the warped IIB string compactifications of \refs{\GKP}.  These are solutions to leading order in $\alpha'$, which are thus found as solutions to the type IIB supergravity action, supplemented with local terms that summarize the effects of branes and orientifolds.

The supergravity action (in Einstein frame) for the type IIB  string theory
is
\eqn\IIB{\eqalign{ S_{\rm IIB} =&{1 \over 2\kappa_{10}^2} \int d^{10}x
\sqrt{-g} \left( {\cal R} - {\partial_M\tau \partial^M
\overline{\tau} \over 2 ({\rm {Im}\, \tau)^2}} -
 {G_{(3)}\cdot\overline{G}_{(3)} \over
12\, {\rm Im}\, \tau }- {\tilde{F}_{(5)}^2\over 4\cdot 5!}\right)\cr
 &+ {1\over 8i\kappa_{10}^2} \int
{C_{(4)}\wedge G_{(3)}\wedge \overline{G}_{(3)} \over {\rm Im}\,
\tau}\ +\ S_{\rm loc}}}
where $G_{(3)}= F_{(3)} - \tau H_{(3)}$ is the combined three-form flux, with $G_{(3)}= dC_{(2)}$, $H_{(3)}= dB_{(2)}$
\eqn\taudef{\tau=C_{(0)} + i e^{-\phi}\ ,}
 and
 \eqn\combfive{
 {\tilde F}_{(5)} =
F_{(5)} - {1\over 2} C_{(2)}\wedge H_{(3)} + {1\over 2}
B_{(2)}\wedge F_{(3)}\ ,}
with $F_{(5)}=dC_{(4)}$.
In addition to the equations of motion obtained from the above
action, the condition
\eqn\selfdual{{\tilde F}_{(5)} = {*}{\tilde F}_{(5)}}
 must
be imposed by hand.

Ref.~\GKP\ focuses on solutions with maximal symmetry in four dimensions,
\eqn\metricansatz{ds_{10}^2 = e^{2A(y)}\tg_{\mu\nu}(x) dx^\mu
dx^\nu + e^{-2A(y)} \tilde{g}_{mn}(y) dy^m dy^n}
which are the general solutions  to these equations
with D3- and D7-brane as well as O3 plane sources.
These obey a BPS-like condition,
\eqn\bps{{1\over 4}(T_m^m - T_\mu^\mu)^{\rm loc} \geq T_3 \rho_3^{\rm loc}}
where $\rho_3^{\rm loc}$ is the D3 charge density of the
localized sources.
The most general fluxes consistent with maximal 4d symmetry are three-form flux
$G_{(3)}$ with all components in the compact directions, a dilaton/axion that varies over the compact manifold $\tau=\tau(y)$, and five
form flux of the form
\eqn\fiveform{\tilde{F}_{(5)} =  (1 + {*})[d\alpha(y) \wedge dx^0
\wedge dx^1 \wedge dx^2 \wedge dx^3]}

Under these conditions, one can find the general solution at leading order in $\alpha'$.  It is specified by the following data:

\item{a)} The underlying geometry is given by an orientifold of a Calabi-Yau manifold, or more generally by an F-theory compactification.  In the first case, the data is the Calabi-Yau metric,
\eqn\cydata{ {\tilde g}_{mn} = g_{mn}^{\rm CY}}
together with the orientifold projection.  This in general will have $h^{2,1}$ complex moduli $z^\alpha$ and $h^{1,1}$ Kahler moduli $\rho^i$.
In the more general case, the underlying geometry given by the F-theory solution $X$, which projects to the six-dimensional data of a (non Calabi-Yau) metric ${\tilde g}_{mn}$, the positions/charges of the resulting D7 branes, and a varying axidilaton, $\tau(y)$.

\item{b)} Closed three-form fluxes $F_{(3)}$, $H_{(3)}$, satisfying the usual quantization conditions (to be described explictly in the next section).

\item{c)} The six-dimensional positions of a collection of space-filling D3 branes.

This data must satisfy the Gauss-law constraint that the total D3 charge in the compact space vanishes,
 \eqn\gauss{{1
\over 2 \kappa_{10}^2 T_3} \int_{M_{6}}  H_{(3)}\wedge F_{(3)}\ +\
Q_3^{\rm loc}=0\ ,}
where $Q_3^{\rm loc}$ summarizes the D3 charge of the D3 branes and either O3 planes, or that induced on the D7 branes as discussed in \GKP.  In the latter case, one finds a contribution
\eqn\dsevthree{Q_3^{D7} = -{\chi(X)\over 24}\ .}

For such a choice of data, \GKP\ argues a supergravity solution can be found with vanishing four-dimensional curvature.  (Many of the following formulas will be literally true only for the orientifold case, but have more-or-less obvious generalizations to the F-theory case.)
The  Bianchi identity
\eqn\bianchi{d{\tilde F}_{(5)} = H_{(3)}\wedge F_{(3)} + 2
\kappa_{10}^2 T_3 *_6\rho_3^{\rm loc}}
(where $*_6$ denotes the dual with respect to the metric $g_6$)
takes the form
 \eqn\bianchiscalar{ \tilde\nabla^2 \alpha = ie^{2A}
{G_{mnp}({*}_6 \overline{G}^{mnp}) \over 12 Im \tau} +
2{e^{-6A}}\partial_m\alpha \partial^{ m} e^{4A}
 + 2\kappa_{10}^2 e^{2A} T_3 \rho_3^{\rm loc}\ ,}
where tildes are used throughout the paper to denote covariant quantities constructed with the metric $\tg_{mn}$.
Combining this with the trace of Einstein's equations gives
\eqn\biei{
\tilde \nabla^2 (e^{4A} -\alpha)  =
{e^{2A} \over 24 \,{\rm Im}\,\tau}  \Bigl|i G_{(3)} - {*}_6G_{(3)}\Bigr|^2
+ {e^{-6A}} |\partial(e^{4A} - \alpha)|^2
 + 2{{\kappa_{10}^2} e^{2A}
\biggl[{1\over 4}(T_m^m - T_\mu^\mu )^{\rm loc} - T_3 \rho_3^{\rm loc}
\biggr]\ .}}

The left hand side of \biei\ vanishes when integrated over the manifold, and positive definiteness therefore requires vanishing of the individual terms on the right.  This implies
\eqn\efa{\alpha=e^{4A}}
and that the flux
$G_{(3)}$ is imaginary self-dual (ISD):
\eqn\ISD{\ast_6 G_{(3)} = i G_{(3)}\ .}
Note that on a Calabi-Yau manifold, the flux is automatically primitive,
\eqn\primit{J\wedge G_{(3)} =0\ .}
For a primitive flux the condition \ISD\ has a solution for each choice of Kahler moduli.    However,  this condition then fixes\refs{\GKP} the complex structure moduli $z^\alpha$ and the dilation $\tau$, or, in the F-theory context, the complex structure moduli of $X$, which include the D7-brane positions.  Vanishing of the last term in \biei\ implies saturation of the pseudo-BPS condition \bps\ for the sources.
Finally, given this data, the five-form and warp factor are determined by solving the equation
\eqn\warp{ -\tilde \nabla^2 (e^{-4A}) =
{G_{mnp}{\bar{G}}^{\widetilde{mnp}} \over 12Im\tau } +  2
\kappa_{10}^{2} T_{3}  \tilde\rho_3^{\rm loc}\ ;}
on the background geometry.
For a collection of D3 charges of strength $N_i$ at point $y_i$, the source becomes
\eqn\dthsour{\tilde\rho_3^{\rm loc} = \sum_{i} N_{i} {1 \over \sqrt{{\tg}}}
\delta^{6} (y-y_{i})\ }
(where the context here and elsewhere should make it clear that $\tg=\tg_6$);
notice that O3 planes give the negative contribution necessary for \warp\ to be consistent.

We close this review with a word about approximations.  The equations we write are valid only to leading order in $\alpha'$ and beyond this receive corrections.  At the same time, we include sources (fluxes, branes) whose contributions to the geometry are strictly speaking suppressed by a power of $\alpha'$ relative to the leading geometry.  We wish to do this as some of the physically interesting effects -- such as warping of the metric -- only contribute at this order in $\alpha'$.  The justification is  that fluxes and brane charges may be large (though apparently not parametrically) in the compactifications of \GKP.  Thus there is   good motivation for keeping such effects that are only suppressed  by a power of $K\alpha^{\prime}$, for a typical large flux quantum $K$, while ignoring effects of order $\alpha'$.

\newsec{Perturbations of flux compactifications -- kinematics}

 To begin an analysis of dynamics of flux compactifications, we begin by discussing massless or light perturbations of these configurations.  The analogous discussion for traditional Calabi-Yau compactifications is a straightforward exercise in cohomology and dimensional reduction.

While analysis of perturbations of a general warped compactification has additional complications due to the warping, the study  of the perturbations of the solutions of \GKP\ is significantly simplified by the relation to the underlying Calabi-Yau manifold.  In this section we examine these perturbations more carefully.  We begin by discussing the light perturbations present in these solutions; in addition to four-dimensional gravity, one finds a rich spectrum of scalars and vectors.  Given the particular importance of stabilizing the moduli, and their possible role in cosmology, we will particularly focus on the scalar spectrum, and investigate that
first for static perturbations in the case of vanishing three-form flux then in the presence of three-form flux.

\subsec{The light spectrum}

The light fluctuation spectrum of the IIB compactifications of \GKP\ is largely inherited from the light fluctuation spectrum of the underlying Calabi-Yau orientifold, or F theory compactification.  Therefore we begin with a discussion of that spectrum.  We will focus on the orientifold case; the F-theory case is largely a straightforward generalization.

\subsubsec{Covering space}

First recall that  the Calabi-Yau manifold is in general endowed with a holomorphic $(3,0)$ form $\Omega$.  The other non-trivial cohomology is represented by $h^{2,1}$ forms $\chi_\alpha$ of type $(2,1)$, and $h^{1,1}$ forms $\omega_i$ of type $(1,1)$, together with their duals.   All of these forms may be taken to be harmonic.  The resulting four-dimensional supersymmetry multiplets are as follows.

The RR axion $C_{(0)}(x)$ and the ten-dimensional dilaton $\phi(x)$ give the complex field $\tau(x)$, as in \taudef.  Likewise, $B_{(2)}(x)$ and $C_{(2)}(x)$ dualize into two pseudoscalar axions that can be thought of as components of a single $N=1$ superfield.  These axions and $\tau$ together give the bosonic components of an $N=2$ multiplet that may be thought of either as a hypermultiplet or as a double-tensor multiplet\refs{\BrandtUW\TheisJJ-\DallAgataYR}.

Deformations of the complex structure give rise to scalar fields $z^\alpha(x)$, also in $N=1$ chiral multiplets.  At the $N=2$ level, these combine with vectors $V^\alpha_\mu(x)$ from perturbations
\eqn\Cvec{C_{(4)} = V^\alpha(x)\wedge\chi_\alpha }
(self-duality halves the number of components)
of the four-form potential to give bosonic components of $h^{2,1}$ $N=2$ vector multiplets.

Deformations of the Kahler structure give rise to scalar fields $r^i(x)$.  These combine with scalars $a^i(x)$ arising from perturbations
\eqn\Cscal{C_{(4)} = a^i(x) *_6 \omega_i + D_{(2)}^i(x) \wedge \omega_i}
of the four-form potential to give bosonic components of $N=1$ scalar superfields, which we denote by complex fields $\rho^i$.  (Self duality fixes the $D_{(2)}^i$ in terms of the $a^i$.)  Likewise, deformations of the two-form potentials of the form
\eqn\twoscal{B_{(2)} = b^i(x) \omega_i\ ,\ C_{(2)} = c^i(x) \omega_i}
give the components $(b^i,c^i)$ of $N=1$ scalar superfields.  These combine with the $\rho^i$ to give the bosonic components of $h^{1,1}$ $N=2$ hypermultiplets.

Next, there is a deformation
\eqn\metdef{ C_{(4)} = V\wedge \Omega}
of the four-form potential that corresponds to a vector $V_\mu(x)$; this combines with $g_{\mu\nu}(x)$ to give the $N=2$ gravity multiplet.

\subsubsec{Orientifold-projected spectrum}

As described in \GrimmUQ, the orientifold projection breaks the $N=2$ supersymmetry to $N=1$ and eliminates some of the $N=2$ states.  The cohomology breaks into eigenspaces under the holomorphic involution $\sigma$ (which here we consider to be of the type that induces O3 and O7 planes),
\eqn\cohomspl{ \sigma H^{(p,q)}_\pm = \pm H^{(p,q)}_\pm\ }
with corresponding dimensions $h_\pm^{p,q}$.  The remaining $N=1$ supermultiplets are the gravitational multiplet, $h_+^{2,1}$ vector multiplets, and $h_-^{2,1}+h_+^{1,1} + h_-^{1,1} +1$ chiral multiplets with bosonic fields $z^\alpha$, $\rho^i$, $(b^i,c^i)$, and $\tau$, respectively.\foot{Note that the mode corresponding to the Goldstone boson of \refs{\GHK} is projected out.}
Finally, the positions of mobile spacefilling D3 branes (or D7 branes in the F-theory case) correspond to massless moduli, and are described by $N=1$ chiral multiplets.

A critical aspect of the warped IIB compactifications of \GKP\ is the presence of three-form flux, that is expectation values for $F_{mnp}(y)$ and $H_{mnp}(y)$.  The resulting four-dimensional action for the perturbations with such flux backgrounds has recently been systematically studied, to leading order in the large-volume expansion, in \GrimmUQ.  Working at this order amounts to neglecting the effects of the warping induced by the presence of D3 branes and fluxes, as we will explain in more detail.  Since the effects of the warping are important for various phenomena (in particular lowering mass scales), part of the goal of this paper is to begin to go beyond the no-warping approximation.

 \subsec{Moduli space: $G_3=0$}

  We next turn to a description of moduli in the warped case but with $G_3=0$.
 With three-form flux turned off, the moduli of the underlying Calabi-Yau orientifold compactification and the value of the axidilaton $\tau$, together with the brane positions (or more generally moduli of the F-theory compactification $X$)  determine the moduli of the GKP solutions, which in this case reduce to the solutions of Chan, Paul, and Verlinde \refs{\HVer,\CPV}.  This follows directly from the construction:  once one specifies $\tau$, a given point in Calabi-Yau moduli space and the brane configurations, a corresponding warped compactification follows automatically by solving the equations of the previous section.

Specifically, consider a perturbation of the underlying Calabi Yau
\eqn\CYpert{ \tilde{g}_{mn} \to \tilde{g}_{mn}  + \delta
\tilde{g}_{mn} \ .}
As discussed above, the general $\delta \tilde{g}_{mn}$
corresponds to an element of $H^{(1,1)}$ or $H^{(2,1)}$ of the
Calabi Yau, the Kahler and complex structure deformations,
respectively. All of these perturbations except for the universal
Kahler modulus can be made traceless by an appropriate choice of
basis. We describe the traceless perturbations in a single
framework, and shall treat the universal Kahler modulus
separately.  For vanishing three-form flux, fluctuations of $\tau$
decouple.

\subsubsec{Traceless metric deformations; brane locations}

For a traceless deformation of the Calabi Yau metric in the absence
of three-form flux, the equation determining the warp factor \warp\
also implies a change in the warp factor $\delta e^{-4A}$, given by solving
\eqn\CHtrless{ \tilde{\nabla}^{2} \left(\delta e^{-4A}\right) = - \tnab_m\left(\delta\tg^{mn} \partial_n e^{-4A}\right)}
which can be done explicitly in terms of the scalar Green function.
As a result, the corresponding small variation in the solution takes the form
\eqn\CHmetric{ ds_{10}^2=(e^{2A}+ \delta e^{2A})dx_4^2
+  (e^{-2A}+\delta e^{-2A})
(\tilde{g}_{mn}+\delta \tilde{g}_{mn})  dy^m dy^n}
\eqn\CHAfive{ \tilde{F}_{m \mu \nu \rho \sigma} =
\partial_{m} \left( e^{4A}+ \delta e^{4A}
\right)
\delta_{\mu \nu \rho \sigma}}
\eqn\CHBfive{ \tilde{F}_{mnopq} = - (e^{-8A}+ \delta e^{-8A})
\sqrt{\tilde{g}}\delta_{mnopqr}
({\tilde g}^{rs} + \delta {\tilde g}^{rs}) \partial_{s}(e^{4A}+\delta e^{4A})
\ }
where all variations of $A$ are determined from the solution to \CHtrless.

Likewise, a variation in the positions of the branes also leads to a variation in the warp factor,
\eqn\warpvar{ \tilde{\nabla}^{2} [\delta e^{-4A}] = -2\kappa_{10}^2 T_3\delta \tilde\rho_3^{\rm loc}\ }
and corresponding variation in the ten-dimensional metric and five-form flux.

\subsubsec{Universal Kahler deformation}

The universal Kahler modulus, which in a non-warped compactification corresponds to an overall scaling of the compact metric, gives rise to an additional subtlety.
Note that the conditions a)-c) of section 2 are invariant under
\eqn\scalemetric{ \tilde {g}_{mn} \rightarrow \lambda \tilde
{g}_{mn}\ .}
Under such a scaling \warp\ requires that
 \eqn\warpscale{e^{2A} \rightarrow \lambda e^{2A}}
Thus the solution transforms to
\eqn\metrichange{ds_{10}^2 = \lambda e^{2A(y)} \eta_{\mu\nu}
dx^\mu dx^\nu + e^{-2A(y)} \tilde{g}_{mn} dy^m dy^n}
\eqn\compfivechange{ \tilde{F}_{m \mu \nu \rho \sigma} = \lambda^2
\partial_{m}e^{4A} \delta_{\mu \nu \rho \sigma}}
\eqn\noncompchange{ \tilde{F}_{mnopq} =
-e^{-8A}\sqrt{\tilde{g}}\delta_{mnopqr}\partial^{\tilde r}e^{4A}}
The scaling of the underlying CY is precisely canceled by
that of the warp factor and the internal manifold remains
unchanged. The scaling instead changes the volume of the four large dimensions.  This could be compensated by an overall Weyl rescaling, but we find it expedient to identify the Kahler modulus via a different (but related) procedure.

Note also that a simple scaling of
the internal metric $(g_{mn}=e^{-2A}\tilde{g}_{mn})$
 \eqn\againscale{{g}_{mn} \rightarrow  \lambda {g}_{mn}}
is not a zero-mode. Such a scaling can be thought of as a scaling
of the underlying CY without any change in the warp factor, and
this violates \warp. The situation here
is analogous to IIB string theory on $AdS_{5} \times S_{5}$; the
radius of the five sphere and AdS are not moduli of the solution.
Their scale is set by the D3 brane charge.

                To  isolate the mode corresponding to the universal Kahler modulus, note  that the equation determining
the warp factor, \warp, only determines $e^{-4A}$ up to a constant {\it shift}.  This means
we have a one
parameter family of solutions given by
\eqn\cmetric{ ds_{10}^2 =  [{e^{-4A_0(y)}}+ c]^{-1/2}
\eta_{\mu\nu} dx^\mu dx^\nu + [{e^{-4A_0(y)}} +
c]^{+1/2}\tilde{g}^0_{mn}dy^m dy^n }
\eqn\cnoncompfive{ \tilde{F}_{m \mu \nu \rho \sigma} =
\partial_{m}[{e^{-4A_0(y)}} + c]^{-1} \delta_{\mu\nu\rho
\sigma}}
\eqn\ccompfive{ \tilde{F}_{mnopq} = -[{e^{-4A_0(y)}}
+c]^{2}\sqrt{\tilde{g}}\delta_{mnopqr}\partial^{\tilde
r}[{e^{-4A_0(y)}} + c]^{-1}}
where we have chosen a fiducial metric $\tg^0_{mn}$ which we can take to have unit volume, and
${e^{-4A_0(y)}}$ is a particular solution to \warp\ which we assume asymptotes to $\calo(1)$ in regions away from sources of D3 charge.  Near singular sources, $e^{-4A}$ diverges.  For example, in the vicinity of a point $x_0$ where $N$ D3 branes are located, the warp factor  in \cmetric\ behaves as
\eqn\dthreewarp{e^{-4A} \approx c+ {4\pi\alpha^{\prime2} N\over |x-x_0|^4}}
in the local flat coordinates $x$.

Changing  $c$ changes the volume of the internal manifold, so it  naturally plays the role of the universal Kahler modulus.  Note that the $c$ dependence of the solution
is known exactly.
For locations far away from the sources, $e^{-4A}$ approaches
a constant and the perturbation behaves like an overall scaling of
the manifold.  In particular, if one
excises neighborhoods of such points and finds a large remaining volume over which the variation is small,
\eqn\choice { {\Delta [e^{-4A}] \over c} <<1}
then we can cleanly identify this region as the ``bulk," distinct from the AdS throats in the vicinity of the removed points.  For points close to the sources, $e^{4A}$ is small
and the dependence on $c$ is suppressed, in keeping with the
fact that the size of these regions is totally determined by the local D3 charge.

The large volume limit is obtained by  $ c \to \infty$.
In this limit, the volume of the compact space  behaves as
\eqn\vcomp{V\sim c^{3/2} V^0\ }
in terms of the fiducial volume $V^0$,
from which we see that, for large $c$, this parameter is related to the radius of the space by
\eqn\radc{c\sim R^4\ .}

As we have already noted, a 4d Minkowski solution exists for each value of the complex, Kahler, and D3 position moduli.  This means that the four dimensional effective potential vanishes for these modes.  We will see this directly in sec.~5.

\subsec{Frames}

While the parameterization \cmetric-\ccompfive\ transparently reveals the origin of the universal Kahler modulus, we often work in other frames differing by a rescaling \scalemetric, \warpscale.  Note that the metric $g_{mn}$ -- and hence the volume of the compact space -- are frame independent.

\subsubsec{Ten-dimensional Einstein and string frames}

The action \IIB\ is given in ten-dimensional Einstein frame.  However, note that the four-dimensional part of the metric \cmetric\ does not asymptote to the usual flat metric as $c\rightarrow\infty$.
We can normalize it so that it does by taking rescaling parameter
\eqn\lstring{\lambda_E= \sqrt c}
so that
\eqn\metstr{e^{-4A_E}= 1 + {e^{-4A_0}\over c}\ ,\ \tg^E_{mn} = c^{1/2} \tg^0_{mn}\ .}
Also, recall that the relation to ten-dimensional string frame is given by
\eqn\EtoS{g^E_{MN} = e^{-\phi/2} g^S_{MN}\ .}

\subsubsec{Four-dimensional Einstein frame}

The four-dimensional Einstein frame, which is defined to be the frame in which the four-dimensional Planck mass is constant, is most appropriate for four-dimensional analysis and its treatment via a four-dimensional superspace formalism.  The conversion factor between the four- and ten-dimensional Planck masses is given by the warped volume
\eqn\vwdef{V_W = \int d^6y \sqrt\tg e^{-4A}\ .}
Notice that the warped volume transforms under \scalemetric, \warpscale\ as
\eqn\warpsc{V_W\rightarrow \lambda V_W\ .}
Einstein frame is thus defined so that the warped volume is a modulus-independent
constant, which we might for example set to
\eqn\vzdef{V^0=\int d^6y \sqrt {\tg^0}\ .}
Then Einstein frame is reached from \cmetric\ through a scaling \scalemetric, \warpscale\ with
\eqn\leins{\lambda_E = {V^0\Big/\left(c V^0 + \int d^6y \sqrt{\tg^0} e^{-4A_0} \right)}\ .}

 \subsec{Moduli space: $G_3\neq0$}

 We now consider the moduli of solutions with three-form flux turned on.  Let $A_\alpha$, $B^\alpha$, $\alpha=1,\ldots,h^{2,1}$ be a symplectic basis of three cycles,\foot{In the orientifold case, note that in accord with the discussion of section 3.1 we restrict attention to cycles with odd intrinsic parity on the covering Calabi-Yau.} with intersection numbers
 \eqn\sympint{A_\alpha\cap B^{\alpha'}=\delta_\alpha^{\alpha'}\ , A_\alpha \cap A_{\alpha'}=B^\alpha\cap B^{\alpha'}=0\ .}
We allow general three-form fluxes, satisfying the quantization conditions
 \eqn\thrquant{\eqalign{ \int_{A_\alpha}F_{(3)}&= (2\pi)^2\alpha' M_e^{\alpha}\ ,\  \int_{B^\alpha}F_{(3)}= -(2\pi)^2\alpha' M_{m\alpha}\cr
 \int_{A_\alpha}H_{(3)}&= (2\pi)^2\alpha' K_e^{\alpha}\ ,\  \int_{B^\alpha}H_{(3)}= -(2\pi)^2\alpha' K_{m\alpha} }}
 with integer  ``electric" and ``magnetic" fluxes $M_e^{\alpha},M_{m\alpha}$, $K_e^{\alpha},K_{m\alpha}$.

 Before turning on the flux, the solutions of the preceding subsection have moduli corresponding to the Kahler and complex structure moduli for the underlying Calabi-Yau metric $\tg_{mn}$, the string dilaton, the positions of the D3 branes, and the positions of the D7 branes.  Turning on the flux then fixes many of these: in the Calabi-Yau orientifold case the $h^{2,1}+1$ complex structure and axidilaton moduli are fixed, leaving as free parameters the Kahler moduli and the D3 positions.  In the more general F-theory case, D7 brane positions are also in general fixed.

The conditions fixing the moduli can be thought of as follows.  Choose a general point in the Kahler and complex structure moduli spaces.  Given the cohomology classes of the fluxes, \thrquant, we may choose unique harmonic representatives of these classes.  Note that the harmonicity condition in general depends on both the choice of complex structure and Kahler class, so the vacuum value of the three-form field $G_3$ varies with both.  
Moduli are fixed by the ISD condition \ISD, which implies
 \eqn\fluxcond{G_{(1,2)}=G_{(3,0)}= 0\ .}
 Since the Hodge dual on three forms varies with the non-universal Kahler moduli, one might be concerned that the ISD condition or equivalently \fluxcond\ also varies, requiring the fixed values of the complex structure moduli and axidilaton to vary with Kahler structure.  However, one can show (see appendix A) that while the vacuum value of $G_3$ varies with Kahler structure, its type remains $(1,2)+(3,0)$ so it remains ISD under variations of the Kahler moduli.  Thus the remaining flat directions correspond to Kahler deformations.

Static perturbations are given by the formulas of the preceding section, together with the corresponding variations in $G_3$.  For given Kahler and complex moduli, the latter are of course uniquely given by specifying their cohomology classes as in \thrquant, together with the harmonicity condition.

 In section 5 we will discuss derivation of the potential for these moduli.  For the moment, we recall that for flux satisfying the ISD condition, this potential vanishes, as described in \refs{\GKP,\DeGi}.
Thus a vacuum moduli space parametrized by the Kahler moduli and D3 positions remains.  As discussed in \refs{\GKP,\DeGi}, this potential should arise from a Gukov-Vafa-Witten superpotential\refs{\GVW,\TaVa,\CKLT}
 \eqn\gvwpot{W=\int \Omega\wedge G\ .}
 Note that since the superpotential is indepent of change of cohomology representative,
 \eqn\cohomch{G\rightarrow G+dA\ ,}
 the superpotential only depends on the complex structure moduli.

                As a function of the universal Kahler parameter $c$, the metric is given
by \cmetric.
In the case where there are no explicit $D3$ branes, the infrared end(s) of this geometry is a smooth geometry approximately given by  \refs{\KlSt}, and in particular $e^{-4A}$ is no longer singular.  This generates a finite hierarchy between the UV and IR ends of the compact space, parametrized by
\eqn\hierarchy{ { [c + {e^{-4A_0(y)}}_{IR} ]^{1/2}
\over [c + {e^{-4A_0(y)}}_{UV} ]^{1/2}}}

              For values of $c$ of the order of ${e^{-4A_0(y)}}_{IR}$
the hierarchy generated by fluxes is lost. The solutions of GKP
with warped throats are expected to be dual to the N=1
nonconformal dual gauge theory of Klebanov and Strassler coupled
to massless bulk fields of the compactification. From the
perspective of the gauge theory, the IR scale has the
interpretation of being the scale of confinement.

 \newsec{Perturbations of flux compactifications -- dynamics}

 One important goal is to better understand the relation between four-dimensional perturbations and ten-dimensional solutions of the equations of motion, and consequently between the ten-dimensional action and the four-dimensional action.  An analysis of perturbative solutions of the ten-dimensional equations of motion is presented in appendix A.  In this section we give a summary of the rather lengthy analysis and some of its consequences.

 \subsec{Spacetime dependent fluctuations}

 We find that there are new subtleties in promoting the constant perturbations described in the preceding section to spacetime dependent perturbations.

Part of the subtlety
is already illustrated at the level of a conventional Kaluza-Klein compactification.  For example, consider compactification on  a Calabi-Yau manifold with ten-dimensional metric
\eqn\standard{ ds_{10}^2 =  \eta_{\mu\nu} dx^\mu dx^\nu +
{g_{CY}}_{mn} dy^m dy^n\ .}
At zero momentum, the volume-changing perturbation  can be parameterized in terms of a small constant $u$:
\eqn\standardpert{ ds_{10}^2 =  \eta_{\mu\nu} dx^\mu dx^\nu +
(1+u){g_{CY}}_{mn} dy^m dy^n\ .}
To give spacetime dependence one would
like to promote
\eqn\promote{ u \rightarrow u(x)}
with $\partial_{\mu} \partial^{\mu} u(x) = 0 $.  The perturbed ten-dimensional metric should
solve the equations of motion at the linearized
level. The perturbation which achieves this is a combination of \standardpert\ together with a spacetime-dependent rescaling of the four-dimensional metric,
\eqn\stanfluct{ ds_{10}^2 =  [ 1 - 3 u(x) ]\eta_{\mu\nu}
dx^\mu dx^\nu + [1 +  u(x)]{g_{CY}}_{mn} dy^m dy^n}
The extra term is necessary to solve the $(\mu\nu)$ Einstein's equation; in the limit of constant $u$ it reduces to the perturbation \standardpert\ combined with a rescaling of the four-dimensional metric bringing the metric to Einstein frame.

\subsubsec{Compensators and variations}

The presence of extra terms in the spacetime-dependent perturbations extends beyond that noted above, and in general means that there can be metric and field perturbations proportional to derivatives of $u(x)$.  These vanish for a static perturbation, but must be present in order for a general perturbation to satisfy the full ten-dimensional equations.    Terms of this form have occurred previously in studies of dimensional reduction and were called {\it compensators} in \refs{\GrayVW}.

In the case of perturbations of the GKP solutions, the static perturbations are given by \CHmetric-\CHBfive\ for traceless perturbations or brane motion, or by the infinitesimal version of \cmetric-\ccompfive\ for the universal Kahler perturbation.
The general form of the dynamic perturbations we must consider is given in eqns.~(A.3), (A.4).  For the metric, in addition to making the constant perturbation in \CHmetric\ time dependent, we must also include compensator terms of the form
\eqn\metcompens{\delta_c ds^2 = 2\partial_\mu\partial_\nu u^I(x) e^{2A}K_I(y) dx^{\mu} dx^{\nu}  + 2e^{2A} B_{Im}(y) \partial_\mu u^I(x) dx^\mu dy^m\ .}
While the compensators $B_I$ and $K_I$ can in principle be gauged away, in general this imposes on $\tgmn$ a non-trivial gauge, distinct from an a priori choice such as
\eqn\transgauge{\tnab^m \delta\tgmn=0\ .}
Likewise, compensator terms for the three- and five-form field potentials must be included; these take the form
\eqn\formcomp{\eqalign{&\delta_c(C_2 - \tau B_2) = du^I\wedge T_I\cr &\delta_c C_4 =
du^I
\wedge S_I^{(3)} + *_{4} du^I \wedge S_I + u^I(x) w_I^{(4)} +
D_{(2)}^I\wedge w_I^{(2)}\ ,}}
where the the compensators $T_I$, $S_I$, $S_I^{(3)}$ are forms on the internal manifold, and we have also included the perturbations necessary to describe axions, with internal forms $w_I^{(2)}$ and $w_I^{(4)}$ and four-dimensional form $D_{(2)}^I$.

The compensators are determined by the equations of motion, but finding explicit solutions for them is in general a non-trivial challenge.
Ab initio one expects these to be needed in order to deduce the form of the four-dimensional effective action from the higher-dimensional action.

\subsubsec{First example: spacetime dependent universal Kahler deformation}

The subtleties of compensators appear even at the level of the universal Kahler deformation, or ``volume modulus."  This is worked out in appendix A, with the result that a perturbation of the form
\eqn\univKpert{\delta\tgmn = -u(x) \tgmn}
produces a variation
\eqn\warpsoli{\delta e^{-4A} = 2 u(x) e^{-4A} + u(x) {\int d^6y \sqrt \tg e^{-4A} \over \int d^6y \sqrt \tg}\ }
%
%
and non-trivial compensator $K$ solving
\eqn\newone{ \tnab^2 K =  e^{-4A}- {\int d^6y \sqrt \tg e^{-4A} \over \int d^6y \sqrt \tg}  \ .}

\subsubsec{Second example -- traceless deformations, $G_3=0$}

The traceless deformations correspond to complex structure or
non-universal Kahler variations of the underlying Calabi-Yau
manifold.  In the case of vanishing three-form flux, these are
massless perturbations.  The ten-dimensional form of these
perturbations are also studied in appendix A.  If we begin with
the gauge \transgauge, we find that there must in general be
nonvanishing compensators $K_I$, $B_I$, and $S_I$.  The equations
determining these are discussed in appendix A.

\subsubsec{$G_3\neq0$}

As we've discussed, non-vanishing three-form flux lifts the complex structure deformations.  The corresponding ten-dimensional perturbations have both compensators, and as shown in the appendix, non-zero excitation of Kaluza-Klein modes.  The Kahler deformations  remain massless (see appendix section A.7) and also have corresponding ten-dimensional solutions with non-zero compensators.

\subsec{Effective actions -- perturbative level}

One important goal is to derive the four-dimensional effective action for perturbations of warped compactifications.  In the case of the IIB compactifications of \GKP, the light spectrum was discussed  in the preceding section.  The general action for the light spectrum is expected to be rather complicated, with various subtleties arising from the warping; a version of it for IIB orientifolds, valid to leading order in the large-volume approximation, and thus neglecting effects of warping, was given by Grimm and Louis in \GrimmUQ.  In the present paper, we are primarily interested in the action for scalar modes, and their coupling to gravity, but now including effects of warping.  In this section we discuss terms arising from the classical action \IIB; we save discussion of non-perturbative effects and brane motion for later sections.

As we've discussed, in the scalar sector the light modes consist of the dilaton field, $\tau(x)$, the Kahler deformations $\rho^i(x)$, the complex deformations $z^\alpha(x)$, and the D3 brane positions $\phi^a(x)$.  Before introducing three-form flux, all of these are massless.

We first clarify the regime of validity of such a four-dimensional effective action.  This requires understanding the systematics of the Kaluza-Klein expansion.  While details of this for general moduli perturbations are given in appendix A, essential features can be inferred from dynamics of a massless scalar in ten-dimensions,
\eqn\masssc{\sq_{10}\phi =0\ .}
In the metric \cmetric, this equation takes the form
\eqn\scexp{e^{-2A} \sq_4 \phi = -e^{2A} \tnab^2 \phi\ .}
Thus for a Kaluza-Klein mode concentrated in a region with a given warp factor, the mass squared is of order
\eqn\scff{m^2_{KK,0}\sim e^{4A}\ }
in the frame of \cmetric, and
\eqn\scein{m^2_{KK} \sim {e^{4A}\over V_W}}
in four-dimensional Einstein frame.  At large $c$, this gives Einstein-frame mass
\eqn\mkkest{m_{KK}\sim Z/c}
where $Z$ is the redshift factor
\eqn\Zdef{Z^2\sim {1\over 1 + {e^{-4A_0}\over c}}\ }
whose relevance was emphasized in \refs{\DeGi}.  For a KK mode concentrated in the bulk, $Z\sim 1$, but a KK mode supported in a strongly warped region can experience a significant lowering of its mass.  Thus the four-dimensional effective theory is strictly valid only in the range
\eqn\fdrange{E\ll {Z_{\rm min}\over c}}
determined by the most-redshifted Kaluza-Klein mode.  Beyond this energy, one needs to account for Kaluza-Klein dynamics.  In Einstein frame, the mass given to the complex structure moduli by three-form flux is of order
\eqn\mflux{m^2_{\rm flux} \sim {N\over c^3 }}
(see appendix A)
where $N$ is a measure of the total three-brane charge.  Thus, for large enough $c$ we see that there is an energy regime in which moduli are fixed and Kaluza-Klein modes are not relevant.

It is clearly important to understand the four-dimensional effective action at scales below \fdrange; this action, and its extension by non-perturbative and other effects, governs the shape and dynamics of the low-energy foothills of the landscape.

The analysis of dynamic perturbations of the fields in principle provides information about this action.  In particular, one would expect the four-dimensional effective action to arise from the IIB action evaluated on a ten-dimensional perturbation.  Unfortunately, there is a subtlety arising from the ambiguities in prescribing an off-shell action for the five form $F_5$.  In addition, at this point our analysis has been explicitly carried out only up to terms of order $\calo( N/c)$ due to the unknown form of the compensators and Kaluza-Klein excitations.  So far, all that we are able to explicitly compute in this approach is the mass matrix for perturbations up to corrections of this size.

Ref.~\DeGi\ performed checks indicating that warping does not
modify the Gukov-Vafa-Witten superpotential \gvwpot, but does
modify the Kahler potential and metric.  It was suggested that the
kinetic action for metric deformations takes the form
\eqn\KDG{ S_{kin}= {1\over 2\kappa_4^2} \int d^4x \sqrt{-\tg_4} \left\{ -3 {\partial_\mu {\bar \rho} \partial^\mu \rho\over |\rho-\bar{\rho}|^2 } - {1\over V_W} \partial_\mu T^I \partial^\mu T^J \int d^6y \sqrt{\tg_6} e^{-4A} \delta_I\tg_{mn}\delta_J  \tg^{mn}\right\}  }
where $\rho$ is the universal Kahler modulus, $T^I$ denote the fields corresponding to the complex structure and non-universal Kahler moduli, and $\delta_I \tg_{mn}$ are the corresponding normalized metric variations.
In principle, analysis of the perturbative dynamics should provide a direct check of  these statements and in particular give us the Kahler potential,  but in practice we are unable to perform this at present.  (The authors hope to return via an alternate approach in subsequent work.)  The direct perturbative analysis of the appendix only provides a check on these formulas to order $N/c$ and thus does not check the form of the warping contribution in \KDG.  Nonetheless, it is important to note both that there can in general be corrections of this size to the Kahler potential and thus the potential for moduli, and that the presence of these corrections isn't necessarily inconsistent with the expression \KDG.

Despite the present limitations on determining the kinetic part of the four-dimensional effective action, a related approach, to which we now turn, allows us to gain substantial useful information about the potential, and via the discussion of \DeGi, also lends credence to warping corrections of the form \KDG.

\newsec{Potentials on the Landscape }

\subsec{Potentials in warped compactifications}

Determining the potential on the landscape of string vacua is clearly a crucial enterprise as it bears on the possibility of finding a phenomenologically realistic vacuum and the question of how our region of the Universe evolves into this vacuum.  As discussed above, computing the full four-dimensional effective action in the case where warping is relevant can be non-trivial.  Nonetheless, we here outline a prescription to determine a useful formula for the potential for a general warped compactification.

Moreover, prior to discovery of non-trivial warped compactifications such as those of \GKP, and the subsequent arguments for the existence of the metastable de Sitter vacua of \refs{\KKLT},  general no-go theorems were proven\refs{\DSH,\MaNu} (with extensions in \refs{\LMW}) for both of these possibilities.  In seeking a more complete understanding of the landscape, it is worthwhile to understand better how such no-go theorems are evaded, which may give guidance towards other solutions.

Suppose that we are seeking the potential due to some spacefilling branes, fluxes, and other effects that don't explicitly break the (local) four-dimensional Poincare symmetries; these could include
such effects as the non-perturbative effects of \KKLT, or  $\alpha'$ corrections.
In general there will be cosmological solutions to Einstein's equations corresponding to moduli rolling in this potential, and these solutions will be spatially homogeneous in three dimensions.  The general ten-dimensional solution of this kind can be put in the form
\eqn\genmet{ds^2 = e^{2A(y,t)}\left[-dt^2 + a^2(t)ds_3^2\right] + 2 e^{2A(y,t)} \beta_{m\mu}(y,t)dy^m dx^\mu + e^{-2A(y,t)} \tg_{mn}(y,t) dy^m dy^n\ ,}
where $\beta$ has only a $t$ component,
\eqn\betares{\beta_{m\mu} = (\beta_{mt}(y,t),0,0,0)\ .}
Note that $a(t)$ could also be eliminated through redefinition of the time coordinate, but we keep this light mode manifest for contact with usual cosmology.
For these solutions, the ten-dimensional stress tensor has a special form: if we neglect velocity terms
({\it i.e.} derivatives with respect to the non-compact coordinates), we find
\eqn\specstress{T^\mu_\nu = - \delta^\mu_\nu U_{10}(y,t)\ \ ,\ \ T^m_n = T^m_n(y,t)\ .}

The form of the Einstein equation for \genmet\ can be inferred from eqn.~(A.14) of the appendix, with result
\eqn\Einscosm{\eqalign{ &G^\mu_\nu = e^{2A}\left[ -2\tnab^2 A + 4
{ (\widetilde{\nabla A})}^2 - \hf {\tilde R}_6 \right]
\delta^\mu_\nu+ e^{-2A} {\tilde G}^\mu_\nu - {e^{2A}\over
\sqrt{\tg_6}} (\tnab^\mu \tnab_\nu -
\delta^\mu_\nu{\tilde\sq})(e^{-4A}\sqrt{\tg_6})\cr &+ e^{2A}
(\delta^\lambda_\nu \tnab^\mu - \delta^\mu_\nu
\tnab^\lambda)\tnab^m \beta_{m\lambda} +  \calo(v^2,\beta^2,\beta
v) = \kten^2 T^\mu_\nu\ ,}}
 where v denotes a general velocity.
Here we have not made explicit terms proportional to (powers of) velocities, or terms beyond linear order in $\beta$.

The potential $U$ is identified through its role in the four-dimensional Einstein equations, with the constraint equation in particular taking the form
\eqn\fdeins{{\tilde G}^t_t = -\kappa_4^2 U + ({\rm velocities})^2\ .}
This means that  $U$ can be read off from the $tt$ component of the Einstein equations \Einscosm, by dropping all velocity terms (note that second time derivative terms cancel).  The result is
\eqn\Aconst{e^{2A}\left[-2\tnab^2 A + 4 { (\widetilde{\nabla A})}^2 - \hf {\tilde R}_6 \right]  - e^{-2A}\kappa_4^2 U = -\kten^2 U_{10} + \calo(\beta^2)\ .}
For small potentials/velocities we can drop the $\calo(\beta^2)$ since it is Kaluza-Klein suppressed. This equation manifestly depends on $y$, though $U$ must be $y$-{\it independent}.  It is to be thought of as an equation determining the warp factor.

From eq.~\Aconst\ we can find a useful expression for the potential, by multiplying by $e^{-2A}$ and integrating over the compact manifold,
\eqn\Potent{ \kappa_4^2 U= {1\over V_W}\int d^6 y \sqrt{\tg} \left[\kten^2 e^{-2A} U_{10} + 4 (\widetilde {\nabla A})^2 - \hf {\tilde R}_6 + e^{-2A}\calo(\beta^2)\right]\ }
where the warped volume is defined in \vwdef.
%
To correctly determine the potential this expression should be
evaluated in Einstein frame, which we recall is the frame where
$V_W$ is independent of the moduli.  An expression valid in an
arbitrary frame then follows from a transformation of the form
\scalemetric, \warpscale. This gives our formula for the potential
in a general warped compactification,
\eqn\Potentg{ \kappa_4^2 U= {V_W^E\over V_W^2}\int d^6 y \sqrt{\tg} \left[\kten^2 e^{-2A} U_{10} + 4 (\widetilde {\nabla A})^2 - \hf {\tilde R}_6 + e^{-2A}\calo(\beta^2)\right]\ .}
Since $V_W^E$ is moduli independent, a convenient choice of units is to set it to unity.

There are other ways one might attempt to derive a potential from the ten-dimensional Einstein equations.  For example, consider a solution where $\tg_{\mu\nu}$ is maximally symmetric: de Sitter, Minkowski, or anti de Sitter.  The Einstein's equations for such a solution are \Einscosm\ with vanishing noncompact derivatives, and
\eqn\Einmn{G^m_n = e^{2A}\left[ {\tilde G}_n^m + 4({\widetilde {\nabla A}})^2\delta_n^m -8 \nabla_nA{\tilde \nabla}^m A\right] - \hf \delta^m_n e^{-2A} {\tilde R}^{(4)} = \kten^2 T^m_n\ .}

It's common to work with the trace-reversed $(\mu\nu)$ Einstein
equation, whose trace gives
\eqn\trtrev{e^{-2A} {\tilde R}^{(4)} - 4 e^{2A} \tnab^2 A = {\kten^2 \over 2}\left(T^\mu_\mu - T^m_m\right)}
This gives the four-dimensional curvature: multiplying  by $e^{-2A}$ and integrating over the compact space gives
%
\eqn\revtwo{  {\tilde R}^{(4)} = {\kten^2\over 2V_W} \int d^6 y \sqrt \tg e^{-2A} \left(T^\mu_\mu - T^m_m\right)\ .}

This expression was useful in Maldacena and Nu\~nez's ``no-go" theorem\refs{\MaNu}.  For both positive tension spacefilling branes and fluxes with purely compact components, $-T^\mu_\mu$ and $T^m_m$ are both positive.  Thus  \revtwo\ or related integrals of \trtrev\ were used to prove the ``no-go" theorem for both non-trivial warped compactifications, and de Sitter vacua.  This theorem was evaded in \GKP\ by the presence of negative tension objects,  O3 planes, or related sources in F-theory. One can explicitly check that both expressions \Potentg\ and \revtwo\ vanish for those solutions.

As pointed out in \Dealwisone, one might be tempted to define the
four-dimensional potential from \revtwo\ by  $\kappa_4^2 U =
{\tilde R}^{(4)}/4$.  However, taken as a formula for the
potential away from a minimum,  this gives a very disparate result
from \Potentg.  Specifically, notice that the terms in \Potentg\
are typically positive definite, whereas those in \revtwo\ are
typically negative definite.

The reason for the discrepancy is that the difference between
eqns.~\Potent\ and \revtwo\ comes from adding a multiple of the
equation of motion for the internal metric.  If the potential is
$U$, this equation takes the form $U'=0$ at an extremum of the
potential.  However, if we move away from such an extremum, for
example by adding a perturbation of the solution or potential, the
stationary solution for the moduli is no longer correct -- they
are rolling -- and the equation for the internal metric also
includes a kinetic term for the moduli. One can see this in the
perturbative expression (A.48) in the appendix.  Since it neglects
this term, \revtwo\ does not extend to a good expression for the
potential away from an extremum.  (Ref.~\Dealwisone\ pointed out
the relevance of such time dependence in the special case of the
radial modulus.)

\subsec{ Simple examples}

In the case of a brane or flux perturbation, the expression \Potentg\ indeed gives expected results.  A spacefilling $p$-brane wrapped on a cycle $\Sigma$ has ten-dimensional action
\eqn\braneact{S_p = - T_p \int_{M_4\times \Sigma} dV_{p+1}}
where $dV_{p+1}$ is the induced volume element, and thus stress tensor
\eqn\branestress{ T^\mu_\nu = -T_p \delta^\mu_\nu \delta(\Sigma)\ .}
This gives a term in the potential of the form
\eqn\branepot{\delta U_p =  {\kten^2\over V_W^2} T_p \int_\Sigma d^{p-3}z\sqrt{\tg_{\rm ind}}
 e^{(7-p)A}\ ,}
where $\tg_{\rm ind}$ is the induced world-volume metric.  In addition there will be corrections due to the shift in warping, found by solving equation \Aconst, and due to curvature.
Likewise for compact $q$-flux, the ten-dimensional action
\eqn\fluxact{S_q= -{1\over 4\kten^2} \int d^{10}x \sqrt{-g} {F_q^2\over q!}}
gives
\eqn\fluxstress{ T^\mu_\nu = - {1\over 4\kten^2} \delta^\mu_\nu {F_q^2\over q!}\ .}
This leads to a contribution
\eqn\fluxpot{\delta U_q= {1\over 4 V_W^2} \int d^6 y \sqrt{\tg_6} e^{-2A} {F_q^2\over q!}\ ,}
together with warping and curvature corrections.
Note that these expressions are positive, as expected.

We can also check that our potential \Potentg\ gives familiar dynamics for the radial dilaton, as described for example in \refs{\fate,\GiMe}, as well as better understanding the relation to the alternate expression \revtwo.
Consider the simpler case where we perturb an unwarped background by a stress tensor due to branes, fluxes, or other sources.   We start by parameterizing the metric as
\eqn\baack{
    ds^{2} = e^{-6D} \eta_{\mu \nu} dx^{\mu} dx^{\nu} + e^{2D}
    g^0_ {mn} dy^{m} dy^{n} \ .}
Here $D$ is the usual radion field which controls the volume of the internal
dimensions and $g_{mn}^{0}$ is a Calabi-Yau metric of fixed volume.  In this equation the
four metric is appropriately scaled so as to keep the Plank mass
fixed. Our goal is to determine the evolution of $D$ due to the sources.

Cast in terms of our standard notation this corresponds to
 $ e^{2A} = e^{-6D} $ and $ \tg_{mn} = e^{-4D} g^{0}_{mn} $.  We assume that the stress tensor takes the restricted form \specstress, and study the dynamics with the entire
potential treated as a perturbation.  For the present discussion we ignore  all but the radial dilaton modulus.
The perturbation from the stress tensor induces a space-time dependent change in $D$;
to linear order we write $D(x) = D+\delta D(x,y)$ and
\eqn\permet{ds^2 = e^{-6D} [1 - 6 \delta D(x,y)] ({\eta_{\mu\nu}+ \delta \tg_{\mu\nu}})dx^\mu dx^\nu + e^{2D}[1+2\delta D(x,y)] g^0_{mn} dy^m dy^n\ .}
We can easily check a metric of this form gives a solution to the linearized Einstein equations in the appendix.\foot{In some cases one must also introduce a $\delta g_{m\mu}$, but it produces corrections suppressed  in the Kaluza-Klein expansion.}  At linear order the $(\mu\nu)$ equation becomes
\eqn\linD{6e^{-6D}\delta^\mu_\nu\tnab^2 \delta D + e^{ 6D} {\tilde G}^\mu_\nu = \kten^2 T^\mu_\nu\ }
which is an equation for $\delta D(x,y)$.  It integrates to give the potential,
\eqn\potD{- U(D) = e^{-12D} \int d^6y \sqrt {g^0} {1 \over 4} e^{6D} T_\mu^\mu\ .}
 It can be shown that this gives potentials agreeing with the results in \fate, for example for branes \braneact\ or fluxes \fluxact.

 This potential drives $D$ to roll.  To check that we have properly identified it, note that the linearized $(mn)$ Einstein equation, contracted with $\tg_{mn}$ and integrated over the manifold, gives
\eqn\boxd{  -24 \sq \delta D = \kappa_4^{2} \int d^{6}y \sqrt{ g^{0} }
e^{-6D}( T^{m}_{m} - {3} T^{\mu}_{\mu} )}
For a source lagrangian of the form
\eqn\vterma{ -\int  d^{10}x \sqrt{ -g_{10} } U_{10}\ , }
with no velocity terms,
we find
\eqn\stressU{ T^\mu_\nu =-U_{10} \delta^\mu_\nu\quad ,\quad T^m_m = -6 U_{10} - {\partial U_{10} \over \partial D}}
and the $(mn)$ equation becomes
\eqn\newmnq{ 24 \sq \delta D = \kappa_4^2{\partial\over \partial D} U(D) }
confirming our identification of $U(D)$.  From these equations we also explicitly see that \revtwo\ is not the correct expression to identify as the potential, and differs from \potD\ by a multiple of \newmnq.

\subsec{Potential for IIB GKP solutions}

An even more non-trivial check on the expression \Potent\ comes
through evaluating it for the flux compactifications of \GKP. This
also gives us an expression useful for investigating the potential
in cases with extra D3 or anti-D3 branes present.  Here there are
contributions to $U_{10}$ from both three- and five-form fluxes,
and from local sources; for the present discussion we assume that
the axidilaton $\tau$ is constant, though the time dependent
generalization is readily included.

The contributions to $U_{10}$ from the five-form can be read from
the discussion of the appendix (subsection A.3); here we work only
to linear order in the compensators $S$ and $K$.  Likewise we can
infer the contributions from $G_3$ and from local sources with
six-dimensional mass density function $\mu(y)$.  This gives the
$tt$ Einstein equation,
\eqn\ttEins{\eqalign{ -&2 \tnab^2 A + 4(\widetilde {\nabla A})^2 -\hf {\tilde R}_6 - e^{-4A} \kappa_4^2 U + {1\over 4} e^{-8A}(\widetilde {\nabla \alpha})^2 -\hf \sq u^I (S_{Im}+K_I\partial_m\alpha)\partial^{\tilde m}\alpha  =\cr &
-{1\over 24 Im \tau}G_3\cdot{\bar G}_3 e^{-2A}-\kten^2 e^{-2A} \mu(y) + \calo(e^{-2A}\beta^2)}}
where as before we set velocities to zero in the equation for the potential.\foot{The surprising appearance of second derivatives arises from our treatment of the five-form; it can be eliminated using the five-form equation of motion.}  We also need the five-form equation, which follows from eq.~(A.36) in Appendix A, with local D3-charge density $\rho_3$:
\eqn\Fiveeqn{\eqalign{& \tnab\cdot(e^{-4A} \tnab\alpha)-e^{-8A} \tnab e^{4A}\cdot\tnab\alpha +
e^{-8A} \sq u^I \tnab e^{4A} \cdot(S_I+K_I\partial\alpha)\cr
&- \tnab\cdot\left[\sq u^I e^{-4A}(S_I+K_I\partial\alpha)\right]
 = {i\over 12 Im \tau} e^{4A}G_3{\widetilde{\cdot*}}{\bar G_3} + 2\kten^2 e^{-2A} T_3\rho_3\ .}}
Superposing these two equations gives an equation for the quantity
\eqn\adefone{a=\alpha- e^{4A}\ }
which takes the form
\eqn\eqnfora{\eqalign{&\hf \tnab \cdot(e^{-4A} \tnab a) + {1\over
4} e^{-8A} ({\widetilde {\nabla a}})^2 - \hf e^{-8A} \tnab^m
a\cdot (S_{Im} + K_I \partial_m \alpha) \sq u^I \cr &- \hf
\tnab^m\left[\sq u^Ie^{-4A}(S_{Im} + K_I \partial_m\alpha)\right]
-\hf {\tilde R}_6\cr &= - {e^{4A} \over 24 Im \tau} G_3 {\tilde
\cdot}({\bar G_3} -i {\tilde *} {\bar G_3}) + \kten^2
e^{-2A}(T_3\rho_3 -\mu) + e^{-4A}
\kappa_4^2U+\calo(e^{-2A}\beta^2)\ .}}

As in \Potent, the integral of this expression gives the potential (normalized for an arbitrary frame),
\eqn\gkppot{\eqalign{\kappa_4^2 U =& {1\over V_W^2} \int d^6y \sqrt{\tg} \Biggl\{ \kten^2 e^{-2A} (\mu(y) - T_3\rho_3)+{1\over 24 Im \tau} e^{4A} \left[G_3\tilde\cdot({\bar G}_3 - i \tilde*{\bar G}_3)\right]\cr&+{1\over 4} e^{-8A} (\widetilde{\partial a})^2 -\hf {\tilde R}_6-\hf e^{-8A} \sq u^I(S_{Im} + K_I \partial_m\alpha)\partial^{\tilde m}a + \calo(e^{-2A}\beta^2)\Biggr\} \ .}}
For the IIB flux compactifications of \GKP, mass sources and D3 charge sources are equal,
\eqn\massfcn{\mu=T_3\rho_3}
and the corresponding terms cancel.  Moreover, in the GKP solutions, $a={\tilde R}_6=0$, and $*G_3 =iG_3$: the potential vanishes as expected.  For complex structure/axidilaton perturbations away from this solution, the terms on the second line of \gkppot\ are subleading -- $\calo(N/c)$ -- in the Kaluza-Klein expansion, and so the potential takes the form
\eqn\vgkp{\kappa_4^2 U_{GKP} = {1\over V_W^2} \int d^6y \sqrt{\tg}\left[{e^{4A}\over 24 Im \tau} G_3^+ {\tilde \cdot} {\bar G}_3^+ + {\rm KK\ corrections}\right]\ ,}
where
\eqn\gparts{G_3^\pm = \hf(G_3 \pm i *G_3)\ .}
This agrees with the expression in \refs{\DeGi}, and with that in \GKP\ in the limit where warping is neglected.

\subsec{``No-go" results and their evasion: the slope dominance condition}

We are now in a position to better understand the no-go theorems for de Sitter space, and how they are evaded.  With the exception of internal curvature
typical sources such as positive-tension branes or internal fluxes all contribute positively to the potential \Potentg.  Moreover, any such contribution vanishes as the internal volume diverges.  This follows from the very general argument of ref.~\refs{\fate}:  as we see in \Potentg, the four-dimensional potential is essentially the ten-dimensional energy over $V_W^2$, and no realistic dynamics can produce a ten-dimensional energy that overcomes the fall-off $1/V_W^2$.  Thus each contribution gives a term $U_i$ with $U_i'<0$ in the supergravity regime where calculations are reliable, so with only positive-energy sources there can be no non-trivial minimum.

On the other hand, a source of negative $-T^\mu_\mu$ will generally give a contribution likewise asymptoting to zero at infinite volume, and thus contributes a term with $U_i'>0$.  Such terms thus play a key role in obtaining de Sitter vacua (which will necessarily be at best metastable by the arguments of \fate).  Thus, a necessary condition for such dynamics to stabilize moduli at a finite volume is that there must be a range of moduli over which such a  positive contribution to $U'$ dominates over the contributions of the other terms -- we refer to this as ``slope dominance."

\newsec{Potentials for Kahler deformations}

As found in \GKP, complex structure moduli become massive from the presence of non-trivial three-form flux.  Deformations of the Kahler moduli remains massless to leading order.  However, as was emphasized in \GKP, these deformations also have a potential when higher-order terms in the $\alpha'$ expansion are taken into account.  One set of such terms is the $\calo(\alpha^{\prime 3})$ terms of \refs{\BHL}.  In addition, in \refs{\KKLT}, Kachru, Kallosh, Linde, and Trivedi argued that non-perturbative effects can also generate a potential that they argued dominates the dynamics for certain values of the parameters.  Finally, \KKLT\ also considered configurations with explicit anti-D3 branes and thus supersymmetry breaking, which they argued produce de Sitter vacua.  We next turn to a discussion of some of these effects in the context of our expanded understanding of the perturbative dynamics of deformations of IIB warped flux compactifications.

\subsec{Perturbed potential: generalities}

A formula for the potential in warped IIB compactifications was given in \gkppot, and shown to agree with the results of \GKP,\DeGi\ in the case where the sources were D3 branes.  However, the formula is applicable to a wider class of sources.  Various effects -- extra anti-branes, non-perturbative effects, and $\alpha'$ corrections -- lead to perturbations of the mass density function $\mu(y)$ that enters this equation.  In some cases -- notably to analyze the proposed solutions of \KKLT -- all we need is the perturbation of the potential to linear order in the change in $\mu(y)$ away from the ``BPS"-saturated expression.  (An important exception to this is the case where we introduce $D3$-$\overline{D3}$ pairs and study inflationary potentials, as in section 7.)

At linear order in $\delta(\mu-T_3\rho_3)$ (and leading order in
the Kaluza-Klein expansion), the formula for the change in the
potential is easily inferred. Eqn.~\gkppot\ gives the linear
expression
\eqn\gkppert{\kappa_4^2 \delta U = {\kten^2\over V_W^2} \int d^6y \sqrt{\tg} e^{-2A} \delta( \mu-T_3 \rho_3)\ .}
We consider this expression for the different perturbations of \KKLT, anti-D3 branes and euclidean D3 brane/gaugino condensation effects, in turn.

\subsec{Potential from anti-D3 branes}

We begin with the effect of adding an anti-D3 brane to a GKP solution.  Notice, however, that simply adding a $\overline{D3}$ is not consistent -- the charge in the compact space will no longer sum to zero.  To maintain charge conservation, one must add D3 branes and anti-branes in pairs, or consider changing the three-form fluxes $M$ and/or $K$ of \GKP\ such that the change in the D3 charge $N=MK$ compensates the charges of a stack of $\overline{ D3}$ branes.  The former case leads to mobile D3 branes and will be considered more closely in section 7.  The latter case can be heuristically thought of as the result of the inverse of the transition described in \refs{\KPV,\FLW} where anti-D3 charge and flux annihilates.

Either case produces a dipolar source for $\rho_3$ and a monopole source for $\mu(y)$.  To linear order in the charge $\rho_3$ of the $\overline {D3}$'s, the effect on the potential follows immediately from \gkppert.  For example, for $L$ $\overline {D3}$ branes located at point ${\bar y}_0$, this expression gives
\eqn\Padt{\kappa_4^2 \delta U = 2LT_3{\kten^2\over V_W^2}  e^{4A({\bar y}_0)}\ .}
Note that this interpolates between the $1/c^2$ dependence
described in \KKLMMT, when the brane is in a highly-warped region,
and the $1/c^3$ dependence of \KKLT\ if it isn't.

\subsec{Non-perturbative potentials}

A full treatment of the dynamics of non-perturbative effects should properly be carried out within the context of the ten-dimensional theory; then in certain regimes these will have a good four-dimensional effective description.  Here we outline aspects of such an analysis for effects due to euclidean D3 branes and gaugino condensation.

\subsubsec{ Euclidean D3 branes }

We begin with the case of euclidean D3 branes wrapping non-trivial four-cycles of the compact manifold.  These were argued in \refs{\Wittinst} to give rise to a non-perturbative superpotential, which \KKLT\ then used to fix Kahler moduli.

Since the four-dimensional effective theory has a limited regime of validity, one would like to understand the ten-dimensional description of these effects.  While a complete analysis presumably requires a non-perturbative understanding of string/M theory, one can nonetheless give an approximate description.

The starting point is the observation that, whatever the ultimate description of string theory is, the integral over fluctuations about a spacetime of the form ${\bbb R}^4\times  M^6$ (or the generalization to a warped product) includes both configurations of the supergravity (string) fields, and configurations of branes.  Thus, accounting for euclidean branes,
we expect that string dynamics can be approximated by a path integral of the form
\eqn\edbrpath{  Z = \int \cald g_{MN} \cald B_{MN}\cald \phi\cdots e^{i S_{IIB}}\sum_n { 1
\over n!}  \left( \int\cald\theta^{a}\cald X^i \cald A^{m}  e^{ -S_{brane} (X^i, A^m, \theta^a)} \right)^{n}   \ .   }
Here $g_{MN}$, $B_{MN}$, $\phi,\ldots$ are bulk supergravity fields, and included in the ellipses should be fermionic and auxiliary fields.  The brane action is written in the Green-Schwarz form;
 $X^i$ and $A^{m}$ are the bosonic brane embedding coordinates and gauge fields,  and $\theta^a$ are the fermionic brane embedding coordinates. The sum incorporates the effects of an arbitrary number of brane instantons, and can be performed to give
\eqn\expo{ Z = \int \cald g_{MN}\cald B_{MN}\cald \phi \cdots e^{iS_{IIB} + \int \cald
X \cald \theta \cald Ae^{ -S_{brane} ( X^i, A^m, \theta^a)}}\ .}

At this point, one can see how D3 instantons modify the classical
supergravity equations arising from the action $S_{IIB}$.  In the
dilute instanton approximation,  the brane terms give an effective
action
\eqn\Dteff{iS_{ED3}[g_{MN}, B_{MN}, \phi,\cdots]  = \int \cald
X \cald \theta \cald A e^{ -S_{brane} ( X^i, A^m, \theta^a)}}
for the supergravity fields where the brane functional integral is
evaluated in the relevant supergravity background.  There will be
corresponding corrections to the 10D equations of motion analyzed
in appendix A.

A full and direct analysis of the ten-dimensional equations of motion, including these instanton effects,  would require inclusion of the auxiliary fields of type IIB supergravity, and in particular the dependence of  $S_{brane}$ and thus $S_{ED3}$ on these fields.  Specifically, determination of the potential induced by non-perturbative effects would follow from finding the dependence of $S_{ED3}$ on the auxiliary fields, and then integrating these fields out.

While we leave the details of such an analysis for future work, we can see how
 \expo\ gives the superpotential of \Wittinst.   In many backgrounds of interest, there will be zero modes of the brane fields $\theta^a$. The integral over brane configurations has the general structure
\eqn\suppot{  \int \cald X\cald\theta\cald A e^{ -S_{brane} (X^i, A^m, \theta^a)} =  { det_{nzm} F \over det_{nzm} B } \int d^{\zeta}
\eta e^{ -{\bar S}_{brane}} }
where the functional determinants arise from the non-zero modes,
${\bar S}_{brane}$ is the action for the minimal area
configuration of the brane, and the Grassman variables $\eta$
parameterize the $\zeta$ zero modes in the given supergravity
background.  The supersymmetry transformations on the odd brane
coordinates take the form
\refs{\AganagicNN\CederwallRI-\BergshoeffTU}
\eqn\branex{\delta \theta^a = \epsilon^a}
where $\epsilon$ is the parameter of the SUSY transformation.  From this, we see that upon reduction to a four-dimensional supersymmetric theory, translations of the zero modes $\eta$ correspond to translations in the usual four-dimensional superspace coordinates $\theta$.  Thus, in cases where there are two zero modes, the expression \suppot\  gives a superpotential contribution to the four-dimensional theory, of the form
\eqn\superp{ W = { det_{nzm} F \over det_{nzm} B }e^{ - S_{brane} } \ .}

As argued in \Wittinst, on Calabi-Yau manifolds, the number of zero modes is determined by the holomorphic Euler character of the four-cycle in question.
In the case where this holomorphic Euler character is one, there are precisely two zero modes and hence a superpotential.  For higher holomorphic Euler character, the leading order dynamics leads to more zero modes, na\"\i vely indicating that \suppot\ cannot give an F-term.
But in this case
 \refs{\Kachruetal,\Kalloshetal,\Saulina} showed that the presence of a flux background lifts some zero modes and can permit a superpotential; the criterion for this was given in terms of a topological index in \Kalloshetal.  More detailed investigation of the dynamics of warped/flux backgrounds could yield other effects that lift zero modes in cases where the holomorphic Euler character is greater than one.

 Moreover, even with more that two zero modes, in the case of a SUSY breaking background such as GKP where the Gukov-Vafa-Witten superpotential \gvwpot\ has non-vanishing value $W_0$,
 such non-perturbative effects can still generate corrections to the potential, of order $|W_0|^2$.

Holomorphicity of the superpotential \superp\ also indicates how to identify holomorphic coordinates corresponding to the Kahler moduli.  Let $D^i$ represent one of the non-trivial four cycles of an underlying Calabi-Yau manifold.  The leading result in \superp\ arises from the classical minimal-area brane configuration.
To leading order in $\alpha'$ the action of the (euclidean) brane wrapping this cycle takes the form
\eqn\brane{S_{brane} = T_3 \int_{D^i} d^{4}z  \sqrt{\tg_{ind}} e^{-4A}-i \mu_{3}\int_{D^i} {\tilde C}_{4} }
where the integral is performed over the minimal-area surface in the appropriate class and ${\tilde C}_4$ is the sum of the potential and the dual potential for $F_5$.  (On shell, with $*F_5=F_5$, the expression simply reduces to the potential.)
Indeed, this motivates a definition of the complex coordinates $\rho_i$ representing Kahler deformations:  if $D^i$ correspond to $h^{1,1}$ independent four-cycles, we can define
\eqn\holoK{ \rho_i =  i\int_{D^i} d^{4}z \sqrt{\tg_{ind}}e^{-4A}  + \int_{D^i} {\tilde C}_{4}\ .}
Notice that this definition is independent of change of frame, \scalemetric, \warpscale.

\subsubsec{ Gaugino Condensation on D7 stacks}

A similar analysis exhibits the origin of the superpotential resulting\KKLT\ from gaugino condensation on stacks of D7 branes.
The starting point is again \edbrpath, where now the brane action is for D7's, the brane configuration is classical rather than euclidean, and there is no instanton sum.  For $N$ D7's, the gauge group is $SU(N)$, and zero modes $\eta,\phi$ of the fields $\theta^a, X^i$ correspond to massless matter of the worldvolume theory.  In the limit where we neglect massive fluctuations of the D7 about the minimal area cycle, the functional integral
\eqn\dsint{\int\cald\theta^{a}\cald X^i \cald A^{m}  e^{ iS_{D7} (
X^i, A^m, \theta^a)} = { det_{nzm} F \over det_{nzm} B } \int \cald \eta\cald \phi \cald A e^{iS_{D7}}\ }
reduces to that of a low energy supersymmetric $SU(N)$ gauge theory with matter fields $\eta,\phi$.  Gaugino condensation is then described directly in four-dimensional terms.  The superpotential's dependence on the size modulus then arises from the form of the effective coupling of the four-dimensional theory,
\eqn\Dsevcon{  { 1 \over g_{YM}^{2} } =  { 2\pi  \int d^4z \sqrt{\tg_{ind}} e^{-4A}\over g_{s}
} \ ,}
and corresponding dependence on $C_4$ likewise arises through the topological coupling
$\int C_4\wedge F\wedge F$.

\subsubsec{Form of potential}

Based on work of \Wittinst, ref.~\KKLT\ argues that the potential that would arise from a calculation along the above lines can be thought of as arising from a superpotential that supplements \gvwpot\ with non-perturbative effects,
\eqn\kkltW{W= \int \Omega\wedge G +\sum_k A_k e^{ia_k\rho_k}= W_0 + W_{NP}\ ,}
where we take $k$ to parametrize four-cycles corresponding to elements of $H^{1,1}$.  In the simplest case of one Kahler modulus the resulting potential takes the form
\eqn\kkltV{\kappa_4^2U_{NP}= e^K\Upsilon}
where $K$ is the full Kahler potential and  where  we define
\eqn\upsdef{\Upsilon = (\rho-{\bar \rho})\left[ \left(\overline{\partial_\rho W_{NP}} W_0 - h.c.\right) + \left(\overline{\partial_\rho W_{NP}} W_{NP} - h.c.\right) - {(\rho-\bar\rho)\over 3} \partial_\rho W_{NP} \overline{\partial_\rho W_{NP} }\right]\ ,}
with straightforward generalization to more Kahler moduli. Thus,
if we were able to calculate the contribution to the stress tensor
from the non-perturbative effects, by comparing with \Potentg\ we
expect it to have the form
\eqn\oneinst{{\kten^2\over V_W^2} \int d^6y \sqrt{\tg} e^{-2A} (T^\mu_\nu) = -e^K \Upsilon \delta^\mu_\nu}
where the piece linear (quadratic) in $W_{NP}$ should come from a one- (two-) instanton effect in our ten-dimensional analysis.

\subsec{De Sitter solutions}

Now we consider the setup of \KKLT, with both anti-D3 branes and non-perturbative effects.  If we work to linear order in both these effects, \gkppert\ makes it clear that we can find the potential by adding \Padt\ and \kkltV,
\eqn\linKKLT{\kappa_4^2 U_{KKLT,1} = \kappa_4^2 U_{NP}+ 2LT_3{\kten^2\over V_W^2}  e^{4A({\bar y}_0)}\ .}
Beyond linear order in $LT_3$ and $\Upsilon$ this formula receives corrections.

From these expressions we can see explicitly how these configurations evade the ``no-go" results for de Sitter solutions, which we outlined in section.~5.4.  The one-instanton contribution in \upsdef\ has a phase that arises from the axion partner to $\rho$.  Minimum energy is thus attained when this axion adjusts itself so that its phase combined with that of $W_0$ makes this contribution real and negative, as assumed in \KKLT.  Consequently for some values of the parameters, the potential can produce a minimum at positive vacuum energy.

\newsec{Higher corrections and regimes of validity}

This section will describe corrections to our effective action formulas, and their relevance to the KKLT approach to finding de Sitter vacua\KKLT.    While we focus on the IIB compactifications of \GKP, many of our statements here as in the rest of the paper generalize.

To summarize the results of the preceding sections, the scalar (moduli) perturbations of the solutions of \GKP\ are governed by a four-dimensional effective action.  This can be derived by study of the ten-dimensional form of the perturbations.  The kinetic terms are expected to take a form such as \KDG, although a precise check of the warping corrections there has not yet been made.  The $G_3$ flux-induced potential, which comes from our very general formula \Potentg, takes the form \vgkp.  This gives masses with sizes \mflux\ to the complex structure moduli.  At
supergravity level, the Kahler moduli remain massless.

\subsec{Corrections -- sources and magnitudes}

As was pointed out in \GKP, higher order corrections will lift the
Kahler flat directions; \KKLT\ provided an explicit example of
this.  Let us begin by enumerating the various corrections and
their magnitudes.

\subsubsec{$\alpha'$ corrections}

The expansion parameters for string corrections are of the form
\eqn\strcorr{\alpha' R_S\quad ,\quad \alpha' p_S^2\ ,}
where the subscript $S$ denotes string-frame quantities.  Converting to four-dimensional Einstein units, we find small expansion parameters
\eqn\Eincorr{\alpha' e^{-\phi/2 + 2A} {\tilde R}_6\quad ,\quad
\alpha' e^{-\phi/2 - 2A}V_W {\tilde p}_4^2\ .}
The arguments of \BHL\ indicate that curvature corrections begin only at third order in $\alpha'$.

\subsubsec{Warping corrections}

Much previous analysis of the solutions of \refs{\GKP,\KKLT} has neglected warping.  Our results allow us to make generic statements about the size of corrections due to warping.  The magnitude of the corrections depends on the location of the ten-dimensional phenomenon being studied, relative to the warped regions.  This is particularly clear for example from the potential formula \Potentg.  At large $c$, corrections due to warping are of order
\eqn\warpsize{{\delta U\over U} \sim {\int d^6y \sqrt \tg {e^{-4A_0} \over c} U_{10}\over \int d^6y \sqrt \tg  U_{10}}\ . }
Consider for example the warping produced by a stack of $N$ D3 branes.  For a ten-dimensional phenomenon with $U_{10}$ relatively uniformly distributed with respect to the metric $\tg_{mn}$, \warpsize\ gives
\eqn\corru{{\delta U\over U} \sim {N\over c} \int_0^1 r^5 dr {1\over r^4} \sim {N\over c}\ .}
This size differs from $\alpha'$ corrections \Eincorr\ by the expected power of $g_sN$, corresponding to the AdS radius of the region close to the D3 branes.  Thus, for situations with large D3 charge, it can make sense to keep warping corrections while neglecting other $\alpha'$ corrections.  A similar estimate of the magnitude of warped corrections applies to the kinetic terms \KDG, in cases where the metric perturbations are not concentrated in highly warped regions.

For phenomena concentrated in highly warped regions, the corrections can be more substantial.  A classic example is the potential due to an anti-D3 brane, \Padt, which depends directly on the warp factor at the brane.  An extreme case is to place such a brane near a stack of D3 branes -- it will sink to the bottom of the throat, and its energy will vanish.  Likewise, corrections to other terms in the action can also be large in cases where the corresponding expressions are concentrated in regions of large warping.

\subsubsec{Kaluza-Klein corrections}

The Kaluza-Klein expansion was briefly discussed in section 4.2, where it was found that the Kaluza-Klein expansion parameter is of the form
\eqn\KKexpan{e^{-4A} V_W {\tilde p}_4^2}
for a phenomenon characterized by four-dimensional Einstein frame momentum ${\tilde p}_4$.
Note that the $\beta^2$ corrections in our general potential formula \Potentg\ are also suppressed in this expansion.  Neglecting warping, the expansion parameter is of order $c^2 {\tilde p}_4^2$, but in the case where Kaluza-Klein modes are excited in regions of large warping the expansion breaks down at much lower energies.

\subsubsec{String loop expansion}

String loop corrections are suppressed by powers of $g_s$.  There can also be string non-perturbative effects, from D-instantons and  euclidean D branes.  If $V_{a,S}$ represents the string-frame volume of the cycle $C_a$ wrapped by the brane, the expansion parameter governing the contribution of such effects takes the form
\eqn\npexp{ e^{-{V_{a,S}\over g_s}}\ .}

\subsec{KKLT models}

It is particular important to understand the role of such corrections in attempts such as that of KKLT\KKLT\ to construct phenomenologically realistic de Sitter vacua.  The original analysis of \KKLT\ argued for the neglect of $\alpha'$ corrections, warping corrections, Kaluza-Klein corrections, and string loop corrections, but kept one class of nonperturbative corrections in $g_s$, due to either euclidean D3 branes or gaugino condensation on D7 stacks.

Beginning with the solutions of \GKP, we know that warping corrections and Kaluza-Klein corrections cannot lift the Kahler directions since, as we've reviewed, they correspond to flat directions (combined with certain complex structure deformations) of the full ten-dimensional theory.  On the other hand, $\alpha'$ corrections, $g_s$ corrections, and non-perturbative corrections are expected to generically give a potential in these directions\GKP.

The leading $\alpha'$ corrections were studied in \BHL.  There it
was argued that these corrections shift the Kahler structure part
of the Kahler potential,
\eqn\Kalpha{K(\rho,{\bar \rho})= -2\ln V \rightarrow -2 \ln\left(V + \hf {\xi\over g_s^{3/2}}\right)}
where $\xi$ is given in terms of the Euler character $\chi$ of the compact manifold as
\eqn\xidef{\xi = -{\chi\over 2} \zeta(3)\ .}
These corrections then lead to a non-zero potential for the universal Kahler modulus, of the form
\eqn\alphapot{ \kappa_4^2U_{\alpha'} \sim e^K \delta K |W_0|^2 \sim e^K |W_0|^2 {\xi\over |\rho-{\bar \rho}|^{3/2} g_s^{3/2}}\ .}

Our discussion of how to evade ``no-go" theorems, in section 5.4, tells us that we must find a contribution to the potential that is negative and gives dominant contribution to the slope in some region of moduli space.  Ref.~\KKLT\ invokes the contributions from euclidean D3 branes, or gaugino condensation on D7 branes, with superpotentials of the form $W_{NP}$ in \kkltW, and giving a potential of the form \kkltV.  Recall that this term is minimized when the axion in $\rho$ adjusts its phase such that the one-instanton contribution proportional to $W_0$ is real and negative.  Thus this is a natural candidate for the mechanism needed to achieve a de Sitter vacuum.

The resulting potential (with KKLT's added $\overline{D3}$ brane) consists of \linKKLT, together with $\alpha'$ corrections, as well as corrections to this potential from Kaluza-Klein modes and warping.\foot{There also could be non-perturbative contributions from, {\it e.g.}, other euclidean branes.}  The latter two do not lift the flat directions by themselves, but we expect they could generically give corrections of order ${\cal O}\left({N\over c}\right)$ to non-zero terms in the potential, or, in cases of dynamics in strongly warped regions, even larger corrections.  (We are also informed of explicit calculations\refs{\BHKIP} of open-string loop corrections at order $\delta K\sim g_s N/c$.)

Let us examine the slope dominance condition from our  ``no-go" discussion.  After the axion partner to $\rho$ has adjusted, and then taking $\rho=i\sigma$, we have a one-instanton potential (in the simplest one-modulus case)
\eqn\oneinstpot{\kappa_4^2U_1 = - 4 e^K \sigma a |AW_0| e^{-a\sigma}\propto -{a|AW_0|\over \sigma^2} e^{-a\sigma}\ . }
Thus the corresponding slope is
\eqn\slopeinst{ \kappa_4^2U' = 4a e^K|AW_0|{e^{-a\sigma}}(2+ a\sigma)\ .}

Validity of the instanton expansion requires
\eqn\instval{a\sigma>1\ .}
From \Eincorr, we find validity of the $\alpha'$ expansion requires
\eqn\alphaval{\sigma\gg{1\over g_s}\ ,}
and recall that validity of the string loop expansion requires $g_s\ll1$.
For the slope \slopeinst\ to dominate over that of the $\alpha'$ corrections \alphapot, we find the condition
\eqn\slopdom{ |A| a {e^{-a\sigma}}(2 + a\sigma) \roughly> {|W_0| \xi \over g_s^{3/2}\sigma^{5/2} }\ ,}
or
\eqn\slopedomtwo{|A| (a\sigma)^2  (\sigma g_s)^{3/2} e^{-a\sigma} \roughly> |W_0|\ .}
This gives us a fine-tuning condition for $W_0$.
For example, for euclidean $D3$ branes, $a=2\pi$.  Moreover, \alphaval\ and for example $g_s\sim 1/10$ then gives the extreme tuning condition
\eqn\exttun{10^{-24}\gg|W_0| \ .}
KKLT found more success with gaugino condensation with large gauge
groups, where $a=2\pi/N$.  This weakens the tuning condition
\slopedomtwo\ considerably, but still requires moderate
fine-tuning of $W_0$.  For example, even $N=60$, so that $a=0.1$,
still requires $|W_0|\ll 1$.

If a $\overline{D3}$ is added as in KKLT, the condition for slope dominance of the instanton contribution over the $\overline{D3}$ potential \Padt\ is
\eqn\domdt{T_3 e^{4A({\bar y}_0)} \roughly<g_s |W_0A| a
(a\sigma)e^{-a\sigma}\ .}
In light of the fine-tuning for $|W_0|$ and \alphaval, this
implies that the warping at the $\overline{D3}$ location must be
significant.

While in the present constructions, these are the only obvious places that large warping is important, in general large warping should be relevant to moduli potentials in any context when the effects generating the potentials are concentrated in regions of large warping; in this case it lowers the energy scales of the corresponding phenomenon.  In particular,  large warping can lead to breakdown of four-dimensional analysis at the redshifted energy scale in \fdrange, much lower than the na\"\i ve  scale $E\sim 1/R^4$.  Note that even in the context of \KKLT\ this possibility could be relevant and further restrict the regime of validity of the KKLT solutions, given the localization of the deformation controlling the conifold to the region at the bottom of a warped throat (see the next subsection).

As a final comment on these constructions, we note that other non-perturbative corrections due to euclidean branes could contribute to the potential once supersymmetry is broken.  For example, in the IIB context, these include D-instantons, euclidean D-strings wrapping two cycles, and even five branes wrapping the entire compact manifold.  However, since the latter two break the supersymmetries of the original compactification, their contributions are expected to be suppressed by fermion zero modes.  They could contribute to the potential once supersymmetry is broken, but one would expect the contributions to be suppressed by the square of the gravitino mass  $\sim|W_0|$.  Taking into account dependence on the size of cycles, this suggests potential contributions due to euclidean $p$ branes of the general form
\eqn\dpots{ U_p \sim |W_0|^2 e^{-{a\over g_s} (g_s\sqrt\rho)^{(p+1)/2}}\ .}
The smallness of $|W_0|$ suggests that these should be suppressed, but such contributions should be further investigated.  Contributions of D-instantons have been discussed in \refs{\CQS}.

\subsec{Statistics on the landscape}

Another place where one should consider warping is in the statistical arguments about string vacua\refs{\MDone\MDtwo\MDthree\MDfour-\MDfive}.  These arguments are based on a particular form for the super and Kahler potentials.  While we don't believe that the Gukov-Vafa-Witten superpotential is modified by warping, the Kahler potential does appear to be.  An important part of the arguments of  \refs{\MDone\MDtwo\MDthree\MDfour-\MDfive} is the independence of the Kahler potential on the fluxes.  However, \KDG\ suggests Kahler potential modifications of the form
\eqn\Kahz{K(z,\bar z) \sim -\log \left( -i\int e^{-4A} \Omega\wedge {\bar \Omega}\right)\ ,}
although, as we have emphasized, there could be more complicated
dependence on the warping as well.  Such expressions depend on the
flux through the warping.  While typically one would expect the
flux corrections to be small  at large radius -- of relative order
$N/c$ -- there could be exceptions.  Notably, consider the
deformation of the conifold studied in \GKP.  Corresponding to
\Kahz, the metric for complex structure deformations would take a
form
\eqn\zmet{{\cal G}_{\alpha {\bar \beta}}\sim \int e^{-4A}  \chi_\alpha\wedge {\bar \chi_\beta}\ }
where $\chi_\alpha$ form a basis for the (2,1) forms.  In particular, as has been found in explicit studies of conifold geometry\refs{\CveticDB,\HerzogXK}, the (2,1) form corresponding to the conifold modulus is concentrated in the vicinity of the shrinking cycle, where the warp factor becomes large.  While a precise statement requires more careful study of the structure of the metric ${\cal G}_{\alpha {\bar \beta}} $ in the context of compact geometry, this suggests
enhanced dependence on the warping, and thus the fluxes (as well as flux-dependent suppression of the mass-scale associated to the conifold modulus $z$).  So it appears  that warping could modify landscape statistics arguments in regions with strong warping.

\newsec{Dynamics of mobile branes}

Another important application for an understanding of the dynamics of warped compactifications is the case of mobile branes.  The idea that the modulus corresponding to a moving brane could serve as the inflaton field has been widely investigated since \refs{\DvTy} (for a review see \refs{\Quev}), but the idea typically experiences difficulties with known interbrane potentials, primarily a failure of slow roll.  Refs.~\refs{\KKLMMT} suggested that warping could lower interbrane potentials and allow slow roll, but found that this na\"\i ve hope was difficult to realize due to contributions to the slow-roll parameter $\eta=U''/U$ from the stabilization mechanism.  At the same time, this raised a puzzle about the correct definition of the holomorphic parameters corresponding to Kahler moduli in the presence of moving branes, and their coupling to D7 brane gauge excitations (the ``rho problem," see refs.~\refs{\Kalloshrho,\Kalloshtalk}).  A possible resolution was presented, in a special case, after a lengthy analysis by Berg, Haack, and K\"ors in \refs{\BHK} (see also \BHKtwo).

In this section we will investigate this complex of ideas using our deeper understanding of the origin of the potential on the landscape.  In particular, our formula \Potentg\ allows us to give a systematic approach to computing interbrane potentials.  This gives a more generic understanding of the origin of the problem with slow roll.  It also gives a clear resolution of the rho problem and the question of moduli coupling to D7 gauge theories.

\subsec{Kinetic terms for D3 branes}

 The dynamics of space filling D3 branes provides promising
candidates for the inflaton field. In this section and the
subsequent one we obtain the kinetic and potential terms
associated with this dynamics by using the DBI action
\eqn\DBI{ S  = - T_{3} \int  d^{4} \xi [-\det(g_{\alpha
\beta})]^{1/2} + \mu_{3} \int {\tilde C}_{4} }
  Here $\xi^{\alpha}$ are coordinates on the world volume of the brane
and $g_{\alpha \beta}$ the pullback of the ten dimensional metric
on the brane, and $\tilde C_4$ includes the dual potential to $C_4$.

                 For space-filling branes, the world volume
coordinates can be taken to same as those describing the
non-compact directions and the brane motion in the compact
directions can be parametrized as
\eqn\ydef{ y^{m} = y_{0}^{m} + y^{m}(x) }
Working in a general frame, this gives the pullback metric
$g_{\alpha \beta}$ as
\eqn\pback{ g_{\alpha \beta} = e^{2A} \eta_{\alpha \beta} +
e^{-2A} \tg_{mn}  \partial_{\alpha} y^{m}
\partial_{\beta} y^{n}\ .   }
The kinetic terms can then be obtained from the first term in
\DBI\ and are given by
\eqn\sclrkina{  -\hf T_{3}  \int
d^{4}x\tilde{g}_{mn}\partial_{\alpha} y^{m}\partial^{\alpha}
y^{n}\ .}
The frame relevant to the four dimensional dynamics is the
Einstein frame, where the warped volume is set to a constant, say  $V^{0}$. An
expression which reproduces this answer but can be evaluated in
any frame is
\eqn\sclrkina{  -\hf T_{3} { V^{0} \over V_{W} } \int
d^{4}x\tilde{g}_{mn}\partial_{\alpha} y^{m}\partial^{\alpha}
y^{n}\ .}

\subsec{Potentials for brane motion}

The structure of the potential for mobile branes is of particular interest, since brane positions have been widely explored as candidates for inflatons, for example in the work of \refs{\KKLMMT} and subsequent developments.

For our basic setup, we assume that we start with a GKP solution, and then add some balanced brane and antibrane charge.  The simplest way to do this is to add one or more anti-branes, and an equal number of branes.  Alternately, the charge of some of the antibranes can be balanced by changing the ISD fluxes (as discussed in section 6.2).  For concreteness we focus on the case of brane-antibrane pairs, with straightforward generalization to the case of shifted fluxes.

Adding a brane-antibrane pair at positions $y_0$ and $\bar y_0$ shifts the sources $\rho_3$ and $\mu$ as follows:
\eqn\deltarho{\delta \rho_3 = {\delta^6(y-y_0)\over \sqrt g} -  {\delta^6(y-\bar y_0)\over \sqrt g}\ ,}
\eqn\deltamu{\delta \mu = T_3 {\delta^6(y-y_0)\over \sqrt g} +T_3 {\delta^6(y-\bar y_0)\over \sqrt g}\ .}
For more pairs we simply superpose such shifts. The equation \gkppot\ for the potential of the GKP solutions was derived with enough generality to incorporate such sources.  The presence of the $\overline{D3}$ explicitly breaks supersymmetry and leads to a nonzero potential.  If we are interested in the potential for $D3$ motion, we must then evaluate $U$ to quadratic order in the perturbation given by \deltarho\ and \deltamu.

Both $a$ and $\mu-T_3\rho_3$ are linear in these sources, and we drop higher order terms in the Kaluza-Klein expansion, which are suppressed by $N/c$.  Thus, to quadratic order in the mobile brane tensions, the potential is
\eqn\quadpot{\eqalign{\kappa_4^2 U_P(\rho_k,y_0,\bar y_0) = 2{\kten^2T_3 e^{4A(\bar y_0)}\over V_W^2} &+ 2{\kten^2T_3 \delta e^{4A(\bar y_0)}\over V_W^2}+{1\over V_W^2} \int d^6y\sqrt\tg \left[ {1\over 4} e^{-8A}(\widetilde{\nabla a})^2\right]  \cr &- 4\kten^2 T_3 {\delta V_W\over V_W} {e^{4A(\bar y_0)}\over V_W^2} \ ,}}
where in this expression $V_W$ and $A$ are the {\it unperturbed}
values.  The first term is the cosmological constant contribution
due to the anti-brane, and the second includes the interbrane
potential, which can be small for large warping, as well as self
energy of the $\overline{D3}$.  The third term gives a
compensating  self-energy correction, and the last term leads to a
shift in the potential for $y_0$ that is proportional to the
original potential. To evaluate these terms we solve for $a$ and
$\delta e^{-4A}$ to linear order.

The quantity $a$ is determined by \eqnfora, which, again to linear order in $a$ and leading order in the KK expansion, becomes
\eqn\alin{ \tnab \cdot(e^{-4A} \tnab a)  = 2\kten^2  e^{-2A}(T_3\rho_3 -\mu) + 2e^{-4A} \kappa_4^2\delta U\ .}
This simplifies in the metric $g$ to
\eqn\newa{ \nabla^2 a = 2\kten^2  e^{4A}(T_3\rho_3 -\mu) + 2 e^{2A}\kappa_4^2\delta U\ .}
The consistency condition from the integral of this expression fixes $\delta U$, which is the linear (first) term in \quadpot.  The quantity $a$ can thus be found in terms of the Green function for the laplacian (with convenient background charge term),
\eqn\grdef{\nabla^2_y G(y,y') = {\delta^6(y-y')\over \sqrt g} - {1\over V}\ .}
The linear order $a$ is thus
\eqn\alinord{ a(y) = \int d^6y' \sqrt{g} G(y,y') \left[2\kten^2  e^{4A}(T_3\rho_3 -\mu) + 2 e^{2A}\kappa_4^2\delta U\right]\ .}

Next, $\delta e^{-4A}$ can be found from the five-form equation, \Fiveeqn.  Eliminating $\alpha$ by \adefone\ and using the $a$ equation \eqnfora, we find
\eqn\deltaefa{ \delta e^{-4A} = -\int d^6y'\sqrt{\tg} {\tilde G}(y,y') \left( 2\kten^2  \delta\tilde\mu - e^{-4A} \tnab e^{-4A} \cdot\tnab a-2 \kappa_4^2 e^{-8A} \delta U\right)\ ,}
where the quantities $\tilde G$, and $\tilde \mu$ are computed using the metric $\tg$.

To summarize, these equations tell us that the source for $a$ is the $\overline{D3}$ density, the source for $e^{-4A}$ is the mass density, and likewise the source for $\alpha$ is the D3 charge density.  Substituting \alinord\ and \deltaefa\ into \quadpot\ and using
\eqn\deltavw{\delta V_W = \int d^6y \sqrt\tg \delta e^{-4A}}
gives an explicit formula for the potential, as a function of the position of the branes.

\subsubsec{Slow roll?}

As expected, the potential is (locally) minimized when the $\overline{D3}$ sits at the bottom of a Klebanov-Strassler throat in the warped compactification.  To stabilize the Kahler moduli, we include the non-perturbative potential \kkltV\ generated from \kkltW,
\eqn\Ueqn{U= U_P+ {1\over \kappa_4^2} \left[e^K\Upsilon + \delta\left(e^K\Upsilon\right)\right]\ ,}
where $\Upsilon$ was defined in equation \upsdef. Consistent with
our other formulas, we expect that $e^K\propto 1/V_W^2$ in this
formula. In \quadpot, the potential for the  motion of the mobile
D3 comes from the second and last terms  -- these are the only
terms  that depend on $y_0$.  The second term gives the interbrane
potential; the redshift from $e^{4A}$ means that this can give a
slow-roll parameter
\eqn\etadef{\eta= {U''\over U}}
that is small.  However, the last term in eqn.~\quadpot\ together
with the $y_0$ dependent terms in $\delta(e^K \Upsilon)$ also
contribute  to $\eta$.  The following argument indicates that this
contribution is proportional to the cosmological constant
$U_{min}$.

To understand the dependence on $y_0$, note  that a good
definition of holomorphic coordinates is the (complexified) warped
volume of the minimal four-cycles, eqn.~\holoK, as these are
precisely the variables occurring in the non-perturbative
superpotential \kkltW.  An alternative way to motivate this is to
recall from section 3 that we've identified the imaginary part of
the universal Kahler modulus as the constant part of $e^{-4A}$,
but such a definition depends on  the decomposition into a
constant part and the background $e^{-4A_0}$.  The full $e^{-4A}$
is the geometrically defined quantity, and since it depends on the
compact coordinates, it must be integrated to define a quantity
that can be identified as a modulus.  For the non-perturbative
superpotential, the relevant integral is then that over a four
cycle.

The presence of the $D3-\overline{D3}$ source shifts the warp-factor and hence the definition of the holomorphic coordinates by a non-holomorphic piece that depends on $y_0$.  That is, the good holomorphic coordinates for the new problem -- with mobile branes -- are related to the original holomorphic coordinates $\rho_{0,i}$ by
\eqn\holoreln{\rho_i  = \rho_{0,i} + i\int_{D^i} d^4 z \sqrt{\tg_{ind}} \delta e^{-4A}\ .}
(The shift from $\tilde C_4$ is typically subleading.)
The dependence of this shift on $y_0$ follows from the first term in \deltaefa, and thus takes the form
\eqn\delrho{\delta\rho_i = -2i\kten^2T_3\int_{D^i} d^4 z \sqrt{\tg_{ind}} {\tilde G}(z,y_0)\ .}

The warped volume also experiences an explicit shift $\delta V_W$
from \deltaefa, \deltavw.  In order to correctly minimize the
potential, we write the warped volume in terms of the {\it new}
$\rho_i$:
\eqn\Vshift{V_W = V^0_W(\rho_i - \delta \rho_i(y_0)) + \delta V_W \approx V^0_W(\rho_i)- \delta \rho_i(y_0)\partial_{\rho_i} V^0_W + \delta V_W\ .}
The explicit dependence $\delta V_W$ on $y_0$ is weak -- for example, it clearly vanishes on the torus $T^6$.  The combined potential \Ueqn\ now fixes the new coordinate $\rho_i$
and the warped volume picks up the $y_0$ dependent piece from the Kahler modulus dependence in \Vshift.  This dependence is generically $\calo(1)$  since \delrho\ is computed in terms of {\it unwarped} quantities, and from \quadpot\ and \Ueqn\ we see that this shift generically makes an $\calo(1)$ contribution to $\eta$ proportional to $U_{min}$.

In short, the potential essentially fixes the holomorphic Kahler moduli $\rho_i$, but at fixed $\rho_i$ it depends on the $D3$ positions not only through the interbrane potential, but also through $V_W$.

A first check on this argument immediately follows.
Recall that in the coincident limit of a D3 brane with a D7 brane, the gauge coupling should vanish\Gano.
We see this behavior quite explicitly in \delrho, as in a small neighborhood the Green function behaves as
\eqn\greensm{{\tilde G}(z,y_0) \sim {1\over (z-y_0)^4}}
and its integral diverges when $y_0$ touches a $D7$ brane wrapped on the cycle $D^i$. The identification \Dsevcon\ thus implies vanishing coupling constant.

Different forms of this argument were made in \refs{\KKLMMT,\BHK,\BuRo,\McAl}. The relationship of our argument to those can be better understood by investigating the resolution of what has been called the ``rho problem."

\subsec{The rho problem and Kahler potential}

The approach of \refs{\KKLMMT} led to a puzzle\refs{\Kalloshrho,\Kalloshtalk} known as the ``rho problem."  The Kahler potential proposed in \DeGi\ that correctly reproduces no-scale structure and the kinetic terms for mobile branes takes the form
\eqn\DeGiK{K(\rho,{\bar \rho}; \phi, {\bar \phi})= -3 \ln\left[-i(\rho-\bar \rho) + k(\phi,\bar \phi)\right]\ }
where $\phi$ are fields corresponding to brane positions $y_0$.
Comparison with known brane kinetic terms suggested that the relation between the holomorphic Kahler modulus and the volume is given by identifying the volume with the argument of the log:
\eqn\rhoprop{-i (\rho-\bar \rho) = V^{2/3} - k(\phi,\bar \phi)\ .}
But then the relation $g_{YM}^2\propto V^{2/3}$ for the gauge coupling in the D7-brane action $Tr(F^2)$ suggests that the gauge coupling is not the real part of a holomorphic function, in contradiction to SUSY.

This question appears cleanly resolved in our approach.  In our definition of holomorphic coordinates \holoK, the gauge coupling \Dsevcon\ is manifestly the real part of a holomorphic function, and with the identification
\eqn\newK{K = -3 \ln V_W\ ,}
eqn.~\Vshift\ also explains the presence of the shift term in the argument of \rhoprop.

However, a small puzzle remains.  Consider a simple example where
the compact space is a product of three tori, $T_1^2\times
T_2^2\times T_3^2$ (or some orbifold thereof).  Then the shift in
the $\rho_i$ that corresponds to $D^i=T_1^2\times T_2^2$, given by
\delrho, only depends on the coordinates of the D3 brane along
$T^2_3$, that is it only depends on the coordinates perpendicular
to the four cycle.  Eqn.~\Vshift\ then does not obviously give the
expected form for $k(\phi,\bar\phi)$ in \DeGiK. Indeed, while it
was not emphasized in \BHK, the corrections found there for
example in the case of $K3\times T^2$ had precisely this property.
(This serves as another check on our analysis; indeed, our
calculation can apparently be thought of as a closed-string dual
to the open string loop calculation done in \BHK.)

While this looks promising for generating a flat direction in the potential, arising from the corresponding shift symmetry as in \refs{\Kalloshshift,\Shandera}, at the same time this does not appear to give a result of the form \rhoprop.

An apparent resolution of this puzzle arises from the formula
\Vshift, which gives a sum of terms corresponding to {\it each} of
the four cycles.  The democracy among the cycles should give an
expression that depends on all components of the D3 position; this
is exactly what happens on the torus.  Indeed, one can see a
possible origin for such a statement in a more general geometry,
at least in the large volume limit.  Parameterize the Kahler class
of the underlying metric $\tg$ in terms of elements of
$H^2(M,{\bbb R})$ as $\tJ = \sum_i t^i J_i$, so that for
two-cycles $D_i$ dual to the four-cycles $D^i$,
\eqn\Jint{\int_{D_i} J = t^i\ .}
In the unwarped limit, the four-cycle moduli $\rho_i$ are related to the $t^i$ by
\eqn\modreln{ \rho_i = \partial_{t^i} {\tilde V}\ .}
At large $c$ we find
\eqn\VWt{V_W = c\int \tJ^3 + \int e^{-4A_0} \tJ^3 = {c\over 3} \rho_i t^i \left[1+\calo\left({N\over c}\right)\right]\ .}
Accounting for dependence of the warp factor on the coordinate of the D3 gives the shift \delrho; we might anticipate that likewise  two-cycle coordinates could be defined in terms of the actual metric,
\eqn\tdef{t^i = i\int_{D_i} d^2z {\sqrt {\tg_{ind}}} e^{-2A}+ \int_{D_i} C_2}
although we don't expect these to be naturally holomorphic for the supersymmetry corresponding to D3 branes.
These likewise would thus receive a $y_0$ dependent shift
\eqn\tshift{t^i = t_0^i +i \int_{D_i} d^2z \sqrt {\tg_{ind}} \delta e^{-2A}\ .}
The induced dependence on the distance to the cycle $D_i$ would introduce complementary dependence to the distance to the dual cycle $D^i$ resulting from \delrho.  This apparently resolves the puzzle, but for the same reason, we expect shift symmetries that would generate flat directions in the potential to be generically lifted.

 \newsec{Conclusion}

This work has improved the systematic understanding of warped compactifications, which we have seen develop a possibly central relevance in the phenomenology and cosmology of string theory.  While our systematic treatment of time-dependent solutions is only at the linearized level, we have been able to address many questions, and in  particular have provided a very general formula for the potential that extends beyond the linear approximation.  This paper has also clarified multiple issues in proper description of moduli of the flux compactifications of \GKP, and in understanding their lift to ten-dimensional dynamics.  Beyond that, it has begun a careful investigation of other effects, including non-perturbative effects, in the ten-dimensional warped context.  Such analysis is critical to understanding in what cases vacua with positive cosmological constant arise, and ultimately will be important to the better understanding of the phenomenology and statistics of such solutions.  It is also important for investigating possible origins of inflation.  There are many open questions that remain in the physics of warped compactifications, but this paper should make some of them more accessible and thus bring us a step closer to a full understanding of their rich dynamics.

\bigskip\bigskip\centerline{{\bf Acknowledgments}}\nobreak

The authors would like to thank S. de Alwis, D. Berenstein, O. DeWolfe, A. Frey,  C. Herzog, S. Kachru, J. Louis, G. Moore, D. Morrison, P. Ouyang, J. Polchinski, S. Sethi, S. Shenker, and S. Trivedi  for valuable conversations, and M. Berg and M. Haack both for several useful conversations and for comments on a draft of this paper.  They also thank the organizers of String Phenomenology 2004 for an opportunity to present preliminary results of this work, and the Fields Institute for its hospitality and support during its Workshop on Gravitational Aspects of String Theory.  This work was supported in part by Department of Energy under Contract DE-FG02-91ER40618, and by a UCSB Regent's Fellowship.

\appendix{A}{Linearized spacetime-dependent perturbations of warped compactifications}

In this appendix  we will derive the linearized equations of motion for fluctuations about  warped compactifications, with particular focus on the IIB case of \GKP.  We begin by studying the general form of the perturbations.  We then write the equations of motion, and finally explain how they are solved.

\subsec{Metric perturbations}
\subsubsec{Time dependent perturbations and compensators}

We begin with a general warped background metric of the form
 \eqn\pbackmet{ds^{2} = e^{2A} \eta_{\mu \nu} dx^{\mu} dx^{\nu} +
          e^{-2A} \tilde{g}_{mn} dy^{m} dy^{n}\ .}
%
%
Time-independent perturbations of this metric arising from moduli take the form
\eqn\ppertstat{\delta A = u^I \delta_I A\ ,\ \delta \tg_{mn} = u^I \delta_I\tg_{mn}\ .}
%
%
where $u^I$ denotes a general modulus parameter, and $\varI A$ and
$\varI \tg_{mn}$ are the corresponding changes in the warp factor
and metric.  If we wish to generalize this to a
spacetime-dependent variation (or a linear combination thereof),
parameterized by $u^I(x)$, we must consider a general perturbation
of the form 
%
%
 \eqn\gppert{\eqalign{&ds^{2} = \left[ e^{2A}  +u^I(x)\varI  e^{2A}\right]\eta_{\mu\nu}dx^\mu dx^\nu +
          \left[e^{-2A}\tg_{mn} + u^I(x)\varI\left(e^{-2A} \tg_{mn}\right)  \right] dy^{m} dy^{n} \cr
      &+ 2\partial_\mu\partial_\nu u^I(x) e^{2A} K_I(y) dx^{\mu} dx^{\nu}  + 2e^{2A} B_{Im}(y) \partial_\mu u^I(x) dx^\mu dy^m  \ .}}
The first line follows from the static deformation \ppertstat, but for a spacetime dependent perturbation the terms in the second line are also allowed.

We can also consider tensor perturbations of the four-dimensional metric,
\eqn\tenpert{ds^2 \rightarrow ds^2 +   e^{2A} f^K(y) \delta_K g_{\mu\nu}(x) dx^{\mu} dx^{\nu}}
labeled by index $K$, which may include Kaluza-Klein excitations with nontrivial compact dependence encoded in $f^K(y)$.

We refer to the quantities $K_I$ and $B_{Im}$ as ``compensators" for the metric.  These
may be eliminated by ten-dimensional diffeomorphisms generated by vectors
 \eqn\kelim{\xi^{\mu} =
-  K_I(y) \partial^{\mu } u^I(x) \ \  , \ \ \xi^{m} =
e^{4A} \partial^{\tilde m} K_I(y) u^I(x) }
 and
 \eqn\belim{\xi^{m} = -e^{4A}B_I^{\tilde m} u^{I}(x)\ , }
respectively.  However,
note that these transformations also induce a diffeomorphism on the internal six-dimensional space, so that in the new coordinates, with vanishing compensators, $A$ and the six-dimensional metric have changed by
\eqn\diffmet{\varI \tgmn\rightarrow \varI \tgmn + \calL_{\xi_I} \tgmn \ \ , \ \ \varI A\rightarrow \varI A + \calL_{\xi_I} A\ , }
where $\calL_{\xi_I}$ is the Lie derivative with respect to the vector
\eqn\liedef{\xi_I^m = e^{4A}\left(\partial^{\tilde m} K_I - B_I^{\tilde m}\right) \ .}
This means that if we fix a familiar gauge such as transverse gauge
\eqn\metgauge{{\tilde \nabla}^m \varI \tgmn=0\ ,}
this does not necessarily agree with the gauge in which the
compensators vanish.  Indeed, we will find that the equations of
motion imply that either we can fix the gauge \metgauge\ or we can
fix a gauge where the compensators $K_I$ and $B_{Im}$ vanish, but
these two gauges are not in general mutually compatible.

\subsubsec{Curvature and Einstein Tensor}

Metric fluctuations are controlled by Einstein's equations, for which we need  the components of the Ricci and Einstein tensors for the metric perturbation \gppert.  Specifically, one can work out the change of these tensors induced by the deformations \gppert, \tenpert\ about the metric \pbackmet.

The components of the change in the Ricci tensor follow from the general formula
 \eqn\pwald{
         \delta R_{MN} = -{ 1 \over 2} \nabla^{P} \nabla_{P}
         \delta g_{MN} - { 1 \over 2} \nabla_{M} \nabla_{N} g^{PQ}
         \delta g_{PQ} + \nabla^{P} \nabla_{(M }\delta g_{N)P}\ .}
The fully non-compact part takes the form
\eqn\pdelrmunu{\eqalign{ \delta R^{\mu}_{\nu} = - \delta^{\mu}_{ \nu}\Bigl[&u^I \varI\left(e^{2A} {\tilde \nabla}^2  A\right) + \sq u^I e^{-2A} \varI A \Bigr] + e^{-2A}\partial^\mu\partial_\nu u^I(4\varI A - \hf \varI \tg)\cr&+e^{2A}\partial^\mu\partial_\nu u^I \tnab^m(B_{Im} - \partial_m K_I ) + e^{2A} \delta^\mu_\nu\sq u^I \partial^{\tilde m} A(B_{Im} - \partial_m K_I)\cr
&+ e^{-2A} f^K(y) \delta_K R^{(4)\mu}_\nu - \hf e^{2A}\left[ \delta_K g^\mu_\nu \tnab^2 f^K + \delta^\mu_\nu \delta_K g^\lambda_\lambda \partial^{\tilde m} A \partial_m f^K \right]\ ,}}
where $\delta_K R^{(4)\mu}_\nu$ denotes the  four-dimensional Ricci tensor resulting from $ \eta_{\mu\nu} + \delta_K g_{\mu\nu}$ and indices are raised on $\delta_K g$ and $\partial_\mu\partial_\nu u^I$ using $\eta^{\mu\nu}$.
The remaining curvatures are
\eqn\pdelrmmu{\eqalign{ \delta R_{m }^{ \mu} = e^{-2A} \partial^\mu u^I \Biggl\{&2 \partial_m\varI A - 8 \partial_m A \varI A - {1\over 2} \partial_m \varI \tg + \partial_m A \varI \tg\cr & -2 \partial^{\tilde p}A \varI \tg_{mp} + \hf {\tilde \nabla}^p\varI \tg_{mp} \cr
&-\hf\tnab^p\left[e^{4A}\left(\tnab_p B_{Im} - \tnab_m B_{Ip}\right)\right] + 2(\partial_m A B_{Ip} - \partial_p A B_{Im}) \tnab^pe^{4A}
 \cr&+ \hf  e^{8A} B_{Im}\tnab^2 e^{-4A}   -  e^{4A} {\tilde R}_m^n B_{In}\Biggr\}\ ,} }
and
\eqn\pdelrmn{\eqalign{\delta R_{n}^{ m} = &u^I \varI\left[ {\tilde R}_n^m + {\tilde \nabla^2} A \delta_n^m - 8 \partial_n A \partial^{\tilde m} A \right] + e^{-2A} \sq u^I \left( \varI A \delta_n^m -\hf \tg^{mk}\varI \tg_{kn} \right)\cr
&+\hf e^{-2A} \sq u^I \Biggl\{\tnab^m\left[e^{4A}\left( B_{In} - \partial_n K_I\right)\right] + \tnab_n \left[ e^{4A} \left(B_{I}^{\tilde m}- \partial^{\tilde m} K_I \right) \right]\cr & -\hf \delta^m_n \partial^{\tilde p} e^{4A} \left(B_{Ip} -\partial_p K_I\right)  \Biggr\} \cr &
 - {1\over 4} \delta g_{K\lambda}^\lambda e^{-2A}\left[ \tnab^m\left( e^{4A} \partial_n f^K\right)+\tnab_n\left( e^{4A} {\tilde \partial}^n f^K\right) -\hf \delta^m_n\tilde\partial^p e^{4A} \partial_p f^K\right]
\ .}}

{}From these, it is easy to deduce the perturbation in the Einstein tensor:
\eqn\pgmunu{\eqalign{ \delta G^{\mu}_{\nu} = &\delta^\mu_\nu u^I \varI\left\{e^{2A}\left[ -2 {\tilde\nabla}^2A +4 ({\widetilde {\nabla A}})^2 -\hf {\tilde R}\right]\right\} + e^{-2A} \left(\partial^\mu\partial_\nu u^I - \delta^\mu_\nu \sq u^I\right) (4 \varI A -\hf \varI \tg)\cr &
+\left(\partial^\mu\partial_\nu u^I - \delta^\mu_\nu \sq u^I\right)e^{2A}\tnab^p(B_{Ip} - \partial_p K_I)\cr& +e^{-2A} f^K\delta_K G^{(4)\mu}_\nu -\hf\left(\delta_K g^\mu_\nu - \delta^\mu_\nu \delta_K g^\lambda_\lambda\right)e^{2A}\tnab^2 f^K \ , }}
\eqn\pgmmu{\eqalign{ \delta G_{m }^{\mu} = \delta R_{m }^{\mu} =& e^{-2A} \partial^\mu u^I \Biggl\{2 \partial_m\varI A - 8 \partial_m A \varI A - {1\over 2} \partial_m \varI \tg + \partial_m A \varI \tg\cr & -2 \partial^{\tilde p}A \varI \tg_{mp} + \hf {\tilde \nabla}^p\varI \tg_{mp} \cr
&-\hf\tnab^p\left[e^{4A}\left(\tnab_p B_{Im} - \tnab_m B_{Ip}\right)\right] + 2(\partial_m A B_{Ip} - \partial_p A B_{Im}) \tnab^pe^{4A}
 \cr&+ \hf  e^{8A} B_{Im}\tnab^2 e^{-4A}   -  e^{4A} {\tilde R}_m^n B_{In}\Biggr\}\ ,} }
and
\eqn\pgmn{\eqalign{ \delta G_{n}^{m} =& u^I \varI \left\{e^{2A}\left[ {\tilde G}_n^m + 4({\widetilde {\nabla A}})^2\delta_n^m -8 \nabla_nA{\tilde \nabla}^m A\right]\right\} -\hf e^{-2A} \sq u^I \tg^{mk} \varI \tg_{kn}\cr & + \delta_n^m e^{-2A} \sq u^I ( -2\varI A + \hf \varI \tg)\cr
& \sq u^I \Biggl( \hf e^{-2A} \left\{\tnab^m\left[e^{4A}\left( B_{In} - \partial_n K_I\right)\right] + \tnab_n \left[ e^{4A} \left(B_{I}^{\tilde m}- \partial^{\tilde m} K_I \right) \right]  \right\} \cr &
-\delta^m_n \tnab^p\left[e^{2A}\left(B_{Ip} -\partial_p K_I\right)\right]  \Biggr)
\cr & + \hf \delta_K g_\mu^\mu \left\{-\hf e^{-2A} \left[ \tnab^m\left(e^{4A} \partial_nf^K\right) + \tnab_n \left(e^{4A} \partial^{\tilde m} f^K\right)\right] + \delta_n^m \tnab^p\left[e^{2A} \partial_p f^K\right] \right\}
\cr&-\hf\delta^m_n f^K e^{-2A} \delta_K R^{(4)}
\ .}}
These will be used in the perturbed Einstein equations in section A.5.

\subsec{Perturbations of $G_3$}

In the warped compactifications of \refs{\GKP}, fluctuations of the metric also couple to fluctuations of the three- and five-form NS-NS and Ramond-Ramond fields.  We first consider the three-form case.

With
\eqn\Gtdef{G_3 = F_3 - \tau H_3 = dC_2 - \tau dB_2\ ,}
we find
\eqn\Gtpert{\delta G_3 = d \delta C_2-\tau d\delta B_2 - \delta\tau H_3 ,}
The general form for $\delta C_2$, $\delta B_2$ (without exciting ``model independent" axions) is then
\eqn\deltcalA{\eqalign{\delta C_2 &= u^I \varI C_2 + du^I\wedge T_I\cr
\delta B_2 &= u^I \varI B_2 + du^I\wedge R_I\ ,}}
where $\varI C_2$, $\varI B_2$ correspond to the static variations of the potential and $T_I$ and $R_I$ are compensators for the three forms.  From this we find
\eqn\Gtpertt{ \delta G_3 = d\left[ u^I\left( \varI C_2 - dT_I\right)\right] -\tau d\left[ u^I\left( \varI B_2 - dR_I\right)\right]  -\delta\tau H_3\ .}
The compensators $T_I$, $R_I$ are determined by the three-form equations of motion.

In order to write Einstein's equations, we also need the stress tensor  for this perturbation.  With stress tensor
\eqn\genstress{T_{MN} = -{2\over \sqrt{-g}} {\delta S \over \delta g^{MN}}\ ,}
we find for the three-form
\eqn\threestress{T_{M}^{(3)N} = {1\over 8 \kappa_{10}^2 {\rm Im}\tau} \left( G_{3MPQ}{\bar G}_{3}^{NPQ} +G^N_{3PQ}{\bar G}_{3M}{}^{PQ}  - {1\over 3} \delta_M^N G_3\cdot {\bar G}_3\right)\ .}
In particular, the linearized $\mu\nu$ component takes the form
\eqn\threestmunu{ \delta T_\nu^{(3)\mu} = -{ 1\over 24 \kappa_{10}^2}  \delta _\nu^\mu \delta \left[{1\over {\rm Im} \tau} \left(G_3\cdot {\bar G}_3\right)\right]}
and we will parameterize the mixed components as
\eqn\threestmix{ \delta T_m^{(3)\mu} = {e^{-2A}\over 2 \kappa_{10}^2} \partial^\mu u^I T_{Im}^{(3)}\ .}

\subsec{Perturbations of $F_5$}
\subsubsec{Form of perturbation}

We next turn to perturbations of the five-form.   This
 can be written in terms of a potential $C_4$ as
\eqn\pfivepar{ F_{5} = dC_{4} + *dC_{4} }
The background four-form is taken as in \refs{\GKP} as
 \eqn\pgkpcpot{ C_{4} =  \alpha(y) \  d^{4}x\ . }
The general form of the perturbation is
\eqn\pcpert{ \delta C_{4} = u^I(x) \varI \alpha  d^{4}x + du^I
\wedge S_I^{(3)} + *_{4} du^I \wedge S_I + u^I(x) w_I^{(4)} +
D^I_{(2)} \wedge w_I^{(2)} \ .}
Here the compensators  $S_I$,  $S_I^{(3)}$ are one and three forms with components only in the compact
directions.  Likewise, $w_I^{(2)}$ and $w_I^{(4)}$ are two and four forms with components only in the compact
directions, necessary to describe the axionic excitations; $D^I_{(2)}$ is a  two form with all
components in the non-compact directions, which is related to $u^I$ by the equations of motion.
Using the metric \gppert, this gives
\eqn\pfivepert{\eqalign{ &\delta F_{5} = \left\{u^I(x)\left[d(\varI \alpha) + e^{8A} \tstar dw_I^{(4)}\right]-
\sq u^I(x) S_I\right\} d^{4}x
                 - *_{4} du^I \wedge d S_I   \cr & - du^I \wedge ( dS_I^{(3)}- w_I^{(4)}) + u^I(x) dw_I^{(4)}
 + \sq u^I(x) e^{-8A}\tstar \left[S_I+K_I d\alpha\right]\cr & - u^I\varI \left(e^{-8A} \tstar d\alpha \right)
             - e^{-4A}  du^I \wedge \tstar \left[dS_I + B_I\wedge d\alpha\right]  \cr &
              - e^{+4A} *_{4} du^I
             \wedge \tstar (d S_I^{(3)} - w_I^{(4)}) + (1+*) d \left(D^I_{(2)} \wedge w_I^{(2)}\right)\ . }}

\subsubsec{Equation of motion and solutions}

The equation of motion for the five-form is
\eqn\eomfive{ dF_5 = {G_3\wedge {\bar G_3}\over {\bar \tau}-\tau} +2 \kappa_{10}^2 T_3 *_6\rho_3^{\rm loc}  \ .}
Working about a background corresponding to one of the solutions of \GKP, the linearized equations then take the form
\eqn\linfive{d\delta F_5 = \delta\left( {G_3\wedge {\bar G_3}\over {\bar \tau}-\tau} \right)+ 2 \kappa_{10}^2 T_3\delta*_6\rho_3^{\rm loc} \ .}
To simplify the discussion, we can set the axionic pieces $w_I^{(2)}$ and $w_I^{(4)}$ to zero, consistent with the equations of motion, and likewise the compensator $S_I^{(3)}$ can be taken to vanish.
We then find from \pfivepert
\eqn\pbiancond{\eqalign{ d\delta F_5= & -u^I(x)
d\left[\delta_I\left(e^{-8A}\tstar d\alpha\right)\right] + \sq
u^I(x) d\left[ e^{-8A} \tstar (S_I +K_I d\alpha)\right]
    \cr&+
    du^I \wedge \left\{  d\left[ e^{-4A} \tstar (dS_I+ B_I \wedge d\alpha)\right] - \delta_I\left(e^{-8A}\tstar d\alpha\right) \right\}
    \cr&+ d \sq    u^I(x)\wedge \left[ e^{-8A} \tstar (S_I+K_I d\alpha)\right]\
    .}}

The source term in \linfive\ then follows from the three-form perturbation, \Gtpertt.  This gives
\eqn\vargwedge{\delta\left( {G_3\wedge {\bar G_3}\over {\bar \tau}-\tau} \right) = d(u^I(x)\beta_I^{(5)})\ ,}
where
\eqn\betafdef{\beta_I^{(5)} = (\varI C_2 - dT_I)\wedge H_3 - F_3\wedge (\varI B_2 - dR_I)}
%
is a five-form on the internal manifold.
Consider the case of a massive perturbation (with obvious restriction to the massless case), which corresponds to
\eqn\masspert{\sq u^I = m_I^2 u^I\ .}
The five-form equation then reduces to
\eqn\fivered{ d\left[e^{-4A} \tstar (dS_I+ B_I\wedge
d\alpha)\right] + m_I^2 e^{-8A} \tstar(S_I+K_I d\alpha) =
\varI\left( e^{-8A} \tstar d\alpha\right)+  \beta_I^{(5)} + 0 \ .}
(The local part is zero, since the branes continue to lie in the
  non-compact directions, i.e they are still point-like on the
  internal manifold )
%
For a given $m_I^2$ and metric compensators $B_I$, $K_I$, which are determined by the Einstein equations, eq.~\fivered\ determines the five-form compensator $S_I$ and the perturbation $\delta_I\alpha$.  A useful equation for determining $\delta_I\alpha$ in the limit when the compensators can be neglected arises from the derivative of \fivered, which gives
\eqn\newfive{\eqalign{u^I&\varI \left( \tnab^2 \alpha - 2 e^{-4A} \tnab^m e^{4A} \tnab_m \alpha\right) - \sq u^I e^{8A} \tnab^m\left[e^{-8A} \left( S_{Im} + K_I \partial_m \alpha\right)\right] \cr&= u^I \delta_I\left(i e^{8A} {G_{mnp} \tstar {\bar G}^{\widetilde{mnp}} \over 12 Im\tau} + 2 \kappa_{10}^2 e^{2A} T_3 \rho_3^{\rm loc}\right)\ .}}

\subsubsec{Energy-momentum tensor}

In preparation for solving Einstein's equations, we need the energy-momentum tensor for the perturbation \pfivepert; again, for the solutions of interest, we neglect axionic excitations and set  $w_I^{(2)}=w_I^{(4)}=S_I^{(3)}=0$. The perturbation in the energy momentum tensor is  given by
\eqn\deltmunu{ \delta T_{\nu}^{\mu} =-{\delta_\nu^\mu} {1\over 4\kappa_{10}^2} \left\{ u^I\varI\left[e^{-6A} (\widetilde {\nabla \alpha})^2\right] -2 e^{-6A} \sq u^I S_{Im} \partial^{\tilde m} \alpha -2\sq u^I K_I e^{-6A} (\widetilde{\nabla \alpha})^2\right\}\ ,}
\eqn\pdeltkmu{\eqalign{ \delta T_{m}^{ \mu} =& {1\over 2
\kappa_{10}^2 }
\partial^\mu u^I e^{-6A} \left[\partial_m S_{Ip} -\partial_p
S_{Im} +\partial_m\alpha B_{Ip} -\partial_p \alpha B_{Im}\right]
\partial^{\tilde p}\alpha\
\ ,}}
and
\eqn\pdeltmn{\eqalign{ \delta T_{n}^{ m}&=-{1\over 2 \kappa_{10}^2} u^I\varI\left\{e^{-6A}\left[\partial_n \alpha \partial^{\tilde m}\alpha - \hf \delta_n^m(\widetilde{\nabla\alpha})^2\right]\right\} \cr&+ {e^{-6A}\over 2 \kappa_{10}^2} \sq u^I \left\{ S_{In} \partial^{\tilde m}\alpha + \partial_n\alpha S_I^{\tilde m} - \delta_n^m S_{Ip} \partial^{\tilde p}\alpha+ 2K_I\left[\partial_n \alpha \partial^{\tilde m} \alpha -\hf \delta_n^m (\widetilde{\nabla \alpha})^2\right]\right\}\ .}}

\subsec{Perturbations of $\tau$}

Before introduction of three-form flux there are also massless perturbations of the complex field $\tau$, defined in \taudef; introduction of flux then makes these massive.  The $\tau$ equation of motion follows from \IIB\ and takes the form
\eqn\taueom{\nabla_M\nabla^M \tau = {\partial_M\tau \partial^M \tau\over i {\rm Im} \tau} - {i\over 12} G_3\cdot G_3\ .}
The linearization of this is
\eqn\taulin{e^{-2A} \sq u^I \delta_I \tau + e^{2A} u^I \tnab^2 \delta_I\tau = -{i\over 6} u^I \delta_I \left(e^{6A} G^+ {\tilde\cdot} G^-\right)\ }
where $G^{\pm}$ are defined in eq.~\gparts.  In the orientifold case where the background $\tau$ is constant, the linearized stress tensor due to $\tau$ vanishes.

\subsec{Equations of motion for perturbations}

We next outline the solution of the perturbed equations of motion
for light modes of the warped IIB compactifications of \GKP.  In
addition to the three- and five- form equations discussed in
sections A.2, A.3 and the $\tau$ equation of section A.4, these
include the perturbed Einstein equation
\eqn\pertEins{\delta G_N^M = \kappa_{10}^2 \delta T_N^M\ .}
For general matter $\phi$ coupled to gravity, the presence of a zero mode corresponds to the existence of deformations
\eqn\zmdef{\delta g_{MN} = u^I \varI g_{MN}(y)\ ,\ \delta\phi = u^I \varI \phi(y)\ ,}
with constant $u^I$, that satisfy the Einstein equations \pertEins.  For example, in the case of warped IIB compactifications \refs{\CPV} without three-form flux, the complex structure and Kahler deformations of the underlying Calabi-Yau, together with the corresponding deformations of $F_5$, are zero modes, as described in section 3.  Turning on flux then lifts the complex structure modes, as explained in \GKP.

\subsubsec{Summary of equations of motion}

A spacetime-dependent perturbation takes the more general form
described in the preceding subsections, with compensators present,
and solution of the equations of motion is consequently more
complicated.  We begin by summarizing the relevant equations we
must solve.

The three-form perturbation is given by \deltcalA.  The equation
of motion for this perturbation, resulting from varying the action
\IIB, then determines the variation of the potential, $\delta_I
{\cal A}_2$, and the three-form compensator $T_I$.  The five-form
perturbation is given by \pcpert.  For a perturbation with
definite value of four-dimensional $p^2 = -m^2$ (and axionic
pieces set to zero), the five-form equation reduces to \fivered\
which determines the variation $\delta_I\alpha$ and the
compensator $S_I$.  The axidilaton perturbation $\delta_I\tau$ is
determined to linear order by \taulin.

The metric perturbation is given in \gppert\ and \tenpert.  These perturbations are fixed by Einstein's equations.  We consider stress tensor perturbations due to perturbations of the three- and five-forms, along with explicit sources such as localized branes.  The $(m\mu)$ equation follows from the Einstein tensor \pgmmu, and the stress tensors \threestmix\ and \pdeltkmu, together with a possible piece from other sources such as branes:
\eqn\mmuein{\eqalign{-\tnab^p &\left[e^{4A} \left(\tnab_p B_{Im} -\tnab_m B_{Ip} \right)\right] + e^{8A} B_{Im} \tnab^2 e^{-4A} -2 e^{4A} {\tilde R}_m^n B_{In} \cr = &e^{4A} \left[\partial_m\left( \delta_I  e^{-4A}\right) -\tnab^p\left( e^{-4A} \delta_I {\tilde g}_{mp} \right)\right] + e^{2A} \partial_m\left(e^{-2A} \delta_I \tilde g\right)\cr & + e^{-4A} \left( \partial_m S_{Ip} - \partial_p S_{Im}\right) \partial^{\tilde p} \alpha + T_{Im}^{(3)} + T_{Im}^{\rm source}\ .\quad\quad\quad (m\mu)}}
Here we have parametrized the source contribution by
\eqn\tlocal{T_m^{{\rm source}\,\mu}= {e^{-2A}\over 2 \kappa_{10}^2} \partial^\mu u^I T_{Im}^{\rm source}\ .}
We can think of  equation \mmuein\ as determining the metric compensators $B_I$.

A simplifying assumption for the source stress tensor is that the $\mu\nu$ and $mn$ components take the form \specstress.
%
This is in general violated by velocities for the moduli (but only
at quadratic order), but will hold near an extremum of the
potential for the moduli.  With this assumption, the equations
simplify.  We will also assume that the only non-zero four-metric
perturbation has $f^K$ unity and $\delta G^{(4)\mu}_\nu \propto
\delta^\mu_\nu$.  This will be valid when the Kaluza-Klein modes
of the four-metric are not excited, and if we restrict to linear
order in perturbations.

With these assumptions the $(\mu\nu)$ Einstein equation has two kinds of terms, proportional to $\partial_\mu\partial_\nu u$ and $\eta_{\mu\nu}$, respectively.  From \pgmunu\ and \deltmunu\ we see that the coefficient of $\partial_\mu\partial_\nu u$ gives the equation
\eqn\metumunu{\tnab^p ( B_{Ip} -\partial_p K_I) = \delta_I e^{-4A} + {1\over 2} e^{-4A} \delta_I {\tilde g}\ . \quad\quad\quad(\mu\nu1)}
We can think of this equation as determining the metric compensators $K_I$.  Likewise, including the contribution of \threestmunu\ and trace-reversing, we find from the coefficient of $\eta_{\mu\nu}$
\eqn\meteta{\eqalign{&u^I \delta_I \left[-{1\over 4} \tnab^2 e^{-4A} + 4 (\widetilde{\nabla A})^2 e^{-4A} -{1\over 4} e^{-12A} (\widetilde{\nabla \alpha})^2   \right] - {1\over 4} e^{-8A} \delta R^{(4)}
\cr & + \sq u^I e^{-8A} \left\{ -{1\over 4} \tnab^m\left[e^{4A}(B_{Im} - \partial_m K_I) \right]
+ {1\over 8} \delta_I {\tilde g} + \hf e^{-4A} (S_{Im}+ K_I \partial_m \alpha) \partial^{\tilde m} \alpha
\right\}
\cr &=u^I\delta_I \left[ {1\over 48 {\rm Im} \tau} e^{-6A} G\cdot{\bar G} - {\kappa_{10}^2\over 8} e^{-6A} \left(T_\mu^\mu-T_m^m \right)^{\rm source}
\right]\quad\quad\quad (\mu\nu2)}}
which we can think of as an equation fixing $\delta_I A$.

Finally, the $(mn)$ Einstein equation follows from \pgmn\ and \pdeltmn:
\eqn\Einsteinmn{\eqalign{ &u^I\delta_I \left\{ {\tilde G}_{mn} + 4 (\widetilde{\nabla A})^2 {\tilde g}_{mn} - 8 \nabla_m A \nabla_n A + \hf e^{-8A} \left[\partial_m \alpha \partial_n \alpha - \hf {\tilde g}_{mn} (\widetilde{\nabla\alpha})^2 \right] \right\}\cr& + e^{-4A} \sq u^I \Biggl\{ -\hf \delta_I {\tilde g}_{mn} + {\tilde g}_{mn} \left(-2 \delta_I A + \hf \delta_I {\tilde g} \right) + \hf\tnab_m\left[e^{4A} (B_{In} - \partial_n K_I)\right] \cr &+ \hf \tnab_n \left[e^{4A} (B_{Im} - \partial_m K_I)\right] - \tgmn e^{2A} \tnab^p \left[e^{2A}(B_{Ip} - \partial_p K_I)\right] \cr &- {e^{-4A}\over 2} \left[ S_{In} \partial_m \alpha + \partial_n \alpha S_{Im} - \tgmn S_{Ip}\partial^{\tilde p} \alpha + 2 K_I \left(\partial_m\alpha \partial_n \alpha - \hf \tgmn (\widetilde{\nabla \alpha})^2\right)\right]
\Biggr\}\cr & - \hf \tg_{mn} e^{-4A} \delta R^{(4)} \cr &= \kappa_{10}^2 \left(\delta T_{mn}^{(3)} + \delta T_{mn}^{\rm loc}\right)\ .\quad\quad\quad (mn)}}
This equation determines $\delta\tgmn$.

\subsec{Massless perturbations: $G_3=0$}

We begin by outlining the solution of these equations before $G_3$ flux is turned on.  In this case, which corresponds to perturbations of the  Chan-Paul-Verlinde solutions\refs{\CPV}, we know we should find massless perturbations corresponding to the complex structure moduli and Kahler moduli.

The equations simplify considerably when $G_3=\sq u=0$. Moreover,
here $\delta G^{(4)\mu}_\nu=0$, and the perturbation of $\tau$
decouples and is constant in $y$, as seen from eqn.~\taulin. In
this case, \newfive\ gives
\eqn\fivesimp{\delta_I \left[\tnab\cdot \left(e^{-8A} \tnab \alpha
\right)\right] = \delta_I \bigg[ 2 \kappa^{2}_{10} e^{-6A}
T_3 \rho^{loc}_{3} \bigg] \ .}
Dividing this by four, subtracting the result from the Einstein equation \meteta, and writing the result in terms of the quantity
\eqn\adef{a= \alpha - e^{4A}}
 gives
\eqn\aeqn{\tnab^2 a + e^{-4A} (\widetilde{\nabla a})^2 =0\ ;}
note that the terms from local sources cancel.
The integral of this over the compact manifold implies that $a$ must be a constant, which can be set to zero.

The Einstein equation \meteta\ then becomes
\eqn\redEins{ \delta_I \left( \tnab^2 e^{-4A} \right) = {\kappa_{10}^2 \over 2} \delta_I \left[ e^{-6A} \left( T_\mu^\mu - T_m^m\right)^{\rm source}\right]\ }
where here $T^{\rm source}$ is due to D3 branes and O3 planes (or more generally D7's as in \GKP).
This determines $\delta e^{-4A}$ in terms of the metric and source variation.  However, this equation is unchanged if
$\delta e^{-4A}$ is shifted by a constant.  The constant is determined by the integral of the Einstein equation \metumunu\ over the compact manifold, which tells us that
\eqn\nowarpch{0 = \int d^6y \sqrt\tg \left(\varI e^{-4A} + \hf e^{-4A} \varI \tg\right) = \varI \int d^6y \sqrt \tg e^{-4A}\ .}
Note that this is precisely the condition that the warped volume given by \vwdef\  be unchanged by the perturbation.

The $(mn)$ Einstein equation is then easily seen to be solved as long as
\eqn\zmeqn{\delta_I {\tilde G}_{mn}=0\ ,}
which is the equation for a zero mode and is satisfied for complex structure and Kahler deformations of the metric.
Finally, the five-form equation \fivered\ together with the $(m\mu)$ component of Einstein's equations, \mmuein,  are then coupled equations determining the compensators $S_I$ and $B_I$.  Given the resulting $B_I$, the Einstein equation \metumunu\ then determines the compensator $K_I$.

Notice in particular that these compensator equations are not in general consistent with the usual transverse gauge for deformations, \metgauge, together with vanishing of the metric compensators,
\eqn\vcgauge{K_I=B_I=0\ .}
In other
words, transverse gauge \metgauge\ and vanishing-compensator gauge
\vcgauge\ are in general {\it distinct gauge choices.}

\subsubsec{Spacetime-dependent universal Kahler perturbation}

These points are illustrated in the case of the universal Kahler deformation, where the equations simplify somewhat.  Specifically, begin with the metric variation
\eqn\Kahvar{\delta \tg_{mn}  = - \tg_{mn}\ .}
Then the warp-factor equation \redEins\  has solution
\eqn\warpsol{\delta e^{-4A} = 2  e^{-4A} + k}
for any constant $k$.  The constant is fixed by \nowarpch, which implies
\eqn\kfix{ k =  {\int d^6y \sqrt \tg e^{-4A} \over \int d^6y \sqrt \tg}\ .}
One can then easily see that equations \fivered\ and \mmuein\
are satisfied with $S=B=0$.
Then \metumunu\ gives
\eqn\newoneb{ \tnab^2 K =  e^{-4A}- {\int d^6y \sqrt \tg e^{-4A} \over \int d^6y \sqrt \tg}  \ .}
This equation fixes the non-zero compensator $K$.

\subsubsec{Spacetime-independent finite universal Kahler deformation}

One can easily check that these deformations integrate to a finite form
\eqn\finiteK{ ds^2 = \lambda(c) \left[c+e^{-4A_0}\right]^{-1/2} \eta_{\mu\nu} dx^\mu dx^\nu + \sqrt{c}\left[ 1+ e^{-4A_0}/c\right]^{1/2} g^{0}_{mn}dy^m dy^n\ ,}
in terms of finite modulus parameter $c$,
where $g^{0}$ is a fixed background Calabi-Yau metric, and $e^{-4A_0}$ is a solution of \warp\ in that metric.  The quantity $\lambda(c)$ in this equation is precisely that defined in eqn.~\leins, which was used to convert to Einstein frame.

Note that  $c\rightarrow \infty$ is the infinite volume limit of the compact space, and in this limit the metric \finiteK\ becomes
\eqn\largec{ ds^2 \rightarrow {1\over c^{3/2}} dx_4^2 + c^{1/2} g^0_{mn} dy^m dy^n\ .}
Thus, in this limit, the Kahler modulus $c$ and volume of the compact space are related as
\eqn\cvlim{V\rightarrow c^{3/2} V^0\ .}

\subsubsec{Traceless metric perturbations}

The remaining zero modes are the solutions of eq.~\zmeqn\ that are not proportional to the metric.  In terms of the background derivatives, this equation takes the form
\eqn\Licheqn{ -\hf \tnab^p\tnab_p \delta_I \tgmn - \hf \tnab_m\tnab_n\left(\tg^{pq}\varI \tg_{pq}\right) + \hf \tnab^p \tnab_m \varI \tg_{np} + \hf \tnab^p \tnab_n \varI \tg_{mp} =\hf{\cal L} \varI \tgmn =0\ ;}
where $\cal L$ denotes the Lichnerowicz laplacian.  This is conveniently analyzed in the gauge
\eqn\gswgauge{\tnab^n \varI\tgmn - \hf \tnab_m \varI \tg =0\ .}
For Ricci flat $\tgmn$, one can readily show that this gauge choice is equivalent to transverse gauge, \metgauge, and consequently
\eqn\traceless{\tnab_m \varI\tg =0\ .}
Thus, in transverse gauge, the complex structure and non-universal Kahler deformations can be taken to be traceless.

As argued above, $\alpha= e^{4A}$.
Then the warp factor equation \redEins\ determines
\eqn\twarp{\delta_I e^{-4A} = \gamma_I(y) + k_I\ ,}
where $k$ is a constant and $\gamma$ is written in terms of the Green function ${\tilde G}(y,y')$ for the scalar laplacian $\tnab^2$ (defined as in \grdef),
\eqn\gammadef{\gamma_I(y) = \int d^6 y' \sqrt \tg {\tilde G}(y,y') \delta_I\tg_{mn} \tnab^m\tnab^n e^{-4A}\ .}
Eq.~\nowarpch\ then gives $k_I$,
\eqn\twarpt{\delta_I e^{-4A} = \gamma_I - {\int d^6y \sqrt \tg \gamma_I\over \int d^6 y \sqrt \tg}\ .}

The five-form equation \fivered\ then becomes
\eqn\newfiveform{d \left[ e^{-4A} \tstar \left( dS_I + B_I d\alpha\right) \right] = - \delta_I \left(\tstar d e^{-4A}\right)\ .}
Moreover, in the transverse gauge \metgauge\ the $(m\mu)$ Einstein equation \mmuein\
can be written in the similar form
\eqn\newmmu{ d \left( e^{4A} \tstar dB_I\right) = -e^{4A} \delta_I\left( \tstar d e^{-4A}\right) + e^{-4A} d \alpha \wedge \tstar dS_I\ .}
This is a coupled set of equations that must be solved for $B_I$ and $S_I$.

As an alternate to transverse gauge \metgauge, we may work in $B_I=K_I=0$ gauge.  In this case the corresponding equations become
\eqn\bzgauge{d \left[ e^{-4A} \tstar \left( dS_I \right) \right] = -
\delta_I \left(\tstar d e^{-4A}\right)\ }
and
\eqn\bztgauge{ \tilde{\nabla}^{p} \varI \tg_{pm} - \nabla_{m}
\varI \tg = -e^{4A} \tstar \left\{ \delta_I\left( \tstar d
e^{-4A}\right) \right\} + e^{-4A}
 \tstar  \left\{d \alpha \wedge \tstar dS_I \right\} .}

To summarize, if we specify a traceless deformation of the metric, $\delta_I \tgmn$, the spacetime-dependent perturbation is given by the metric \gppert\ and five-form \pcpert, with $\delta_I A$ given by \twarpt, $\alpha=e^{4A}$, and the compensators $K_I$, $B_I$, and $S_I$ given by solving the equations \newfiveform, \newmmu, and \metumunu.  We have not yet found a general prescription to solve these compensator equations.  However, we can get a feel for properties of the compensators by solving these equations in a special case.

\subsubsec{Compensator estimates}

While we haven't solved the compensator equations in general, they
can be solved in the toy model of the metric of a stack of D3
branes restricted to finite volume by placing these at the center
of a ball of unit radius in the background fiducial metric.  This
is a good local model for the geometry in the vicinity of an AdS
throat.  This case will in particular illustrate the dependence of
the compensators on the universal Kahler parameter $c$, which is
important in the next section where we investigate massive
perturbations.

Specifically, let $\tgmn$ be the flat metric on the unit-radius ball in six dimensions, and suppose that there are N D3 branes located at the center of the ball.  In this case the warp factor is given by
\eqn\diskwarp{e^{-4A} = c + {{4 \pi N \alpha^{\prime 2}}\over {r^4}}\ .}
Next consider traceless perturbations in
the metric; for illustration, single out the $x$ and $y$
directions on the ball and consider a perturbation $\delta
\tg_{xy}$, which we take to be a constant for simplicity.
We use spherical polar coordinates $r,\theta_i,\phi$, with $i=1,\cdots,4$, and $ 0 \leq \theta_{i} \leq
\pi$, $0\leq \phi<2\pi$, such that the $xy$ plane is parameterized by $r,\phi$, and the metric is
\eqn\coormetric{ ds^2= dr^{2} + r^{2}\left[d\theta_1^2 +
\sin^2\theta_1(d\theta_2^2 + \sin^2 \theta_2(\cdots + \sin^2\theta_4 d\phi^2)\right]\ .}

       In order to solve the compensator equations \newfiveform\
and \newmmu, we first compute  the source term  $ \delta (
\tilde{*} de^{-4A} ) $. The variation in $e^{-4A}$ can be easily
obtained from \diskwarp\ by computing the  change in the radial
distance
\eqn\defa{ \delta e^{-4A} = -\delta \tg_{xy} \sin \phi \cos
\phi { 16\pi N \alpha^{\prime 2} \over r^{4} }\prod_i \sin^2\theta_i\ . }
Combining this with the variation in the hodge dual one obtains
\eqn\sourcechone{\left[ \delta(\tilde{*} de^{-4A}) \right]_{r
\phi \theta_{k} \theta_{l} \theta_{m} } =\left[ \delta(\tilde{*}
de^{-4A}) \right]_{r  \theta_{k} \theta_{l} \theta_{m} \theta_{n} } = 0 }
and
\eqn\sourcechtwo{\left[ \delta(\tilde{*} de^{-4A}) \right]_{
\phi \theta_{1} \theta_{2} \theta_{3} \theta_{4}} =   - \delta \tg_{xy}
\chargeb \Delta\Omega^{(5)} \sin (2 \phi ) \prod_i \sin^2\theta_i }
where $\Delta\Omega^{ 5 }$ denotes the volume element on the five
sphere.  Note that this is independent of $\phi$ in our coordinate
system.

    For the present case, one can argue using symmetries that
the only non-vanishing component of B and S must be in the $r\phi$
direction and their dependence on $\theta_{i}$ must be trivial.
Using such an ansatz  \newfiveform\ yields
 \eqn\rsixone{ -\del_{\phi}
(\Delta\Omega^{5} e^{-4A}r^{3}[ (dS)_{ \phi r } + B_{\phi} \partial_{r}
\alpha ]) = \Delta\Omega^{5} \delta \tg_{xy} \sin (2\phi)  \chargeb\prod_i \sin^4\theta_i\ , }
while \newmmu\ yields
\eqn\rsixtwo{ -\del_{\phi} (\Delta\Omega^{5} r^{3} (dB)_{
\phi r}) = \Delta\Omega^{5} \delta \tg_{xy}\sin (2 \phi)  \chargeb\prod_i \sin^4\theta_i\ , }
 \eqn\rsixthree{ -\del_{r} (\Delta\Omega^{5} e^{+4A}r^{3} (dB)_{ \phi
r}) = - \Delta\Omega^{5} e^{-4A}\partial_{r} \alpha r^{3}[dS]_{ \phi r} }

The compensator solutions are then
\eqn\soluone{ dS_{ \phi r} =   6 \delta \tg_{xy} \cos (2 \phi)
{{ \charge  } \over { e^{-4A} r^{3} } }\prod_i \sin^4\theta_i}
\eqn\soluthree{ B_{\phi} = 0}
and
\eqn\Bsolu{ B_{r} =   3 \delta \tg_{xy} sin (2 \phi ){ \charge
\over r^{3} }\prod_i \sin^4\theta_i \ .}
Note that the compensators are singular at $r=0$.
But these singularities do not appear in the equations of motion
as the compensators are multiplied by a powers of $e^{4A}$ in the
equations of motion, making their contribution finite.

\subsec{Perturbations with $G_3\neq0$}

We next consider warped compactifications with non-vanishing three-form flux, which generically gives a mass to the complex moduli\GKP.  The potential that does this is believed to arise from the Gukov-Vafa-Witten superpotential, although this has not been rigorously derived.  Study of the linearized perturbation spectrum thus gives further information on the problem of deriving the effective potential both due to these fluxes and due to other effects.

Once the deformations receive a finite mass, their form becomes even more complicated.  For this reason, for such massive deformations we can presently only give the solutions in a perturbative expansion.  The parameter governing this expansion is $1/c$, where $c$ is the finite parameter governing a universal Kahler deformation, introduced in \cmetric.  Recall from \radc\ that $c\sim R^4$
where $R$ is the characteristic radius of the compact space.  In the limit $c\rightarrow\infty$, the effects of warping vanish.  Thus including warping involves keeping subleading powers in the $1/c$ expansion.    The $\alpha'$ expansion is also an expansion in $1/c$, but we justify keeping leading terms from warping since they enter in the form $N/c$ where $N$ is a typical D3 charge, which can be large.

Note that, of course, the size of $c$ governs the validity of the
four-dimensional effective theory.  For four-dimensional energies
$E\roughly>1/c\sqrt{\alpha'}$ in the Einstein frame \finiteK\
Kaluza-Klein modes become important\foot{ The conventional
treatment of Kaluza-Klein modes is done in the string frame and
the associated energy scale is $ {1 \over R} $; conversion to the
Einstein frame gives an additional factor of ${ 1 \over R^{3} }
$.}. So our perturbative expansion amounts to taking $c\gg1$ but
finite and studying the physics at four dimensional Einstein
energies
 \eqn\kkvalid{ E \ll 1/c\sqrt{\alpha'} , }
As we will show, the flux-induced masses are of order
$N/c^{3/2}\sqrt{\alpha'}$ , so there is a sensible low energy
approximation where typical Kaluza-Klein effects are negligible
but flux-induced masses are not.  Moreover, for certain physical
quantities, the non-trivial warping can be an important effect.

However, we will find a subtlety in this analysis.  Specifically, if the manifold has significant warped regions, then Kaluza-Klein modes concentrated in these regions will have redshifted four-dimensional masses, as pointed out in \DeGi.  Consider specifically the solutions of \GKP.  These have a bulk region where the warp factor is approximately a constant, attached to an approximately AdS throat, which is then terminated in a smooth geometry at finite warp factor.

Indeed, in the immediate vicinity of the throat, the warp factor is approximately of the form \diskwarp.  Thus the bulk, mouth, and throat regions corresponds to
\eqn\bulkdef{\eqalign{r^4\gg {4\pi  N\over c}\alpha^{\prime2}&\quad\quad{\rm bulk}\cr
r^4\sim {4\pi  N\over c}\alpha^{\prime2}&\quad \quad{\rm mouth}\cr
4\pi N \alpha^{\prime2}e^{4A}_{\rm min}\ll r^4\ll {4\pi  N\over c}\alpha^{\prime2}&\quad \quad {\rm throat}
}}
where as argued in \GKP, the bottom of the throat is determined in terms of flux quanta $K, M$ as
\eqn\minA{ e^{4A}_{\rm min}\sim e^{-8\pi K/3Mg_s}\ .}
The role of these different regions in studying perturbations will become clear in what follows.

\subsubsec{Massive perturbations}

                               For this section we work in 4d units where the
background metric takes the form
\eqn\bkground{ ds^{2} = [ e^{-4A_0}+ c ]^{-\hf} \eta_{\mu \nu}
                         dx^{\mu} dx^{\nu} + [ e^{-4A_0} + c
                         ]^{\hf} \tg_{mn} dy^{m} dy^{n}\ .}
Recall that this differs from the 4d Einstein frame metric of
\finiteK\ by the change of 4d units \leins\ which has to be
performed on our estimates to obtained the canonical 4-d masses.
We also set $\alpha'=1$ for the rest of this appendix.

 In this set up, compactifications of different volume are
obtained by changing $c$ and $ \tg_{mn} $ is taken to be a CY of
${{\it{unit}}}$ volume. We shall describe the massive
perturbations in the presence of flux in terms of eigenfunctions
the Lichnerowicz operator $\cal L$ on this CY, defined in eq.~\Licheqn,
\eqn\evalue{{\cal L} \delta \tg_{mn}  =\lambda^{2} \delta \tg_{mn}\ . }
 The spectrum starts from $ \lambda =0$,
 corresponding to the zero modes,  and is followed by eigenvalues with
spacing of the order of unity corresponding to the KK modes. We
emphasize the fact that these eigenvalues do not depend on $c$.

             As discussed earlier, the fluctuation of the internal
Calabi-Yau, corresponding to a on-shell solution is determined by
the Einstein equations \mmuein-\Einsteinmn\ . We shall see that in order to satisfy
the equations in the presence of flux, is not consistent to excite
just the zero modes: the Kaluza-Klein modes also have to be excited.

Specifically, consider the (mn) Einstein equation \Einsteinmn.   The universal Kahler deformation remains massless in the presence of flux\GKP, so focus on traceless deformations of the metric.  Ignoring contributions from motion of localized sources (which can be treated separately), the variation in the source term on the RHS of the (mn) Einstein equation
follows from \threestress:
\eqn\threeflux{ \varI T^{ (3) }_{mn} ={ 1 \over 8 \kappa_{10}^{2}
 } \varI \left[ {e^{4A}\over {\rm Im} \tau} \left( G_{mpq}
\widetilde{\overline{G}}_{n}^{\ \ pq}  +G_{npq}
\widetilde{\overline{G}}_{m}^{\ \ pq} - { 1 \over
 3 } \tg_{mn} G\cdot  \widetilde{\overline{G}}  \right) \right] \ .}
This expression vanishes for variations that leave the flux
imaginary self-dual or anti-self-dual, and we will further discuss
this zero mode space shortly.  But the I(A)SD conditions
generically fix the complex structure moduli, and so small
oscillations of them will be massive.  For a small variation about
the I(A)SD point, we find
\eqn\threefluxt{ \varI T^{ (3) }_{mn} ={ 1 \over {8 \kappa_{10}^{2}} }   e^{4A} \varI \left[ {1\over {{\rm{Im}}} \tau} \left( G_{mpq}
\widetilde{\overline{G}}_{n}^{\ \ pq}  +G_{npq}
\widetilde{\overline{G}}_{m}^{\ \ pq} - { 1 \over
 3 } \tg_{mn} G\cdot  \widetilde{\overline{G}}  \right)\right]  \ }
to be a non-vanishing source for the Einstein equation \Einsteinmn.

As in the massless case, the compensators $B_I, K_I$, and $S_I$ are fixed by other equations of motion, and \Einsteinmn\ can be thought of as an equation for $\delta\tgmn$.  The perturbation has a piece proportional to the massless solution with $G_3=0$, but also in general can have a Kaluza-Klein contribution.  Specifically, for a zero mode, the first line of \Einsteinmn\ vanishes, but without \threefluxt\ obeying special conditions we are then not guaranteed that the LHS of \Einsteinmn\ has the correct structure to match this source.  Indeed, we see that it cannot in general match, since the functional dependence from powers of $e^{4A}$ is different.  Thus in general the first line should not vanish, implying that Kaluza-Klein modes are excited.

We therefore decompose the perturbation into a zero-mode piece and a Kaluza-Klein piece,
\eqn\perturbdecomp{\delta \tgmn = \delta^0 \tgmn + \delta^{KK} \tgmn\ ,}
and likewise for the other fields $A,\alpha$ and compensators, where the $\delta^0$ pieces are solutions of the $G_3=0$ equations of the preceding subsection.

Let us now consider the large $c$ behavior of the various terms in \Einsteinmn.  If the total three-brane charge carried by the flux is $N\sim MK$ (where $M$ and $K$ are individual flux quanta as defined in \thrquant),
the source term of \threefluxt\ behaves as
\eqn\gtest{ \varI T^{ (3) }_{mn} \sim {N \over {\rm Im} \tau \kappa_{10}^2  c} \ .}
This must balance the terms on the LHS of \Einsteinmn.  The compensators can be estimated from the results of the previous section.  The relevant quantities are all of the same magnitude:
\eqn\compests{S\sim e^{4A}B \sim e^{4A}K \sim { Ne^{4A}\over r^3}\ .}
Their dominant behavior is at the mouth, and so we find
\eqn\compbds{S\ ,\  e^{4A}B\ ,\  e^{4A}K \ ,\ { Ne^{4A}\over r^3}\ \roughly<\  r_{\rm mouth} \sim \left({4\pi  N\over c}\right)^{1/4} .}
The compensator contributions are thus suppressed at large $c$.

The magnitude of the Kaluza-Klein excitations can then be read off from \Einsteinmn\ as they are needed to compensate for the other terms in the equation.
For a perturbation with four-dimensional momentum $p$, these source terms take the form
\eqn\KKsource{\delta {\tilde G}_{mn} \sim -e^{-4A} p^2 \left(\delta^0 \tgmn + {\rm compensators}\right) + \delta T^{(3)}_{mn}
\ .}
If warping is only moderate, we therefore see that the source and hence the Kaluza-Klein perturbations are of order
\eqn\kkneg{ \delta^{KK} \tgmn \sim cp^2 +\calo\left({N\over c }\right)\ .}
These are small if  the momentum and flux-induced mass are much less than the usual Kaluza-Klein mass.  But for large warping, the first term in \KKsource\ can become large; one way to understand this is that the Kaluza-Klein masses can have large redshifts, as emphasized for example in \DeGi.  Thus the condition for the Kaluza-Klein perturbation to be small becomes more stringent.
Note that $\delta A$ and $\delta \alpha$ can also acquire Kaluza-Klein parts of comparable magnitudes.

We can now read off the leading order flux-induced mass matrix and the magnitude of corrections to it.
We do this by multiplying \Einsteinmn\ by $\delta_J^0\tgmn$ and integrating over the internal manifold.  The KK modes are orthogonal to the zero mode under the resulting metric.  The compensator contributions and KK contributions are subleading to the leading answer, which is
\eqn\massform{\eqalign{ -\hf &\int d^6 y \sqrt{\tilde g}e^{-4A}
\delta_I^0 \tgmn \delta_J^0 {\tilde g}^{mn}\  \sq u^I=\cr & {1\over 4
} \int d^6y \sqrt{\tilde g}e^{4A}  \delta_J^0 {\tilde
g}^{mn} \delta_I^0\left[{1\over {\rm Im } \tau} \left( G_{mpq} \widetilde{\overline{G}}_{n}^{\ \
pq}   - { 1 \over
 6 } \tg_{mn} G\cdot  \widetilde{\overline{G}}  \right)\right]
  \ . }}
Specifically, one can see from the above compensator estimates that the integrated  compensator corrections from \Einsteinmn\ are of relative order $N/c$.  This is also true for the corrections from Kaluza-Klein modes in the case of moderate warping.  However, in the case of large warping, there can be larger and even divergent corrections.  For example, these can arise from the term
\eqn\warpcorr{\sq u^I \int d^6y \sqrt{\tg} e^{-4A} \delta_J^0 \tg^{mn} \delta_I^{KK} \tgmn\ .}
For a warp factor of the form \diskwarp, eq.~\KKsource\ implies a {\it divergent} source for the Kaluza-Klein part of the perturbation.  Correspondingly, we can estimate divergent behavior
\eqn\divkk{\delta^{KK} \tgmn \sim {Np^2\over r^2}\ .}
This leads corrections of fractional size
\eqn\KKcontrib{  {N^2p^2\over c} \ln r_{min}}
for the contribution \warpcorr, where $r_{min}$ is a cutoff value of $r$ corresponding to the bottom of the throat.  If the throat is infinite, as in \CPV, this correction is truly divergent.  For the solutions of \GKP, the minimum is at a small $r$ determined by \minA.  Correspondingly, we find a correction of relative order
\eqn\GKPcorr{ {N p^2\over c} {K^2\over  g_s}\ ,}
enhanced by $K^2/g_s$.

 Eq.~\massform,  then, gives us the flux-induced mass matrix.  The warp factor in these expressions takes the form exhibited in \bkground, and thus to leading order simply gives a power of $c$.  Subleading corrections from the warp factor in \massform\ are unfortunately also of order  $N/c$, and thus cannot be distinguished without knowing the KK and compensator contributions.
  The flux-induced masses have magnitude
 \eqn\massmag{m^2_{G,0} \sim {N \over c^2 }}
 in the units of eq. \bkground.  Working in four-dimensional Einstein units, we thus find that the masses have magnitude
 \eqn\massmagt{m^2_{G} \sim {N \over c^3 }\ .}

The Kaluza-Klein modes of the axidilaton $\tau$ are also excited,
as seen from eqn.~\taulin.  We can project out the zero mode by
multiplying that equation by $e^{-2A}$ and integrating, with the
result
\eqn\tauzero{ \int d^6y \sqrt{\tg} e^{-4A} \delta_I \tau \sq u^I = -{i\over 6} u^I \int d^6y \sqrt{\tg} \delta_I \left(e^{4A} G^+ {\tilde \cdot}G^-\right)\ }
giving the $\tau$ block of the mass matrix.  Note that to leading
order in the Kaluza-Klein expansion, $\delta_I\tau$ is simply a
constant.

 \subsubsec{Kahler, super, and scalar potentials}

 In order to investigate the structure of the landscape of string vacua, it would be useful to have a derivation of the four-dimensional effective action describing these fluctuations.
In principle one might have expected the effective action
describing the fluctuations $u^{I} (x)$ to be obtained by
evaluating the quadratic action $ S_{IIB} $ for a field
configuration where the fields $u^{I}(x)$ take on off-shell
values, i.e. $ \sq u(x) \neq m^{2} u(x) $, while all other fields
are restricted to their on-shell values. As we discussed in the
previous section, on-shell solutions involve compensators and
Kaluza-Klein modes, which would therefore contribute to the
effective action.

      Unfortunately there is a subtlety in carrying this procedure out in the case of type IIB,
due to the difficulty of providing an off-shell action formulation for the self-dual five form.  Since contributions from the five-form's dynamics are important, this appears to be an obstacle to evaluating the kinetic term in the effective action.

A heuristic prescription was given in \DeGi, and it may ultimately be possible to generalize and justify this.  A more systematic procedure was given in \refs{\DLS}, and used in \refs{\LMRS} to obtain
the effective action
for a certain class of fluctuations about the $ AdS_{5} \times S_{5} $ background. This procedure
is well suited when the explicit form of the solutions is known as in case of $ AdS_{5} \times
S_{5} $ but it is not clear how to apply it for the present case. We therefore do not evaluate the
effective action explicitly but discuss certain features that are apparent form the structure
the equations of motion.

As emphasized above, effects of warping in \massform\ enter at the same $N/c$ order as other neglected terms, and so can't be checked directly here.
An effective action
neglecting effects of warping was given in \GKP, and a proposal
for modifications due to warping was given in \DeGi. Specifically,
the superpotential there was the unmodified Gukov-Vafa-Witten potential \gvwpot, and warping contributions to the Kahler potential were suggested to produce a kinetic action of the form \KDG.
These certainly agree with \massform\ at  leading order in $N/c$.  Neglecting warping, the kinetic term in \massform\ is the expected form, and the term
proportional to $u(x)$ is equal to
 \eqn\pote{ -{ 1 \over 12 } \delta_{I} \delta_{J}\int d^6y\sqrt{ \tilde{g} } { { G_{mnp}
 \widetilde{\bar{G}}^{mnp} } \over { {\rm{Im}} \tau} } }
which as in the appendix of \GKP\ can be shown equal to
\eqn\gisd{ -
\delta_{I} \delta_{J} \int {G^{+} \wedge {\tilde *} \overline{G}^{+}\over  {\rm{Im}} \tau } \ .}
This is in keeping with the fact that the mass matrix should
be equal to the  second derivative of the potential.  It is tempting to also trust the powers of $e^{-4A}$ in these expressions, especially in light of the agreement between the expression \vgkp\ and the
potential derived in \DeGi\ from the Kahler potential with warping corrections.

Despite this, for the same reason we cannot directly check the
warping dependence in the Kahler potential. Indeed,  the first
non-trivial correction due to warping in \KDG\ is also of order
${\cal O}\left({N\over c}\right)$   and is associated with the
factor of $e^{-4A}$ in the kinetic terms of  \massform.
Supersymmetry imposes strong constrains on the superpotential --
and calculations of the gravitino mass in \DeGi\ provide another
check on it -- so we expect that the subleading corrections to the
scalar potential do indeed arise due to corrections in the Kahler
potential.  But to check corrections to the Kahler potential at
this order, we need to check all ${\cal O}\left({N\over c}\right)$
corrections. The corrections may take the form of explicit
corrections to the potential $K$ or corrections to the definition
of holomorphic coordinates. It is likely that the latter
corrections are present as  the excitations involve a Kaluza-Klein
part, and it is unlikely that the holomorphic coordinates are
completely defined by the zero-mode directions. The fact that the
holomorphic coordinates can depend on the warp factor was also
seen in section 6.3 in case of Kahler moduli.

In summary, we have learned that there are corrections to the Kahler potential of order
 ${\cal O}\left({N\over c}\right)$, but we are not yet able to directly calculate them.  On the other hand, the superpotential appears uncorrected by warping.

\subsubsec{Moduli with $G_3\neq0$}

Non-zero three-form flux therefore provides a potential for the complex structure deformations.
The remaining flat directions can be inferred from the equation for the minimum of the potential.
As discussed in the main text there is a  $h^{1,1}$ dimensional moduli space which can be
parameterized by the values of the Kahler moduli $ \rho^{i} $.  

A priori, zero mode deformations could arise as a general 
linear combination of the complex structure and Kahler deformations, together with a variation of the dilaton,
\eqn\dfr{ (\rho^i, z^{\alpha},\tau ) \to (\rho'^{i}, z'^{\alpha},\tau' ) = (\rho^i +
\d \rho^i , z^{\alpha} + \d z^{\alpha}, \tau + \d \tau )\ . }
Such a deformation will solve the  $(mn)$ Einstein equation \Einsteinmn\ if $\delta_I T^{(3)}_{mn}$, given by \threefluxt, vanishes.  For given flux quanta \thrquant, this is satisfied by deformations 
which leave the flux ISD.

We are therefore seeking the deformation space such that $G_3$, which is closed (we here consider only constant $\tau$) remains ISD, \ISD.  Notice that such a closed ISD form is automatically harmonic.
Since the hodge dual in the ISD condition \ISD\ changes under a change of Kahler structure, one might be concerned that the ISD condition is no longer satisfied for a pure Kahler deformation, \dfr\ with $\delta z^\alpha = \delta \tau=0$.

In order to show that pure Kahler deformations do leave the ISD condition unchanged,\foot{We thank G. Moore and D. Morrison for providing crucial elements of the following argument.} we begin by recalling some properties of the basic harmonic forms on a Calabi-Yau manifold.\foot{These properties then translate into corresponding statements for an orientifold of the Calabi-Yau.}  Given a complex structure, we can always find a holomorphic $(3,0)$ form $\Omega$, whose construction is independent of a Kahler structure.  Moreover, through the relation, due to Kodaira (see \CandelasPI), 
\eqn\domega{{\partial \Omega\over \partial z^\alpha} = k_\alpha \Omega +\chi_\alpha\ ,}
where $k_\alpha$ depends on $z^\alpha$ but not on the coordinates of the manifold, one can likewise define closed $(2,1)$ forms $\chi_\alpha$ without making reference to a Kahler structure.  The holomorphic form $\Omega$ is harmonic in any Kahler metric.  The forms $\chi_\alpha$ are not necessarily harmonic, but the harmonic representatives of the cohomology classes they specify can be found by adding an exact piece.  These harmonic forms, which we denote $\chi_\alpha^\rho$, do depend on the choice of Kahler moduli, and we have
\eqn\harmto{\chi_\alpha^\rho = \chi_\alpha + d \gamma_\alpha^\rho}
for some $\rho^i$ dependent two-forms $\gamma_\alpha$.  Moreover, on a Kahler manifold, the forms $\chi_\alpha^\rho$ are also $(2,1)$.  

Now begin with a configuration such that $G_3$ is ISD, and consider changing the Kahler moduli to $\rho^{i\prime}$.   In the new Kahler structure, $G_3$ is no longer necessarily ISD nor harmonic.  However, we can always find a harmonic form $G_3'$ in the same cohomology class,
\eqn\gtprime{G_3' = G_3 + d  A_2\ .}
To prove that $G_3'$ is also ISD, note that ${\bar \Omega}, \chi^\rho_\alpha$ form a basis for the harmonic ISD forms, and likewise their complex conjugates a basis for imaginary anti-selfdual forms.  Thus, $G_3'$ will be ISD if 
\eqn\isdcond{\int {\bar \Omega} \wedge G_3'=0\quad\quad ,\quad\quad \int \chi_\alpha^{\rho'}\wedge G_3' =0\ .}
The first condition follows immediately from \gtprime\ and integration by parts, together with the statements that $G_3$ is ISD and $\Omega$ is closed.  
Moreover, from  \harmto\ we see that 
\eqn\harmtoo{\chi_\alpha^{\rho'} = \chi_\alpha^\rho + d (\gamma_\alpha^{\rho'} - \gamma_\alpha^{\rho})\ .}
Thus
\eqn\twoonecond{\int \chi_\alpha^{\rho'} \wedge G_3' = \int \chi_\alpha^{\rho}\wedge G_3}
also likewise follows, and vanishes since $G_3$ is ISD.  So $G_3'$ is also ISD, demonstrating that changes of Kahler moduli with fixed complex structure and dilaton $\tau$ correspond to the $h^{1,1}$ flat directions.  Note that this argument relies on the underlying manifold being Kahler, in order to argue that the $\chi_\alpha^\rho$ are of the correct type.

\appendix{B}{Relation to other work}

We have seen from our discussion that the derivation of the Kahler and
scalar potentials given in \GKP\ and \DeGi, while giving the
correct results to leading order, were somewhat heuristic. In addition to these there has been considerable other work in the literature devoted to studying such potentials, and time dependent warped solutions.  The correct description of the dynamics of moduli has produced some confusion, and
we believe our work has now clarified some of that confusion.

In particular, \refs{\Dealwisone,\Dealwistwo} raised objections to
previous derivations of the potential \vgkp.  In \Dealwisone, one
objection was that the potential derived there was negative
definite.  However, the definition leading to the objection was
the incorrect definition of the potential based on \revtwo, not
\Potentg.  As we have explained, the expression based on \revtwo\
only gives the correct answer in the case where one is at a
minimum of the potential so the moduli are static. However, our
more general expression \Potentg\ is suitable away from a minimum,
and gives a potential agreeing with that found from the derivation
in \DeGi.  Ref.~\Dealwisone\ argued that the problem originated in
time dependence of the radial modulus away from the minimum, and
argued that in the case of non-trivial warping there is no clear
derivation of the potential due to the excitation of all
Kaluza-Klein modes.  While we have found that Kaluza-Klein modes
are excited by general spacetime-dependent perturbations, we have
also found that they are small in a controlled approximation, and
thus make only small corrections to an otherwise unambiguous
potential in the low-energy effective theory.  Our results do
agree with the claim of \refs{\Dealwisone,\Dealwistwo} that in a
generic gauge actual solutions of the equations of motion are not
of the form of the ``factorized Ansatz."

Other works have investigated time-dependence of warped
configurations.  For example, \refs{\chendasgupta} considered
$D3-D7$ inflation models, and attempted to construct solutions
lifted up to ten dimensions.  They recognized that their metric
Ansatz
\eqn\cosmoans{ ds^{2} = e^{A(y,t)} \eta_{\mu \nu} dx^{\mu}
dx^{\nu}  + e^{B(y,t)} g_{mn}(y)dy^{m} dy^{n}\ ,}
was potentially over-restrictive. While it captures some of the
ten-dimensional dynamics of such configurations, our analysis of
the general linearized perturbation shows that the metric for a
ten-dimensional warped cosmology cannot in general be put in this
form.

       Finally, \refs{\Buchel} investigates the problem of deriving  the four-dimensional effective action from ten dimensions.  While expressions similar to our potential formula \gkppot\  are  found, the formalism of this paper is only applicable to deriving the potential at a stationary point and does not yield a solution of the ten-dimensional equations for rolling moduli.  Thus the only known example where these results can be compared gives the value $U=0$.  Moreover, this derivation is based on an incorrect parametrization of the universal Kahler modulus,
       corresponding to \againscale.

\listrefs

\end